\documentclass[usenatbib,usegraphicx,useAMS,usenatbib]{mn2e}

\newcommand{\hii}{{H{\scriptsize II} }}
\newcommand{\hi}{{H{\scriptsize I} }}

\newcommand{\kms}{km\,s$^{-1}$}

\usepackage{morefloats}

\title[THOR Pilot OH Survey]{A Survey for Hydroxyl in the THOR Pilot Region around W43}
\author[Walsh et al.]{Andrew J. Walsh$^{1}$\thanks{E-mail:
andrew.walsh@curtin.edu.au}, Henrik Beuther$^{2}$, Simon Bihr$^{2}$, Katharine G. Johnston$^{2,3}$,
\newauthor Joanne R. Dawson$^{4,5}$, J\"{u}rgen Ott$^6$, Steven N. Longmore$^{7}$, Q. Nguy\^{e}n Luong$^{8,9}$,
\newauthor Ralf S. Klessen$^{10,11,3}$, Sarah Ragan$^{3}$, Naomi McClure-Griffiths$^{13}$, Andreas Brunthaler$^{14}$,
\newauthor James Urquhart$^{14}$, Karl Menten$^{14}$, Frank Bigiel$^{10}$ Friedrich Wyrowski$^{14}$
\newauthor and Michael Rugel$^{2}$
\\
$^{1}$International Centre for Radio Astronomy Research, Curtin University, GPO Box U1987, Perth WA 6845, Australia\\
$^{2}$Max Planck Institute for Astronomy, K\"{o}nigstuhl 17, 69117 Heidelberg, Germany\\
$^{3}$School of Physics and Astronomy, University of Leeds, Leeds, LS2 9JT, UK\\
$^{4}$Department of Physics and Astronomy and MQ Research Centre in Astronomy, Astrophysics and Astrophotonics,\\
Macquarie University, NSW 2109, Australia\\
$^{5}$Australia Telescope National Facility, CSIRO Astronomy and Space Science, PO Box 76, Epping, NSW 1710, Australia\\
$^{6}$National Radio Astronomy Observatory, P.O. Box O, 1003 Lopezville Road, Socorro, NM 87801, USA\\
$^{7}$Astrophysics Research Institute, Liverpool John Moores University, 146 Brownlow Hill, Liverpool L3 5RF, UK\\
$^{8}$National Astronomical Observatory of Japan, Chile Observatory, 2-21-1 Osawa, Mitaka, Tokyo 181-8588, Japan\\
$^{9}$EACOA Fellow at NAOJ, Japan \& KASI, Korea\\
$^{10}$Universit\"{a}t Heidelberg, Zentrum f\"{u}r Astronomie, Institut f\"{u}r Theoretische Astrophysik, Albert-Ueberle-Str. 2, D-69120 Heidelberg, Germany \\
$^{11}$Department of Astronomy and Astrophysics, University of California, 1156 High Street, Santa Cruz, CA 95064, USA\\
$^{12}$Kavli Institute for Particle Astrophysics and Cosmology, Stanford University, SLAC National Accelerator Laboratory, Menlo Park, CA 94025, USA\\
$^{13}$Research School of Astronomy \& Astrophysics, The Australian National University, Canberra ACT 2611, Australia\\
$^{14}$Max Planck Institut f\"{u}r Radioastronomie, Auf dem H\"{u}gel 69, D-53121, Bonn, Germany\\
}
\begin{document}

%\date{Accepted 1988 December 15. Received 1988 December 14, in original form 1988 October 11}

%\pagerange{\pageref{firstpage}--\pageref{lastpage}} \pubyear{2002}

\maketitle

\label{firstpage}

\begin{abstract}
We report on observations of the hydroxyl radical (OH) within The
H{\sc I}, OH Recombination line survey (THOR) pilot region. The region
is bounded approximately between Galactic coordinates l=29.2 to 31.5$^\circ$
and b=-1.0 to +1.0$^\circ$ and includes the high-mass star forming region W43.
We identify 103 maser sites, including 72 with 1612\,MHz
masers, 42 showing masers in either of the main
line transitions at 1665 and 1667\,MHz and four showing 1720\,MHz masers.
Most maser sites with either
main-line or 1720\,MHz emission are associated with star formation,
whereas most of the 1612\,MHz masers are associated with evolved stars.
We find that nearly all of the main-line maser sites are co-spatial with
an infrared source, detected by GLIMPSE.
We also find diffuse OH emission, as
well as OH in absorption towards selected unresolved or partially
resolved sites. Extended OH absorption is found towards the well known
star forming complex W43 Main.
\end{abstract}


\section{Introduction}
\subsection{OH masers in star forming regions}
The star formation process in our Galaxy, including how both low- and high-mass stars
are formed, is important to study as it determines how we view galaxies
and how stars and planets are created. Decade long efforts have made significant
progress, although the full details are yet to be
realised. In particular, the formation of high-mass stars is difficult to study
as they form in complex cluster environments that are rare.
Although the formation of low-mass stars either in clusters or in isolation
is more accessible, it is still difficult to define a self-consistent
detailed model that accounts for the wide variety of observational phenomena
associated with the formation of stars of all masses
\citep{maclow04,mckee07,zinnecker07}.\\

In essence, our difficulties in understanding the star formation process are
largely associated with our limited ability to study catalogued examples, each
of which can only tell us about how star formation looks at one specific
location. With a large ensemble of star formation examples to study, there is
the promise that all the examples can be put together to form a coherent
picture.  However, the cataloguing process necessarily introduces biases into
any ensemble, which must be taken into account. Therefore, we need to study the
full population of star forming sites throughout the Galaxy, in order to form
a full picture of the star formation process. Such a grand catalogue is
difficult to realise and so in recent years an alternate approach has been
explored.
This approach is to conduct untargetted surveys of the Galaxy, using many
different Galactic Plane surveys, each of which focus on a particular aspect
relevant to star formation and then complement the data with theoretical
models to help us astrophysically interpret the data. The untargetted approach
allows the observer to
find the full population of objects within their survey, rather than picking
out selected targets from another survey, with its own biases. Many different
Galactic Plane surveys are required in this approach because each untargetted
survey may only observe a select aspect of the star formation process. A
classic example of this is the Infrared Astronomical Satellite (IRAS) point
source catalog \citep{iras}, which proved extremely useful in identifying
all the bright and red far-infrared sources within the Galaxy \citep{wc89} which
are typically associated with high-mass star formation. However, early stages
of high-mass star formation take place in infrared-dark clouds
(IRDCs -- \citealt{rathborne06}) and so
were not detected by IRAS. Thus, IRAS could only be used to identify star
formation sites that were reasonably well evolved, but still embedded in their
natal molecular clouds.\\

Currently, there are many untargetted Galactic plane surveys that are relevant
to star formation. There are continuum surveys from short to long wavelengths
including: the near-infrared (eg. 2MASS -- \citealt{skrutskie06}, UKIDSS --
\citealt{lawrence07}), mid-infrared (eg. MSX -- \citealt{egan96}, GLIMPSE --
\citealt{churchwell09,benjamin03}, MIPSGAL -- \citealt{carey09}), far-infrared
and sub-millimetre (IRAS -- \citealt{iras}, Hi-GAL -- \citealt{molinari10},
ATLASGAL -- \citealt{schuller09}), millimetre (eg.
BGPS -- \citealt{ginsburg13}) and centimetre (eg. CORNISH --
\citealt{purcell13}, VGPS -- \citealt{stil06}, SGPS -- \citealt{haverkorn06}, CGPS
-- \citealt{taylor03}). These continuum surveys allow a wide range of physical
conditions to be probed, from ionised gas to hot gas and dust to cold dust. In
addition to the continuum surveys, there are also spectral line
surveys, which can be used as diagnostics of particular aspects of the star
formation process: CO (eg. \citealt{dame01,jackson06,burton13}), CS (MALT-45 --
\citealt{jordan13}), NH$_3$ (HOPS -- \citealt{purcell13}), OH (eg. SPLASH
-- \citealt{dawson14}) and H{\sc I} (eg. SGPS -- \citealt{mcg05}). In addition
to the above-mentioned spectral line surveys are spectral line surveys of 
strong masers: class I methanol
masers (eg. MALT-45 \citealt{jordan13}), water masers (eg. HOPS --
\citealt{walsh12}), class II methanol masers (eg. MMB -- \citealt{green09})
and OH masers (eg. SPLASH -- \citealt{dawson14}). All these surveys can help
build a detailed picture of the star formation process whilst minimising
inherent biasses in sample selection.\\

\subsection{OH masers and evolved stars}
OH masers are commonly found in the later stages of stellar evolution,
particularly during the asymptotic giant branch (AGB) phase and shortly
thereafter. The AGB phase occurs for stars with masses 1-8\,M$_\odot$ where
the main supply of nuclear fuel (hydrogen) in the core is exhausted. The
outer atmosphere is expelled from the surface of the star in a dense stellar
wind that expands isotropically at typical velocities of 5-30\,\kms, forming
an envelope. The expanding circumstellar envelope can harbour masers from SiO,
H$_2$O and OH, with SiO masers found closest to the star and OH masers found
furthest away, which is probably due to the differing excitation temperatures
of the masing transitions \citep{reid81}.\\

OH masers that occur in the circumstellar envelopes are typified by a
double-horned profile, where the outer edges, showing the strongest maser
emission, are indicative of red-shifted and blue-shifted gas moving behind
and in front of the central star, respectively. Thus, the OH maser profile
may be used to characterise the kinematics of the circumstellar envelope.
This distinctive profile also allows the easy identification of OH masers
associated with evolved stars.\\

Eventually, the star evolves from the AGB stage into a planetary nebula (PN),
through a very short-lived ($\sim\nobreak30 - 10^4$ years; \citealt{blocker95}) post-AGB
phase. During this evolution, the masers that originate in the circumstellar
envelope are extinguished. The SiO masers turn off first, followed by H$_2$O
masers, leaving the OH masers to disappear last. 
Most PNe are observed to be highly asymmetric. It is not entirely
understood how the symmetric AGB star and attendant envelope evolve into an
asymmetric PN. A possible method is that high velocity bipolar jets are formed
during the post-AGB phase that impinge on the nearly-symmetric circumstellar
envelope, where small density variations create the asymmetric PN
\citep{sahai98}. Both H$_2$O and OH masers are seen to trace such jets in
the post-AGB phase (eg. \citealt{imai02,zijlstra01}). These high velocity
jets are exceedingly rare, due to their extremely short lifetimes.\\

Another exceedingly rare class of objects associated with evolved stars are
OHPNe \citep{uscanga12}. These objects show radio continuum emission associated
with a PN, but also show OH maser emission. It is likely that this phase
occurs when the PN phase has just started, with the inner part of the
circumstellar envelope already ionised, but the outer parts harbouring OH
masers.\\

It is therefore important that we increase the numbers of known examples
of these rare, short-lived phases through maser surveys so that we can study
these important phases of stellar evolution.\\

\subsection{A Pilot OH Survey for THOR}
The work described here is part of The H{\sc I}, OH, Recombination line survey
of the Milky Way (THOR).
THOR is a Northern Hemisphere Galactic plane survey that includes continuum
emission, as well as spectral lines of H{\sc I}, radio recombination lines and
the ground-state OH transitions. THOR surveys a large fraction of the Northern
Galactic plane ($l = 15^\circ - 67^\circ, |b|\leq1^\circ$), using the Karl G.
Jansky Very Large Array, with a spatial resolution of about 20\arcsec. Apart
from OH lines, THOR also maps H{\sc I}, continuum in the 1-2\,GHz range
and multiple radio recombination lines. The work reported in this paper focuses
on data from the THOR pilot survey, covering $l = 29.2-31.5^\circ,
|b|\leq1^\circ$ and includes the W43 and W43-South high-mass star formation
complexes. H{\sc I} data \citep{bihr15} and continuum
data (Bihr et al. {\em in preparation}) from the pilot region will be
published elsewhere, as will a full survey description (Beuther et al.
{\em in preparation}). The focus of this work is on OH emission and
absorption. The OH radical has four ground-state transitions around
1.6-1.7\,GHz. All four transitions are known to exhibit strong maser emission,
which is easily detectable in a survey such as THOR. SPLASH is a complementary
survey to THOR. SPLASH concentrates on emission from the hydroxyl radical
(OH) in the Southern
Hemisphere, where this work is part of a Northern Hemisphere survey.
A detailed introduction
on the OH ground-state transitions can be found in the SPLASH pilot paper
\citep{dawson14}.\\

\subsection{A note on maser terminology}
In the literature, some confusion exists over terminology used to describe
various aspects of maser emission. In particular, the term ``maser spot''
has multiple definitions. For example, \citep{walsh01} give a definition in
terms of methanol masers as: ``A methanol maser spectrum typically consists
of many peaks. Each peak corresponds to a well-defined position in the sky,
referred to as a maser spot. Maser spots are usually seen to cluster on a
scale of less than 1 arcsec. These clusters are referred to as maser sites.''
However, an alternate definition is provided by \citet{imai00}, who in
turn cite \citet{gwinn94} as: ``A maser spot is a single velocity component
of the maser emission. A maser feature is a group of the maser spots with
similar positions (within several hundred microarcseconds) and with successive
velocities (with spacings 0.3--3\,\kms). Spots thus reflect somewhat
instrumental factors, especially the velocity spacing of each channel.''\\

The important difference between these definitions is that a maser spot in the
first definition is essentially the same as a maser feature in the second
definition. But there is no equivalent to the maser spot used by \citet{imai00}.
Whilst different definitions may be common in fields of research, it appears
that both definitions of maser spot are in common usage within the maser
community. Furthermore, much work is published where maser spot and
maser feature are used interchangably. Thus there is a strong need to
consolidate terminology amongst researchers to avoid future confusion.\\

The term ``maser hot spot'' was introduced by \citet{litvak71} in response to
high spatial resolution observations by \citet{moran68}, \citet{burke70}
and \citet{johnston71}.
\citet{moran68}, for example, observed OH masers in W3 and found ``knots
of radiation'', corresponding to single peaks in the spectrum, where the
``ratio of separation to size is typically 100 to 1.'' The sizes
of the maser hot spots in these observations were around $10^{14}$\,cm
($\sim10$\,AU), whereas
typical separations were around $10^{16-17}$\,cm. The terminology
``maser hot spot'' by \citet{litvak71} was introduced to highlight that the
masers originated in small regions of the sky where it was thought a single
parcel of gas was responsible for the exceedingly high brightness temperatures
that were measured. This terminology was quickly contracted by subsequent
authors to ``maser spot'' and became common usage. \\

The observations by \citet{gwinn94} identified that a maser spot (using the
definition of \citealt{litvak71}) may break up into substructures that are
found to have slightly (but significant) different positions for each
channel in the data that was analysed. Thus, the terminology was changed to
highlight that a maser feature corresponds to a physical entity that 
typically corresponds to a single peak in a spectrum and a maser spot became
the name assigned to the substructure evident between single channels
of a spectrum and within a single maser feature. However, this new terminology
means that a maser spot is governed by instrumental effects (ie. the
channel resolution), rather than corresponding to a single parcel of gas
at a well defined position that gives rise to a single peak in the spectrum.
Thus, the term maser spot loses its meaning in the context of a hot spot.\\

In this work, we use the term maser spot in keeping with its original
definition as a contraction of maser hot spot which constitutes a single
peak in the spectrum. Such maser spots are expected to
have sizes around $10^{14}$\,cm and are occasionally grouped into regions
with sizes around $10^{16-17}$\,cm, which we call maser sites. However we
note that, given our spatial resolution and typical distances to masers,
maser spots will be unresolved in our work. Indeed, most maser sites will
also remain unresolved in this work. We remind the reader that our definition
of ``maser spot'' is in common usage as either ``maser feature'' or
``maser component'' in other work.\\

\section{Observations and data reduction}
\subsection{VLA Observations}
We mapped a $2^\circ \times 2^\circ$ field around W43 ($l$ = 29.2$-$31.5$^\circ$,
$|b|\leq1^\circ$)
during the 2012A semester with the Karl G. Jansky Very Large
Array (VLA) in New Mexico in C-configuration (Project 12A-161).
The phase centre of the observations was $18^h 46^m 45\fs4,
-02^\circ 16^\prime 19^{\prime\prime}$.
The primary beam at 1667\,MHz has a FWHM of 27\arcmin.
A total of 59 pointings were used to cover the 4 square degree mosaic.
With a broad frequency coverage between 1 and 2\,GHz, the primary beam
sampling is a compromise between sampling density and area coverage. For
the Pilot region detailed in this work, we used a mosaicing scheme consisting
of a hexagonal grid with 17.9\arcmin~spacing. This results in excellent data
quality with very little variation in sensitivity in the inner survey
region. However, for the full survey, we improved the sampling to
15\arcmin~spacing , using a square grid so that the data are Nyquist
sampled at the \hi~line frequency (1.42\,GHz).\\

Each pointing was observed $4 \times 2$ minutes, resulting
in an overall observation time of 10 hours.
The observations were conducted in two blocks each with 5 hours observing
time in April 2012. We chose the quasar 3C286 as a flux
and bandpass calibrator and the quasar J1822-0938 as a complex
gain calibrator, which was observed every 13 minutes. Observing at
L-band and using the new WIDAR correlator, we were able to
simultaneously observe the H{\sc I} line, the 4 OH lines (1612, 1665,
1667 and 1720\,MHz) and 12 H$\alpha$ radio recombination lines.
The continuum, detected in eight spectral windows between
1 and 2\,GHz, was observed in full polarisation. The spectral lines were only
observed in LL and RR polarisations. However, we only present stokes I
information on the OH lines in this work. For the OH
lines, we used a bandwidth of 1\,MHz with a channel width
of 3.906 kHz. This results in a velocity range of about 180\,km\,s$^{-1}$
and a channel spacing of 0.73\,km\,s$^{-1}$ at 1612\,MHz.\\

We use a synthesised beam of $20\arcsec \times 20\arcsec$. With this beam, we
estimate the absolute positional accuracy is
$\Theta_{\rm BEAM} / 2 {\rm SNR}$ \citep{fom99} where $\Theta_{\rm BEAM}$ is the
beam size (20\arcsec) and SNR is the signal to noise ratio. Under typical observing
conditions, we conservatively expect the absolute positional accuracy to be no
better than 10 per cent of the beam (ie. 2\arcsec). With a typical rms noise level of
19\,mJy/beam per 0.73\,\kms~channel (see Section 3.1 for further details),
this means that for any point source stronger than $5\sigma$, the absolute accuracy
will be 2\arcsec. For weaker point sources, the accuracy will be lower and will
follow $\Theta_{rm BEAM} / 2 {\rm SNR}$. We also note that the noise level increases
where there is a nearby strong maser source in the same channel. This is because
the data have a limited dynamic range of about 900 in the presence of strong
sources.\\

\subsection{Calibration}
To edit and calibrate the data, we used CASA
(version 4.1.0)\footnote{http://casa.nrao.edu/} with
a modified VLA pipeline 1
(version 1.2.0)\footnote{https://science.nrao.edu/facilities/vla/data-processing/pipeline}.
The pipeline does automatic flagging for eg. zeros or shadowing of antennas.
Additional flagging for Radio Frequency Interference (RFI) and bad antennas
was done manually. The pipeline also applies the bandpass, flux and
complex gain calibrations to the data. We do not use Hanning
smoothing and do not recalculate the data weights at
the end of the pipeline. We subsequently perform
further flagging and editing by hand. More details on the calibration of the pilot study
are also given in \citet{bihr15}. A full description of our quality check
method will be presented in our forthcoming overview paper
(Beuther et al., {\em in preparation}).\\

\section{Maser Identification}

Maser sites were identified in a similar fashion as the methods described in
\citet{walsh12}. We initially used the {\sc duchamp} software \citep{whiting12}
package to
identify maser candidates. The data cubes contain pixels (5\arcsec) that are
approximately four times smaller than the FWHM size of the restoring beam
(20\arcsec) and the channel
width is 0.73\,\kms. Based on these specifications for the data cubes, we
used {\sc duchamp} to search for emission at twice the rms noise level,
which is typically 19mJy/beam, in the
full cubes, with the following minimum thresholds for detection: no less than
12 pixels in a single channel, no less than 25 voxels and no less than 2
consecutive channels over which the emission was above the 2$\sigma$ level. This
effectively constrains a detection to occur at above the 2$\sigma$ level in each
of as many pixels would fill half a beam for each of two consecutive channels
in the cube. By varying these input parameters and comparing the results, we
found this to be the best compromise to maximise the number of real detections
made, whilst simultaneously limiting the number of false detections. We also
found that, within reasonable limits, the number of real and false detections
did not change greatly when varying the input parameters slightly. This gives
us confidence that our input parameters are reasonably robust.\\

During our initial {\sc duchamp} searches, it became clear that such automated
source finding was limited by the changing noise levels in the data. In
particular, the edges of the survey area had higher noise levels in all
channels, but also, the limited dynamic range (approximately up to 900)
close to very strong
masers significantly increases the noise, in the images containing strong emission.
The effect on {\sc duchamp} results was that many more false detections
would be made. In order to mitigate the effects of changing noise levels,
a noise map was created for each data cube, based on averaging data over
a velocity range where no masers were expected (-110 to -70\,\kms). This
full data cube was divided by the noise map, effectively creating a
signal-to-noise cube, which was then searched for sources in {\sc duchamp}.\\

\begin{figure}
\includegraphics[width=0.45\textwidth]{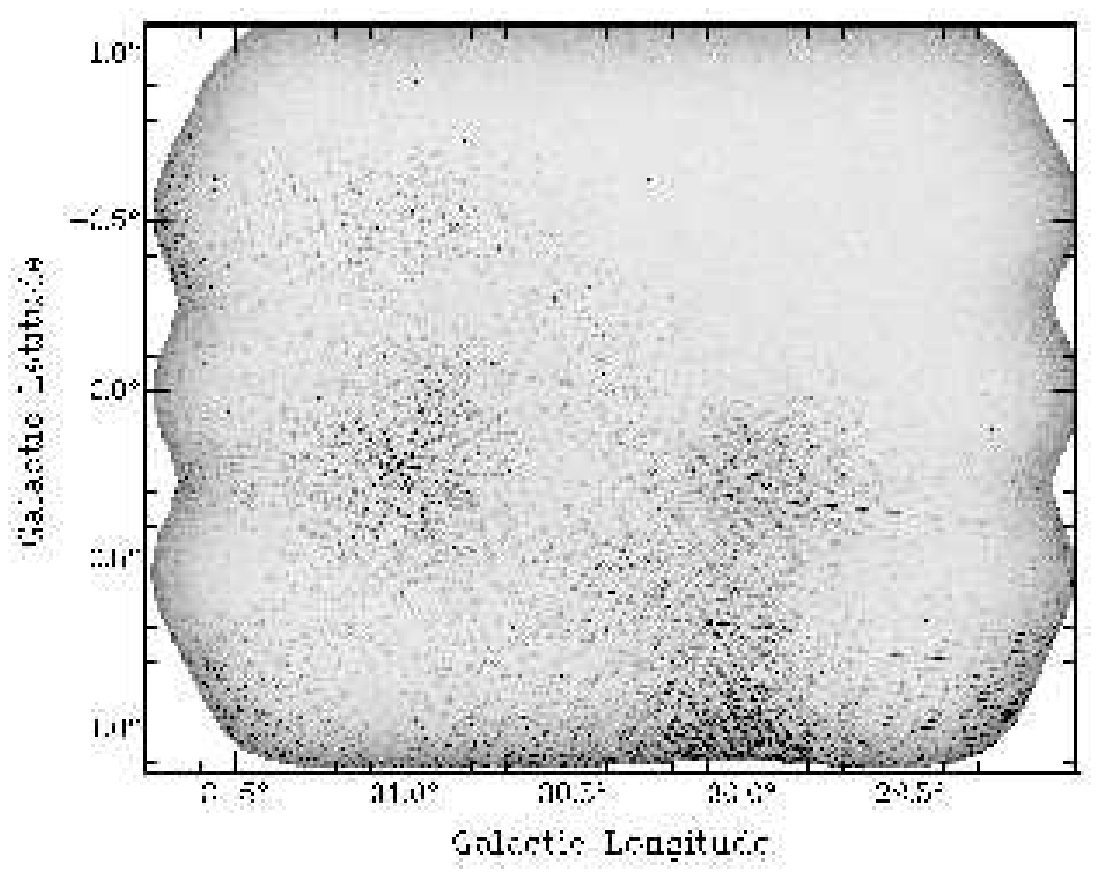}
\includegraphics[width=0.45\textwidth]{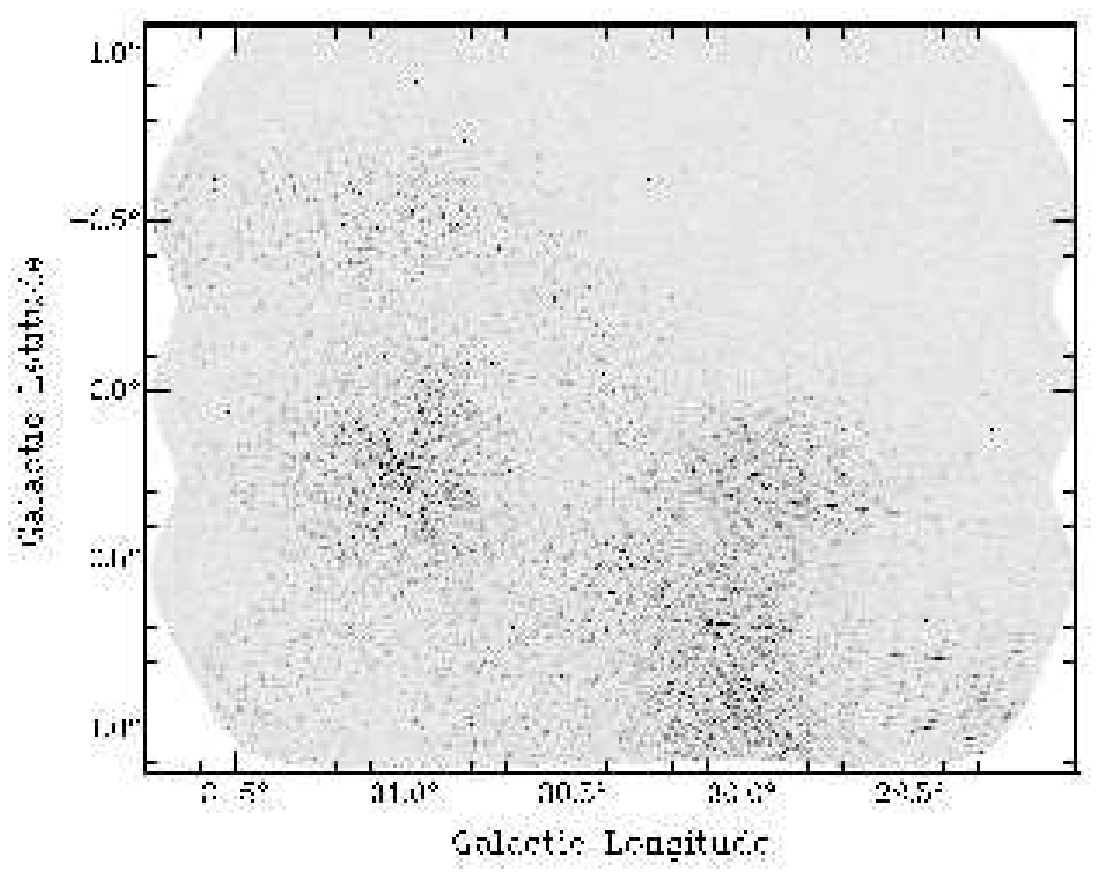}
\caption{The upper panel shows the peak temperature map for the 1612\,MHz
data (as an example). Bright masers can be seen as unresolved dark points. However, the
map is dominated by noisy regions that are associated with the brightest
masers and interference. The edges of the map also show increased noise
levels. The lower panel shows a signal to noise map where the effects of
noise at the edge of the map have been greatly reduced. The effects of bright
masers and interference have also been reduced, but not to such a great extent.
In the full data cube, the effects of bright masers and interference only
manifest in a small number of channels, which can be individually flagged.}
\label{1612noise}
\end{figure}

Figure \ref{1612noise} shows a peak flux density
map for the 1612\,MHz data cube (top) and a signal-no-noise
peak flux density map (bottom). The top map shows increased noise levels
at the edges, as well as sidelobe patterns from the strongest masers, with
the strongest located at G31.102-0.219. The signal-to-noise map shows
improvement, particularly along the edges of the survey area. However, the
sidelobes from strong masers remain. These maser sidelobes occur in only
one or two channels of the data cube. We find that within a primary beam
(about 0.25$^\circ$) of a strong maser, any other potential masers are hidden
within the sidelobes. However, this is only the case for one or two channels
that are strongly affected. At other channels, we are able to search for
potential masers that are close (in the plane of the sky) to a strong
maser, but at a different velocity.\\

In order to confirm detections by {\sc duchamp} as real masers, we visually
inspected each {\sc duchamp} detection. Nearly all false detections can be
identified as sidelobes from nearby strong masers. In addition to identifying
masers through confirmation of {\sc duchamp} detections, we also visually
inspected the data cubes and moment maps, using the HOPSfind method of which
more details can be found in
\citet{walsh12}. This method was particularly successful for the 1612\,MHz
data cube as 1612\,MHz masers, when associated with evolved stars,
often display a characteristic double-horned profile
in their spectra (signifying an expanding shell containing the masing gas),
which is easy to identify by eye, even at low signal-to-noise levels.\\

For each identified maser spot, we use the {\sc miriad} task ``imfit'' to fit
the emission and derive positions and relative uncertainties, as well as peak
flux densities. We derive the velocity minima and maxima for maser spots by
visually inspecting the data cubes and determining over which channels
emission is seen.\\

\subsection{Discussion of Completeness}
The completeness limit of the VLA observations will vary depending on the noise
level. In a region of the data that overlaps in velocity with a bright
maser and is spatially close to the maser, the sidelobe noise level will be high and the
completeness will have a correspondingly low value. Given that OH masers
associated with star formation are likely to be clustered within the same star
formation site, it is also likely that the number of masers
will be slightly underestimated because these high
noise regions surrounding bright masers will contain a higher density of
masers. Therefore, it is not possible to accurately compute a single
completeness level, but it makes more sense to compute completeness levels as
a function of changing noise levels.\\

\begin{figure}
\includegraphics[width=0.45\textwidth]{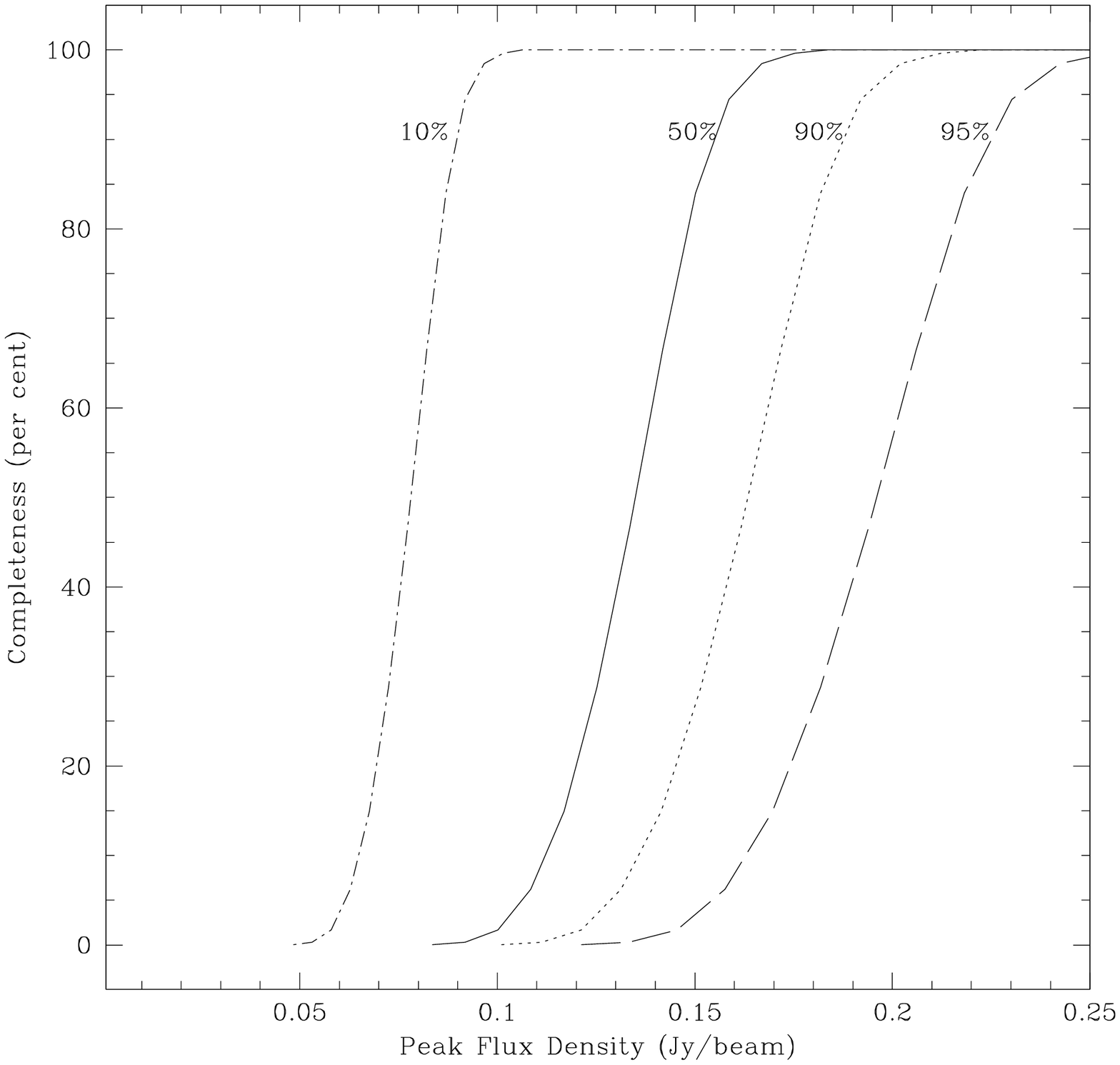}
\caption{Completeness distributions are shown for four representative noise
levels. The dot-dash line shows the distribution where 10 per cent of all the data has a better noise level than 11\,mJy/beam. The solid line shows the distribution where 50 per cent of all the data
has a better noise level than 19\,mJy/beam, the dotted line shows the distribution
where 90 per cent of all data has a noise level below than 22\,mJy/beam and the
dashed line shows the distribution where 95 per of all data has a noise level
below than 27\,mJy/beam.}
\label{complete}
\end{figure}

The noise levels at which 10, 50, 90 and 95 per cent of the data are below this
level are 11, 19, 22 and 27\,mJy/beam per 0.73\,km\,s$^{-1}$ channel, respectively.
We use these noise levels as
representative of the data and apply these noise levels to a Monte Carlo
simulation. In the simulation, we take representative data with noise at each
of these levels and artificially inject point sources into the noisy data. The
point sources have a line width typical of masers: 2\,km\,s$^{-1}$. We then use
our same source-finding methods described above and count the number of
recovered point sources. We can then plot four distributions of the
completeness levels, as shown in Figure \ref{complete}.\\

The completeness distributions show that nearly all data is complete at around
the level of 0.25\,Jy/beam, but that 50 per cent of the data is complete at around
the level of 0.17\,Jy/beam.\\

\section{Results}
\begin{figure*}
\begin{tabular}{ccc}
\hspace{-1.5cm}
\begin{tabular}{c}
~~~~G29.098-0.706~~~~~~~~~~~~~~~~ES\\
\includegraphics[width=0.25\textwidth]{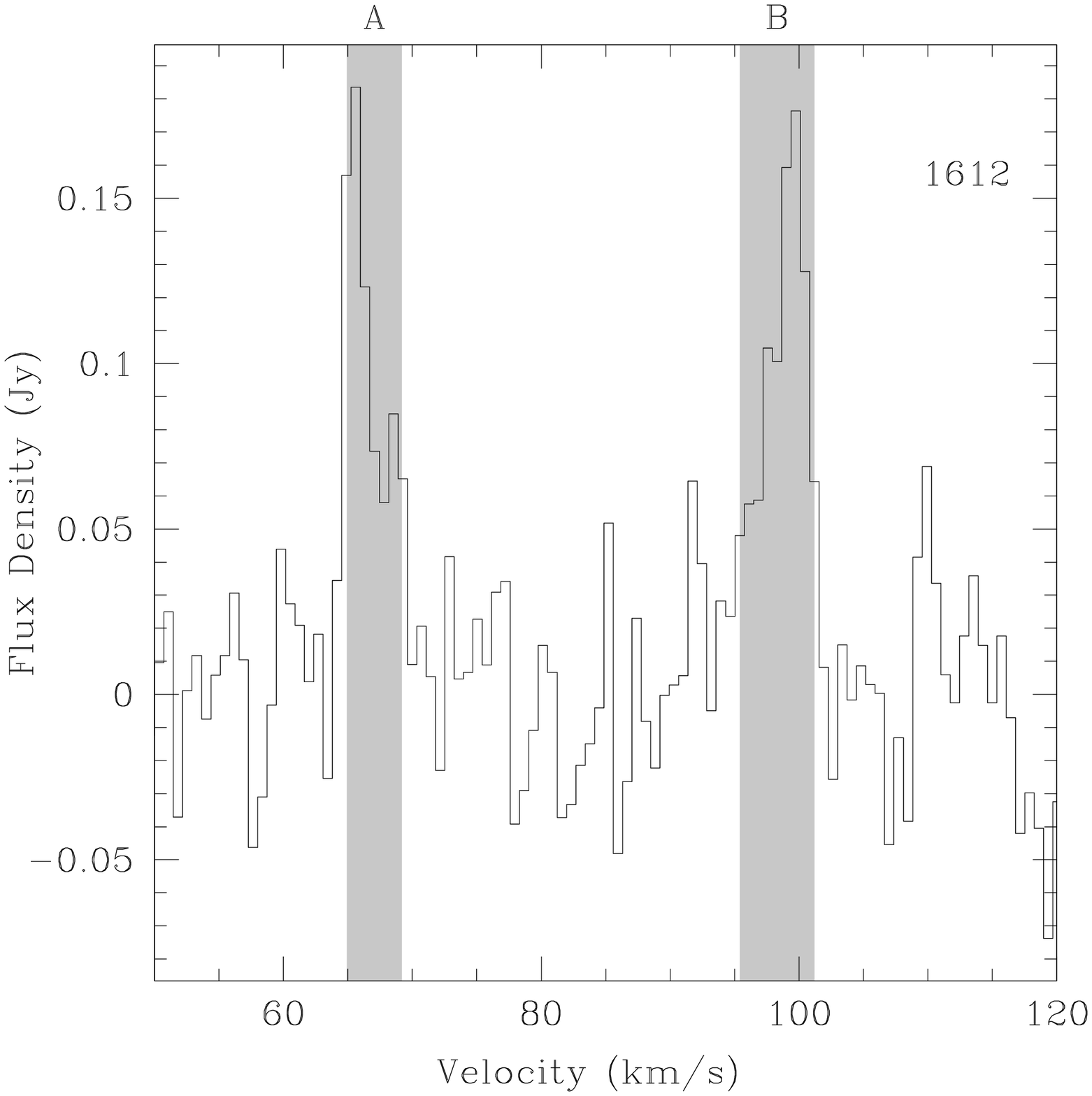}\\
\includegraphics[width=0.31\textwidth]{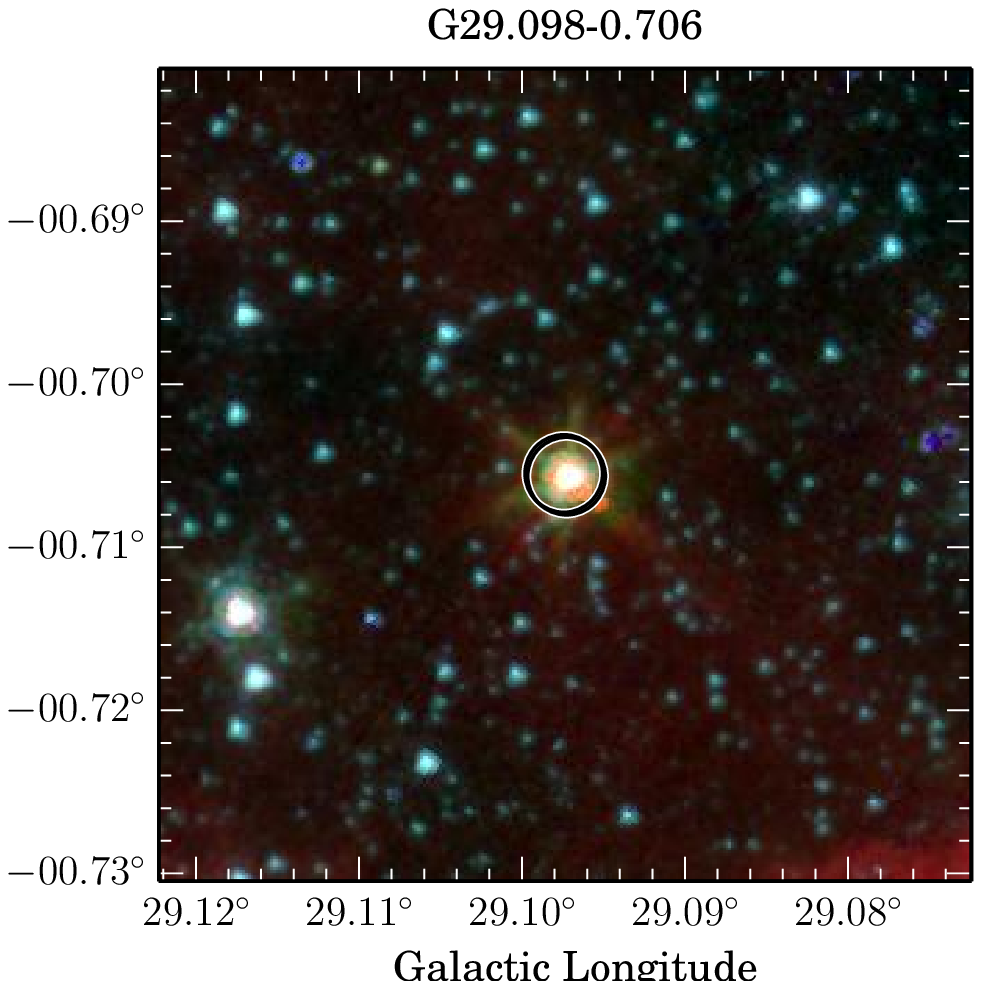}\\
\end{tabular}&
\hspace{-1.5cm}
\begin{tabular}{c}
~~~~G29.164+0.533~~~~~~~~~~~~~~~~ES\\
\includegraphics[width=0.25\textwidth]{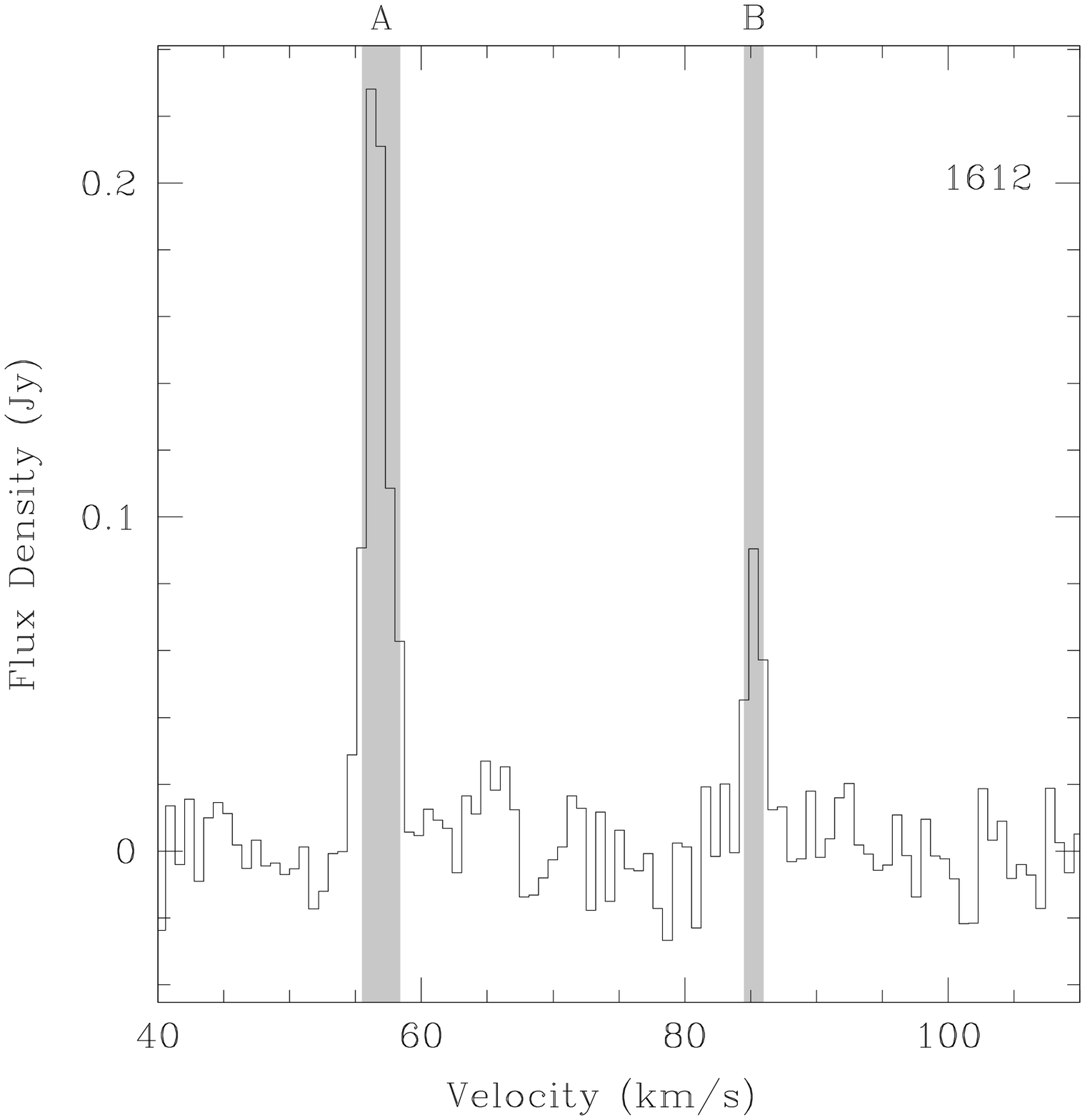}\\
\includegraphics[width=0.31\textwidth]{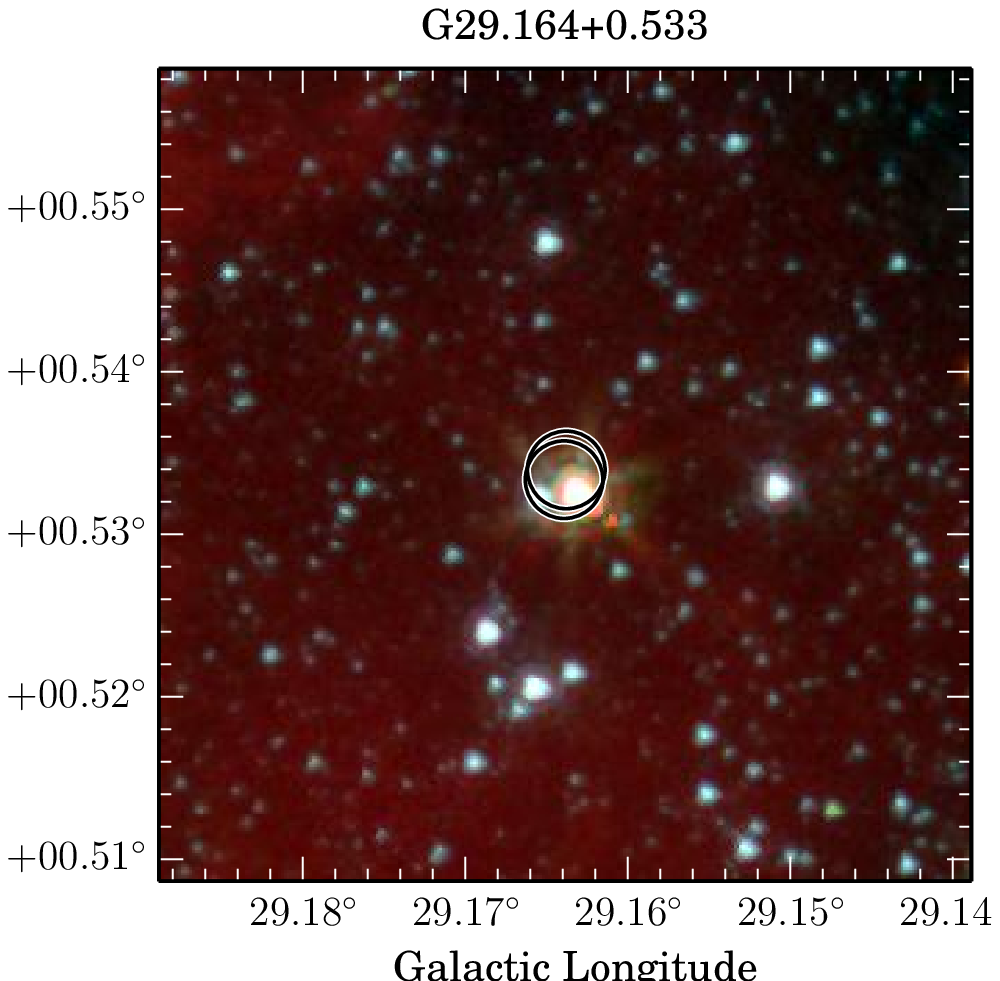}\\
\end{tabular}&
\hspace{-1.5cm}
\begin{tabular}{c}
~~~~G29.170+0.022~~~~~~~~~~~~~~~~~U\\
\includegraphics[width=0.25\textwidth]{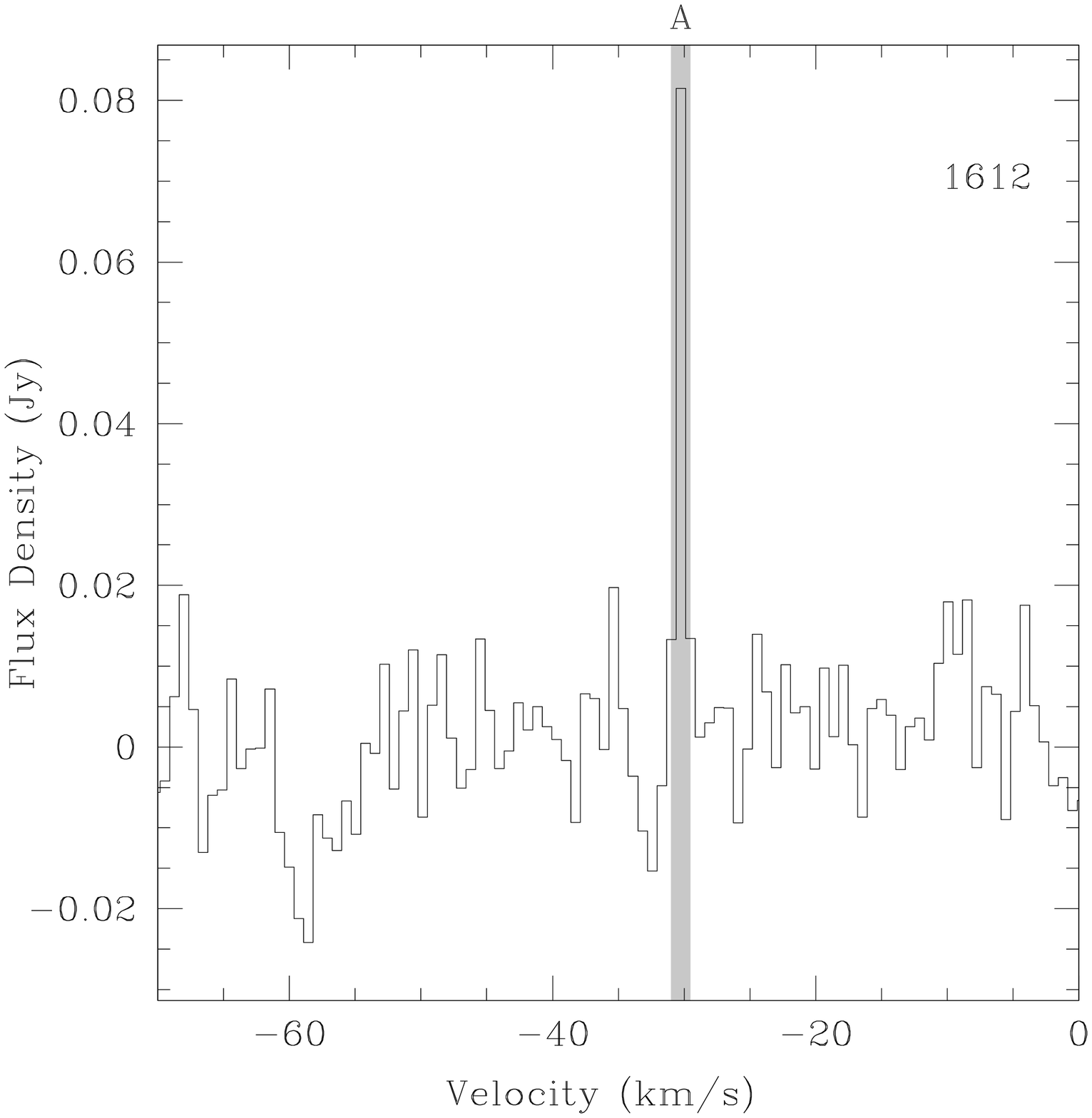}\\
\includegraphics[width=0.31\textwidth]{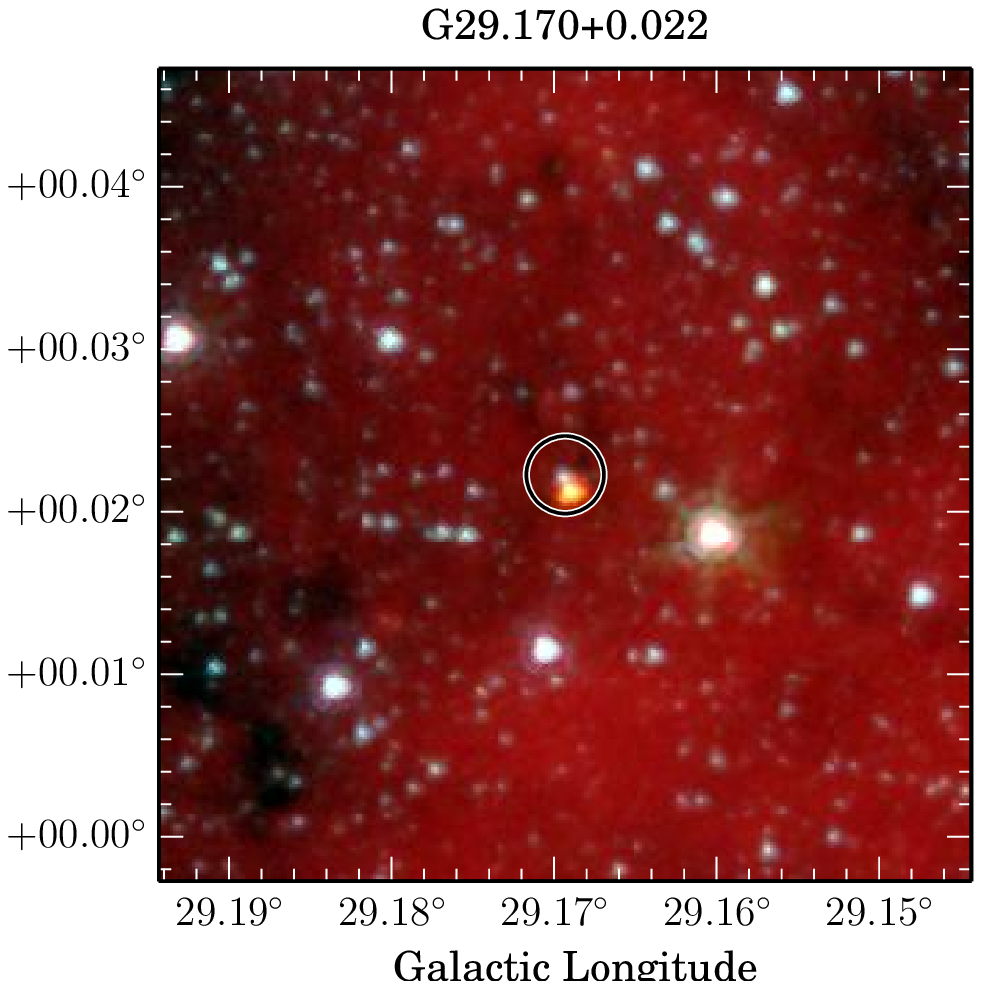}\\
\end{tabular}\\
\hspace{-1.5cm}
\begin{tabular}{c}
~~~~G29.246+0.777~~~~~~~~~~~~~~~~ES\\
\includegraphics[width=0.25\textwidth]{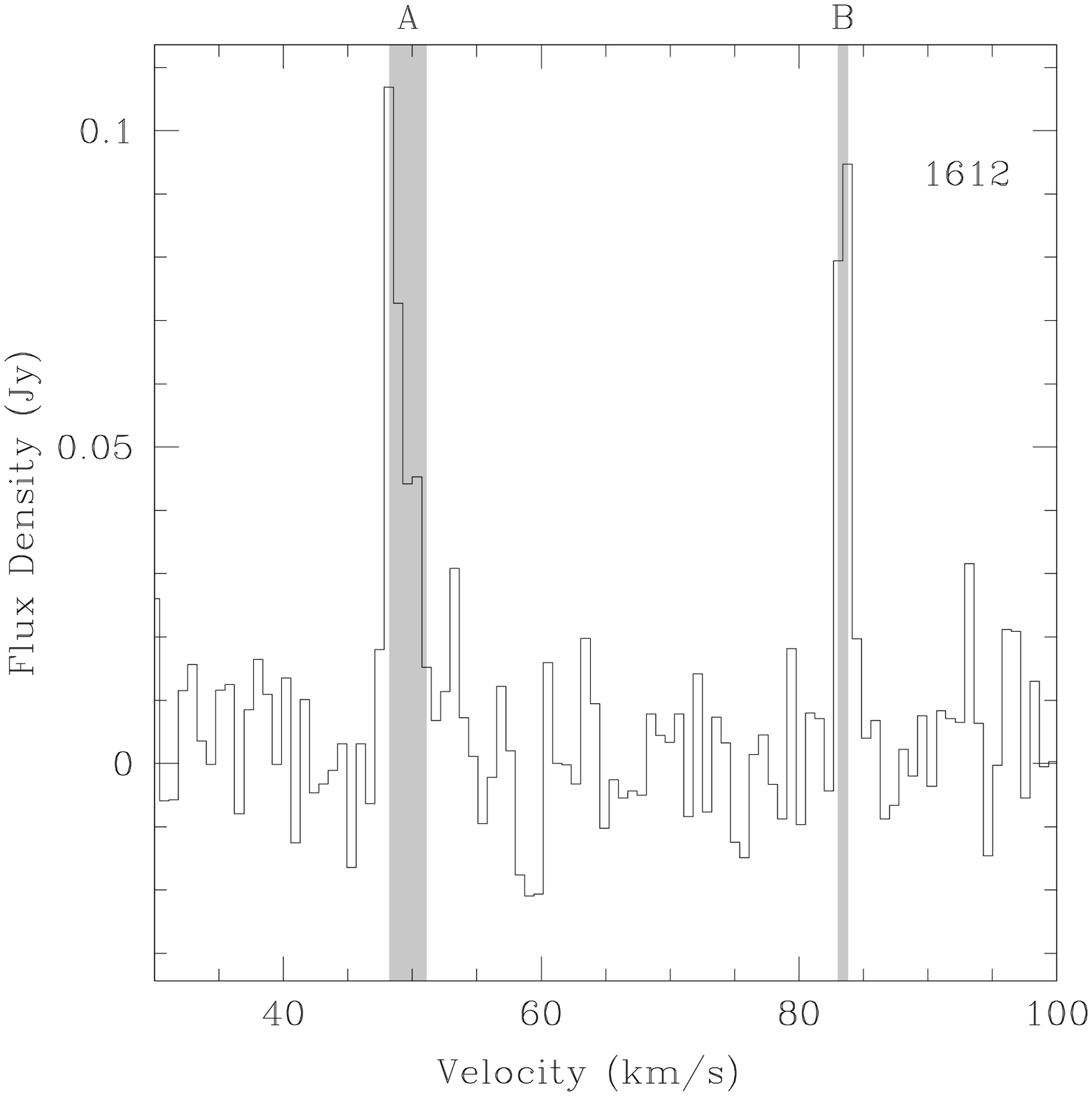}\\
\includegraphics[width=0.31\textwidth]{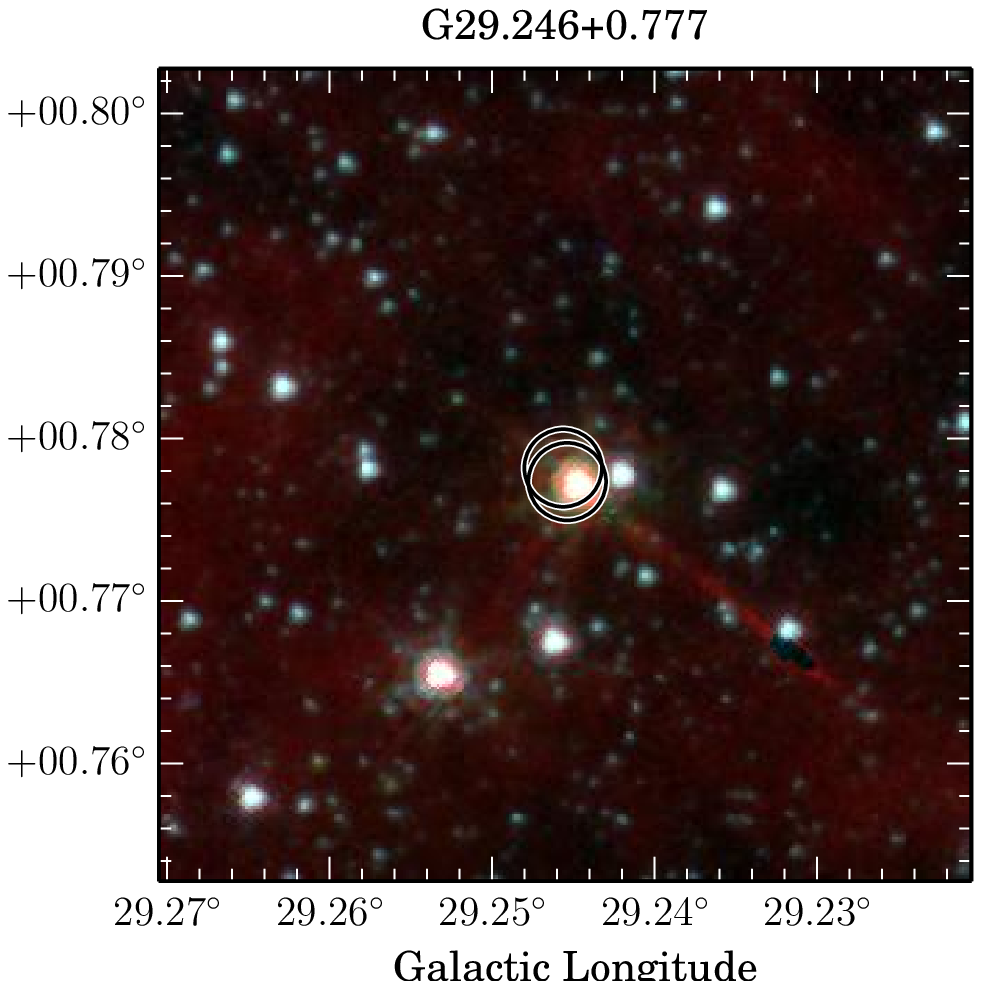}\\
\end{tabular}&
\hspace{-1.5cm}
\begin{tabular}{c}
~~~~G29.247-0.114~~~~~~~~~~~~~~~~ES\\
\includegraphics[width=0.25\textwidth]{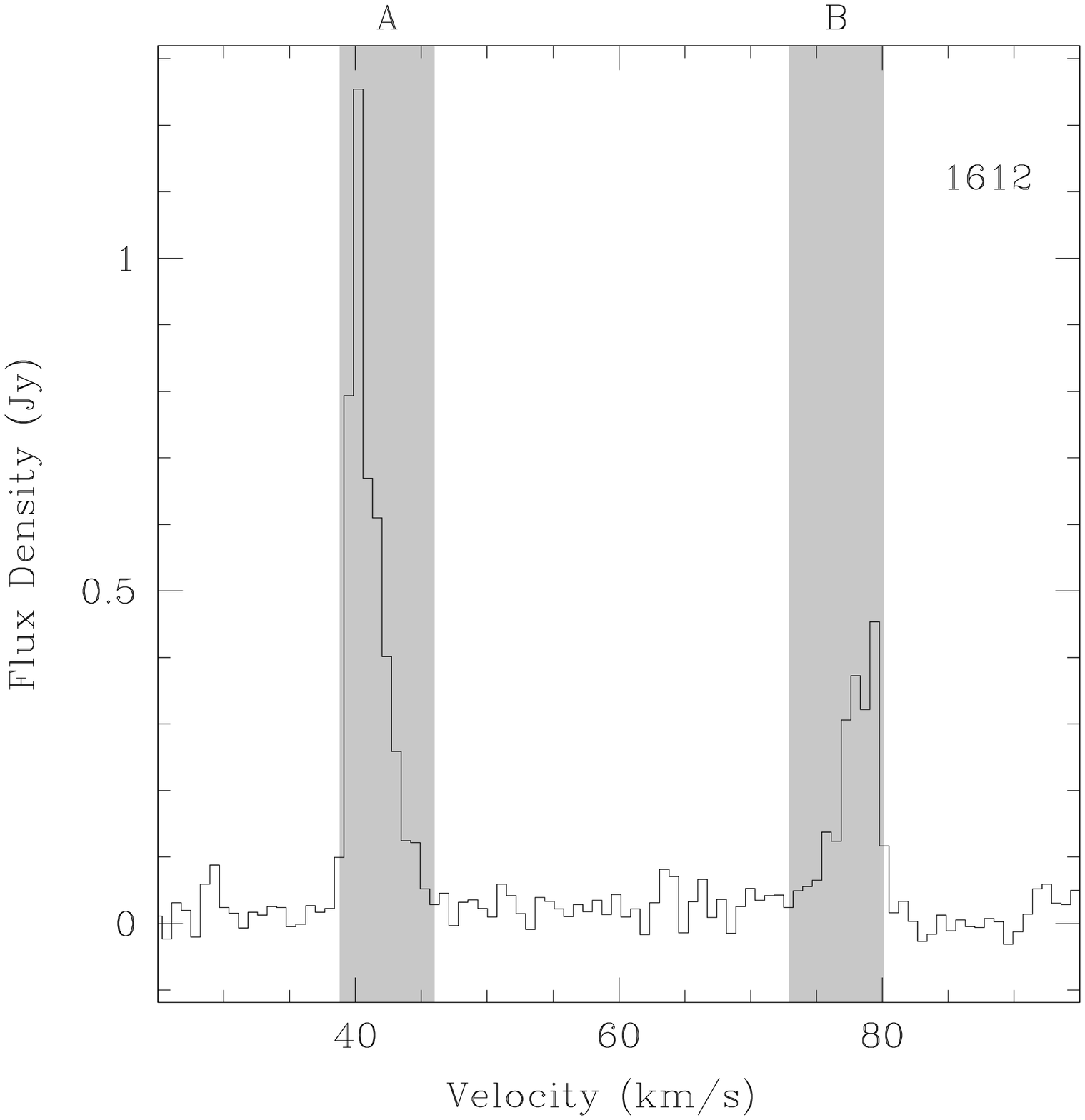}\\
\includegraphics[width=0.31\textwidth]{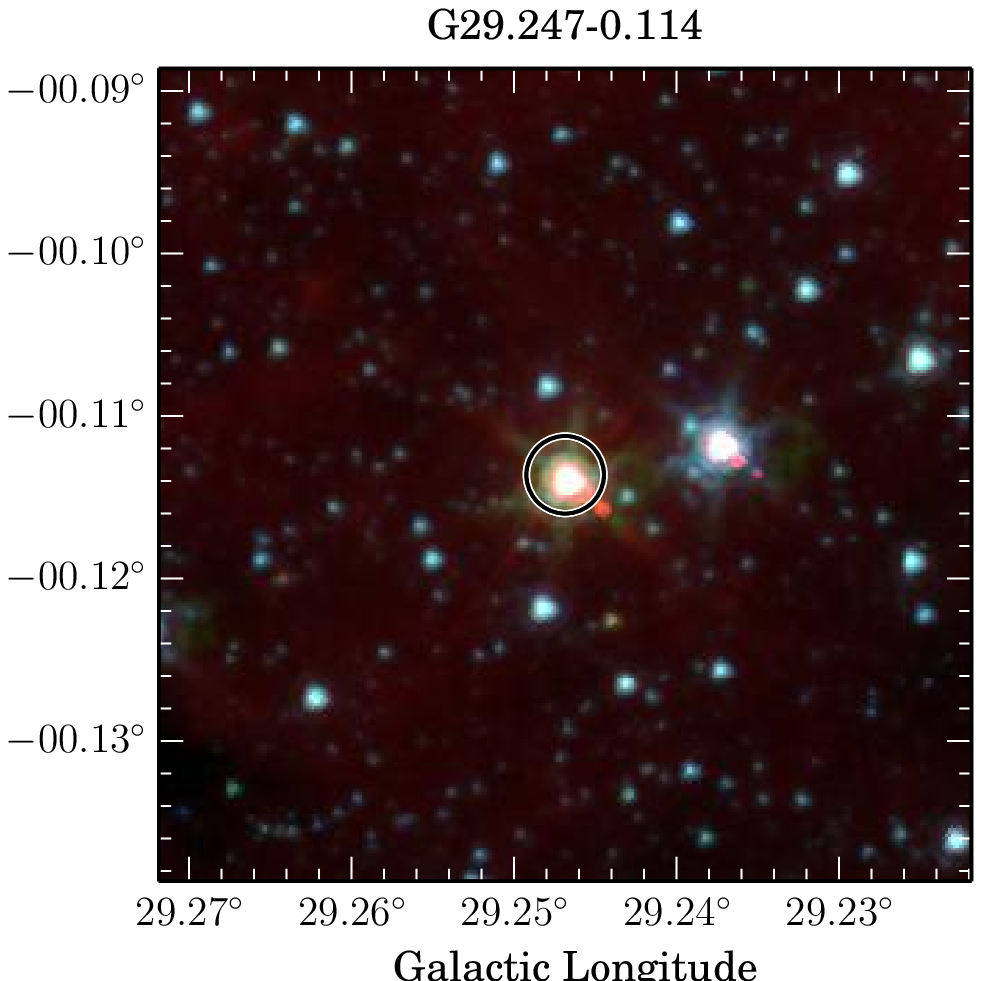}\\
\end{tabular}&
\hspace{-1.5cm}
\begin{tabular}{c}
~~~~G29.273-0.006~~~~~~~~~~~~~~~~~U\\
\includegraphics[width=0.25\textwidth]{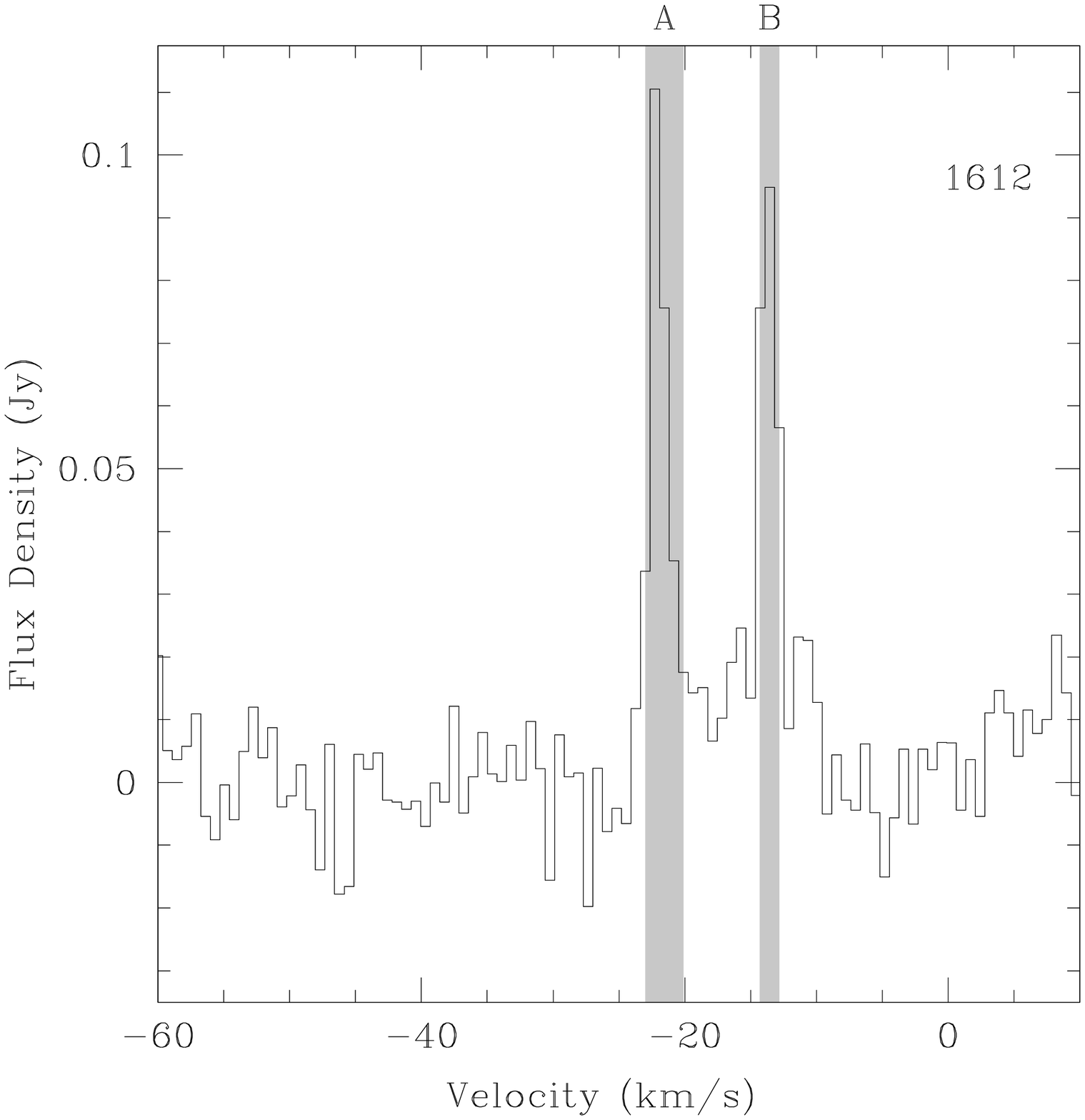}\\
\includegraphics[width=0.31\textwidth]{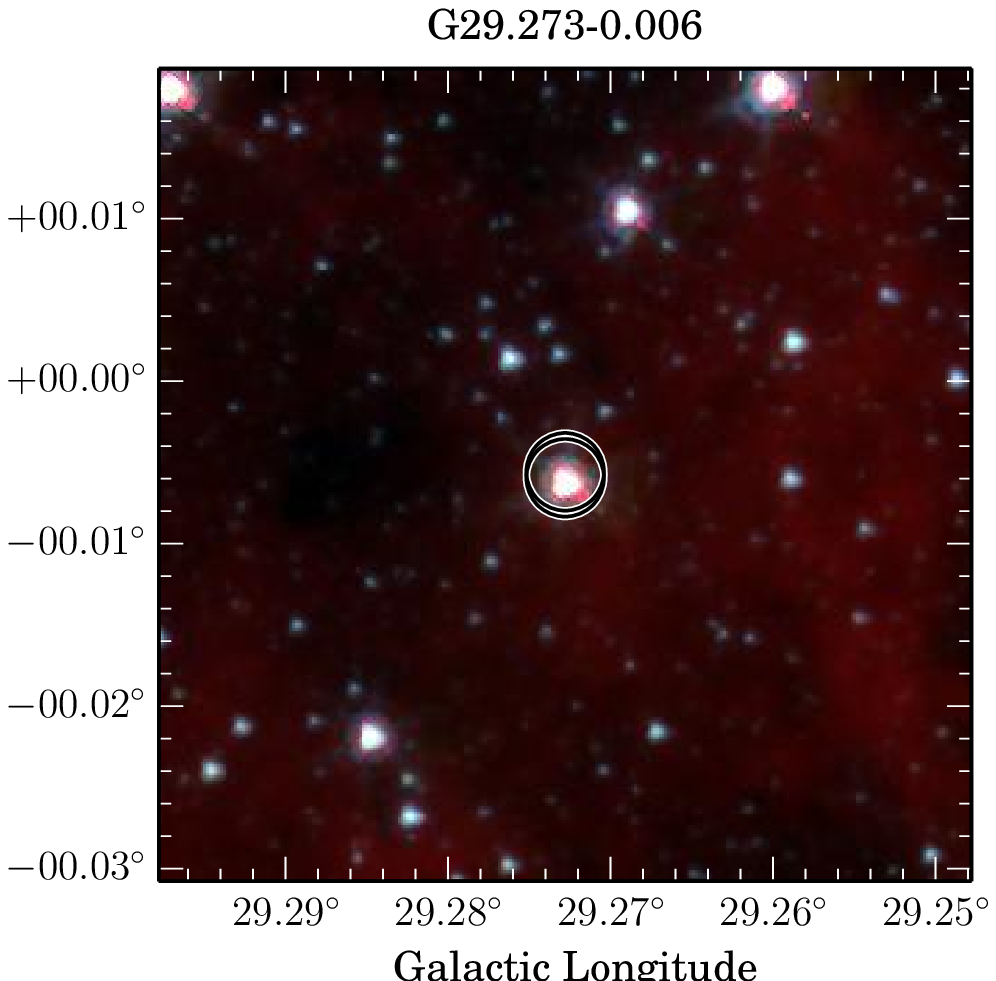}\\
\end{tabular}\\
\end{tabular}
\caption{The environments of detected masers. The upper panel
shows the spectrum
of OH emission with the horizontal axis showing V$_{\rm LSR}$ in \kms~ and the
vertical axis showing flux in Janskys integrated over the source. Shaded areas indicate the velocity range
over which emission is detected for each maser spot. GLIMPSE 3-colour images are
shown: blue = 3.5 $\mu$m, green = 4.5 $\mu$m and red = 8.0 $\mu$m.
Symbols represent the positions of maser spots. 1612\,MHz
masers are shown as circles, 1665\,MHz masers as squares, 1667\,MHz masers as
diamonds and 1720\,MHz masers as triangles. The axes are in Galactic
coordinates. Designations for each maser site are shown in the upper-right
corner, where ES = evolved star, SF = star formation and U = unknown origin.
Note that the size of the symbols does not represent the relative or
absolute position uncertainties of the masers.
The full Figure can be found online.}
\label{images}
\end{figure*}

The detected masers are detailed in Table \ref{maintable}
and their positions and spectra are shown in Figure \ref{images}. We find
a total of 276 maser spots (128 at 1612\,MHz, 72 at 1665\,MHz, 72 at 1667\,MHz,
and 4 at 1720\,MHz). The brightest maser spot has a peak flux density of
124.7\,Jy and the weakest 19\,mJy.\\

\begin{table*}
\caption{Details for maser spots detected in this work. The first column lists the name assigned to
each spot, which is based on the Galactic coordinates and the OH transition which shows the emission.
Columns 2 to 7 give the position of the maser spot. Column 8 lists the peak flux density, based on
the strongest channel. Column 9 lists the
rms noise level, determined from the spectrum. Column 10 lists the integrated flux.
Columns 11, 12 and 13 list the peak velocity of maser emission and the minimum and maximum velocities
over which emission is detected, respectively. The relative uncertainty in the postion is given in
columns 14 and 15. Note that the absolute position uncertainty is about 2\arcsec.
The last column lists comments on each maser site. The full table can be found online.}
\label{maintable}
\begin{tabular}{lccccccccccl}
\hline
Name                &     RA       & Dec          & \multicolumn{3}{c}{Flux}    & \multicolumn{3}{c}{Velocity (\kms)} & \multicolumn{2}{c}{Position}& Comments\footnotemark[1]{}\\
                    & (J2000)      &     (J2000)  & Peak & rms & Integrated  & Peak & Min. & Max. & \multicolumn{2}{c}{Uncertainty ($^{\prime\prime}$)}\\
                    & ($^{\rm h\,m\,s}$)& ($^{\circ~\prime~\prime\prime}$)& (Jy) & (mJy) & (Jy.km\,s$^{-1}$) &      &      &      & Lon. & Lat.\\
\hline
G29.098-0.706-1612A & 18 46 57.108 & -03 44 02.97 & 0.18&30& 0.54 & 66.6 & 64.9 & 69.2 & 2.1 & 2.3 & ES - D,BS\\
G29.098-0.706-1612B & 18 46 57.122 & -03 44 03.72 & 0.18&30& 0.65 & 98.8 & 95.4 & 101.2 & 2.2 & 2.3&\\
\\
G29.164+0.533-1612A & 18 42 39.431 & -03 06 33.22 & 0.228&13 & 0.509 & 56.8 & 55.5 & 58.4 & 0.99 & 0.99 & ES - D,BS\\
G29.164+0.533-1612B & 18 42 39.289 & -03 06 32.65 & 0.090&13 & 0.140 & 85.4 & 84.5 & 86.0 & 2.6 & 2.5&\\
\\
G29.170+0.022-1612A & 18 44 29.300 & -03 20 16.73 & 0.079&9 & 0.081 & -30.2 & -31.0 & -29.5 & 3.5 & 3.8 & U - RS\\
\hline
\end{tabular}
\footnotetext{1}{$^1$Masers are associated with evolved stars (ES), star formation (SF), or unknown (U).
D means that the 1612\,MHz spectrum shows a double-peaked feature between 14 and 47\,\kms~wide,
which is associated with evolved stars. BS, RS and wkS mean that in the GLIMPSE images, a Bright,
Red or weak Star is seen co-spatial with the maser site, respectively. nPL means near a pulsar.
IRDC means the GLIMPSE image shows an infrared dark cloud co-spatial with the maser site and so
the maser site is associated with star formation. Where the literature reports a previous
observation, a reference is given to identify the association for the maser site:
Bel13 -- \citet{bel13}, Blo94 -- \citet{blo94}, Cas95 -- \citet{cas95}, Cod10 -- \citet{cod10},
Dea07 -- \citet{dea07}, Deg04 -- \citet{deg04}, DiF08 -- \citet{dif08}, Fel02 -- \citet{fel02},
He05 -- \citet{he05}, Hil05 -- \citet{hil05}, Ima13 -- \citet{ima13}, Kwo97 -- \citet{kwo97},
Kur94 -- \citet{kur94}, Lou93 -- \citet{lou93}, Mot03 -- \citet{mot03}, Per09 -- \citet{per09},
Pes05 -- \citet{pes05}, Ros10 -- \citet{ros10}, Sev01 -- \citet{sev01}, Tho06 -- \citet{tho06},
Urq09 -- \citet{urq09}, Wal98 -- \citet{wal98}, Win75 -- \citet{win75}.}
\end{table*}

\begin{figure}
\includegraphics[width=0.45\textwidth]{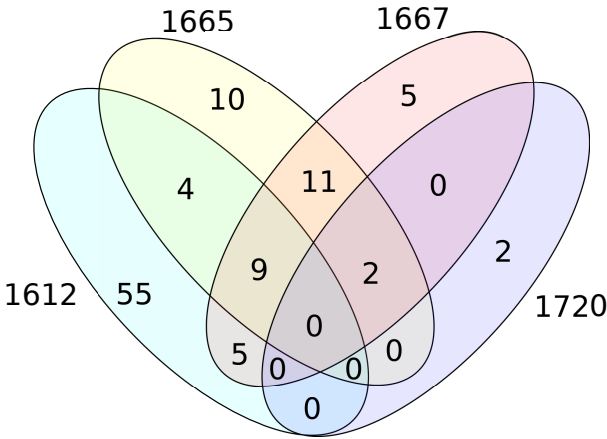}
\caption{A Venn diagram showing the occurrence of maser species for
the 103 identified OH maser sites.}
\label{venn}
\end{figure}

The maser spots can be grouped into maser sites. Maser sites occur at well defined
positions on the sky (usually smaller than 1\arcsec~and very small compared
to our beam size; \citealt{forster00}) and may consist of maser
spots from multiple transitions. We identify a total of 103 maser sites.
Figure \ref{venn} shows a Venn diagram of the number of
maser sites that exhibit emission in the four masing transitions. We find that
approximately half (53 per cent) of the maser sites contain only the 1612\,MHz
transition and approximately 10 per cent of the maser sites fall into each of the
following categories: sites with 1612+1665+1667\,MHz emission, 1665+1667\,MHz
emission and with 1665\,MHz emission only. The remaining categories are
typically sparsely populated. We note that the relative number of detections
in all categories is consistent with results from SPLASH \citep{dawson14}
who have performed a pilot untargetted survey of ground-state OH masers in the
Southern Hemisphere.\\

We show the spectra and infrared environments of each maser site in Figure
\ref{images}. For each maser site, the upper panel shows the spectrum or
spectra, where maser emission was detected. The shaded regions in each
spectrum indicate the range of velocities over which maser emission was detected
and the top of each spectrum shows a letter to label maser spots. Note that
the shaded areas were determined by inspection of the full data cube and include
channels only where emission could easily be identified at a single location.
Thus, there are secondary peaks that lie outside
the shaded regions. In such cases, we have determined that these peaks
are either noise spikes or sidelobe emission from nearby (but unrelated) strong
maser peaks. Thefore the
shaded areas should be used as a guide to identify the velocity range over which
we consider there to be real emission arising from this maser site. 
The lower panel shows a GLIMPSE 3-colour image (based on 3.5, 4.5 and 8.0\,$\mu$m for
blue, green and red colours, respectively) and symbols to represent the
positions of the maser spots. The GLIMPSE resolution is between 1\farcs7
and 2\farcs0 and the astrometric accuracy is typically 0\farcs3.\\

\subsection{Comments on individual sources of interest}

{\em G29.574+0.118}.
This maser site contains both 1612 and 1667\,MHz masers and
appears to be located close to a candidate magnetar
(AX 1845.0-0258). The position of the candidate magnetar is offset
56\arcsec~from the maser site \citep{olausen14}. The candidate magnetar is a
pulsar with a 7\,s second period and this pulsar is likely associated with the
G29.6+0.1 SNR \citep{gaensler99}. The SNR extends over an area approximately
5\arcmin~across and the maser site is found within the bounds of the SNR.
The GLIMPSE image shows there is an infrared stellar object at the position of
the maser site, which does not show a strongly rising flux at longer wavelengths,
as might be expected for an evolved star or star forming region. This star may
therefore be an unrelated field star. Nothing obvious is seen at the position
of the pulsar (G29.564+0.106). There appears to be a filamentary IRDC within an
arcminute of the maser site (above the maser site in Figure \ref{images}), but
does not overlap with the maser site. The maser spectra show emission in both the
1612 and 1667\,MHz lines, with the 1667\,MHz spectrum showing two unusually
broad maser spots, spanning 9.8 and 18.2\,\kms. We have classified this maser
site as of unknown association.
It is not clear whether or not there is a physical association between
the maser site and the candidate magnetar or SNR. If there is an association,
it would certainly be interesting to invertigate further. One possible method
to decide if there is an association or not is to search for periodic
variability in the maser signal that matches that of the pulsar. Such
periodicity in OH masers has been reported for the 1720\,MHz line
\citep{weisberg05}.\\

{\em G31.128-1.015}.
This maser site shows emission in the 1667\,MHz line. The emission is broken
up into four maser spots that together span a velocity range of 145.3\,\kms.
The data indicate that emission is present over this entire velocity range.
We have assigned this site to unknown origin. The GLIMPSE image shows the maser
site is co-spatial with a bright and red star. The Galactic latitude of the
site is just over 1$^\circ$, which indicates that the site is probably
close to us, which would favour an evolved star interpretation.\\

We note that
both G30.394-0.706 and G30.944+0.035 exhibit similar characteristics of
broad emission in the 1667\,MHz line, although the velocity ranges are not
as extreme. Both these maser sites also show emission in the 1612 and 1665\,MHz
lines. The 1612 and 1665\,MHz lines are also unusually broad in G30.394-0.706.
Both are assigned to an evolved star origin.\\

\section{Discussion}
\subsection{The molecular cloud complex W43}
\begin{figure*}
\includegraphics[width=\textwidth]{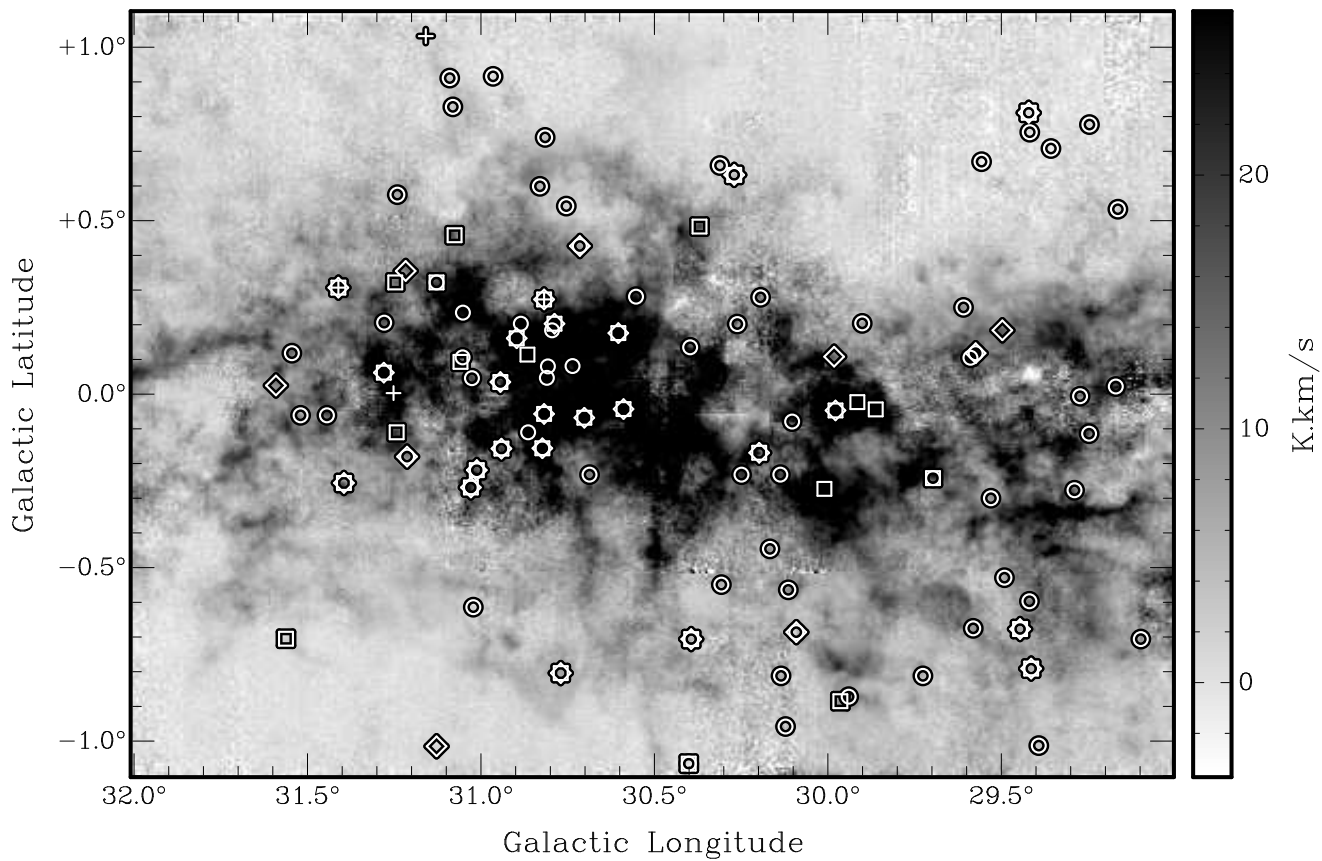}
\caption{The greyscale image shows the distribution of $^{13}$CO (1--0)
velocity-integrated gas that includes the W43 complex and a second complex
typified by gas at around 50\,\kms.
Maser sites are shown with the following symbols: 1612 - circles; 1665 -
squares; 1667 - diamonds ; 1720 - plus signs.}
\label{coplot1}
\end{figure*}

\begin{figure}
\includegraphics[width=0.45\textwidth]{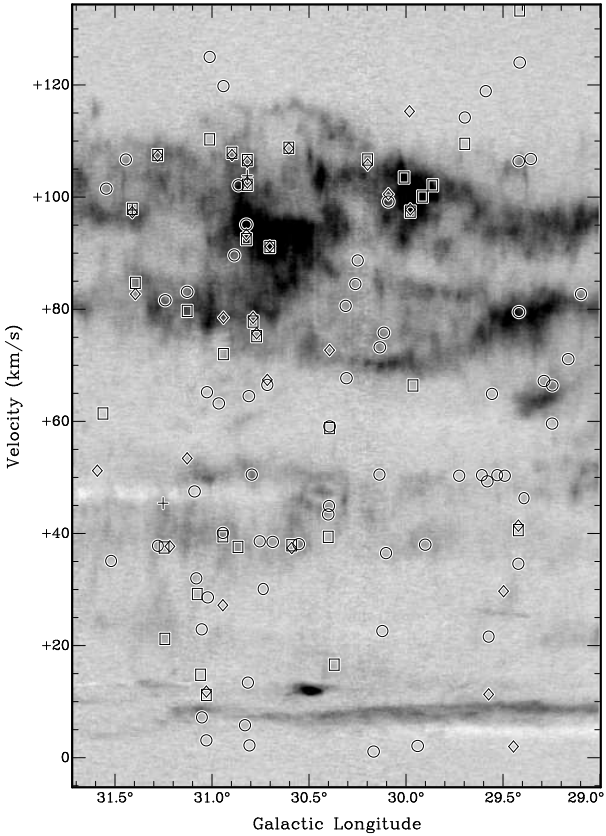}
\caption{Position-velocty diagram using the same $^{13}$CO (1--0) data as
Figure \ref{coplot1}. Maser sites are shown with the following symbols: 1612
- circles; 1665 - squares; 1667 - diamonds; 1720 - plus signs.
The main-line masers (1665/1667) are more closely associated with the
concentrations of CO gas than the satellite-line masers (1612/1720), both in
position and velocity. The W43 complex is usually located between velocities
of [+80; +120]\kms, whereas there is a secondary cloud complex in the velocity
range [+35; +55]\,\kms. Features in the velocity range [0; +20]\,\kms~are most
likely foreground clouds along the line of sight.}
\label{coplot2}
\end{figure}

The portion of the Galactic plane (l=29-32, b=$\pm1$) considered in this study
is host to the molecular cloud complex W43, often referred to as the W43
mini-starburst complex due to its high star formation activity
\citep{mot03,nguyenluong11}. Along the line of sight, there is another
molecular cloud complex having $V_{\rm LSR}\,\sim\,50$\,\kms, with
emission seen from +35--+55\,\kms, which
is different from 80-120\,\kms -- the LSR velocity of the W43 complex.
\cite{beuther12} suggested that these two complexes are interacting due to
their location at the end of the Galactic Bar \citep{nguyenluong11}.
In Figure \ref{coplot1} we show the $^{13}$CO (1--0) velocity-integrated
intensity map and in Figure \ref{coplot2}, the latitude-integrated
position-velocity (pv) diagram and overlay
the positions of the OH maser sites. The maser sites are generally scattered
around the integrated map and the pv diagram. However, there are some differences
between the maser transitions. 1612\,MHz masers are widely scattered within the 
field of view and within the pv diagram, with no close association with W43 or
the feature at a velocity of 50\,\kms. This is because 1612\,MHz masers are
mainly associated with evolved stars and not with the molecular gas.\\

The four 1720\,MHz maser sites, as discussed in Section 5.2.1, are associated
with star formation, two of which are within the W43 molecular cloud complex
and one associated with the 50\,\kms~feature (see Figures \ref{coplot1} and \ref{coplot2}). The 1665/1667\,MHz
maser sites seem more closely clustered to the spatial and velocity range
of W43 and 50\,\kms~complexes, especially around the location of W43-Main,
where large-scale shocks are strong \citep{nguyenluong13} and where the molecular
gas show double peaks profiles \citep{beuther12}. The 1665/1667\,MHz masers
around W43-Main seem to have a velocity spread within the range 50-120\,\kms.
This indicates the dynamic nature of this region and may be more evidence
of the interaction between the two complexes.\\

\subsection{Association of maser sites}
Here we use the word ``associate'' and its derivatives to imply that two
objects are (or were) in close enough proximity that the appearance of
one object today is considered to be (or have been) significantly affected
by the other object. For example, an OH maser may be considered associated
with a star formation site in the sense that the star formation site
creates the physical conditions conducive to the OH maser (ie. without
the star formation site the maser would not exist). Alternately, an OH maser
may be considered associated with a star formation site if the star formation
site merely increases the available pumping photons for the OH maser
(ie. without the star formation site the maser could still exist, but with
a significantly reduced intensity).\\

OH masers are known to be associated with multiple astrophysical objects
including evolved stars, star formatoin and supernova remnants (SNRs).
Evolved stars tend to be prominent in the 1612\,MHz transition, star formation
sites are more prominent in main-line masers \citep{reid81} and SNRs
occasionally show 1720\,MHz masers \citep{wardle02}. However, these transitions
are not exclusively associated with these objects. For example, all four
transitions are seen as masers towards the same star-forming region (eg.
G351.775-0.536 \citealt{caswell13}).\\

In order to associate each maser site in Table \ref{maintable} with an 
astrophysical object, we use a search radius of 20\arcsec~and then search
through the literature, using 
{\sc simbad}\footnote{http://simbad.u-strasbg.fr/simbad/}. In the last column
of Table \ref{maintable}, we identify associations (where possible) using
the literature. If it is not possible to identify associations in this way,
then the spectra and GLIMPSE images are used to make an association. If the
spectrum shows a double-peaked profile in the 1612\,MHz transition,
then we assign the maser site to an evolved
star origin. Usually, such a maser site is also accompanied by a bright star
in the GLIMPSE image. If the GLIMPSE image shows an infrared dark cloud, then
the maser site is assigned to a star formation association. However, if the
maser site shows both a double-peaked 1612\,MHz maser spectrum and is projected
in front of an infrared
dark cloud, then the association remains unknown. Using this method, we identify
47 maser sites as associated with evolved stars, 30 associated with star
formation and 26 with unknown associations.\\

We caution that Miras (and semi-regular stars with thinner circumstellar
envelopes than OH/IR stars, may be devoid of 1612\,MHz emission altogether
and/or show stronger main line emission (the so called ``type I Miras'') with
not so well-defined double-peaked profiles in some cases. This is especially true
if the classification is based on a single observation, as we have done, as the
spectrum can vary with time \citep{etoka01}. We expect that some of the unknown
associations are indeed evolved stars of this type, since we require a previous
evolved star association or a double-peaked profile.\\

\subsubsection{1720\,MHz maser associations}
As mentioned above, sources that are \textit{only} masing on the 1720\,MHz
transition are normally associated with SNRs, while some high-mass star forming
regions with main-line (1665/1667\,MHz) masers may also show 1612 and/or 1720\,MHz
masers (see, eg. \citealt{arg00}).
We detect four 1720\,MHz maser sites in the pilot survey region.
Three of these (G30.818+0.273, G31.252+0.003 and G31.411+0.307) are associated
with star formation and the other maser site (G31.159+1.032) is not associated
with a known object. More details on these associations are given below:\\
{\em G30.818+0.273} is associated with a methanol maser site
\citep{cas95}, with the most accurate known position for the methanol
maser site \citep{wal98} offset less than one arcsecond from the
1720\,MHz position and overlapping in velocity. This is strong evidence for
a star formation association. This maser site also includes main-line
masers (1665/1667\,MHz) which also supports a star formation association.\\
{\em G31.159+1.032} does not appear to be associated with anything in the
literature, with no other OH masers detected in any of the other transitions,
or any association in the GLIMPSE image and so the origin remains unknown. The
1720\,MHz detection is made at an unusually high velocity (181.2\,\kms
-- higher than the terminal velocity at this longitude of 110\,\kms) and
is weak (0.22\,Jy, equivalent to a 5.5\,$\sigma$ detection). This peak is
strong enough for us to classify it as a detection in this work, but further
observations should be carried out to confirm (or otherwise) this 
as a bona fide maser site
since it is possible that the peak is an unusually high noise spike.\\
{\em G31.252+0.003} is co-spatial with a sub-millimetre continuum source
\citep{dif08}, as well as co-spatial with a red star that appears embedded
in a low-contrast IRDC in the GLIMPSE image. In this context, we define
``co-spatial'' as overlapping in the plane of the sky.
Therefore, it is likely associated
with star formation. However, we note that only 1720\,MHz maser emission is
found here. If this is a site of star formation, then it is of a very rare
class where 1720\,MHz masers are seen, but not the main-line masers.\\
{\em G31.411+0.307} is associated with a methanol maser site \citep{cas95},
with the most accurate known position for the methanol maser site
\citep{wal98} offset two arcseconds from the 1720\,MHz position and
overlapping in velocity. This is strong evidence that this maser site
is associated with star formation. This is also supported by the
detection of main-line masers at this site.\\

Three out of four 1720\,MHz maser sources are associated with star formation,
with only G31.159+1.032 remaining as an unknown
association. Two of the star formation maser sites are also associated with
main-line masers, while the third (and the unknown association) only shows
1720\,MHz masers. Sources that only show maser emission in this single line
are almost exclusively associated with SNRs. G31.252+0.003 may be a notable
exception to this, as we associate it with star formation, based on the
presence of a sub-millimetre continuum source and reddened stellar object,
which appears to be embedded in an infrared dark cloud.
Another exception is the FU Ori star V 1057 Cyg toward which \citet{lo74} first
detected 1720\,MHz (only) OH maser emission. In this violently variable
star, as in SNRs, the maser pump may be related to shock excitation.\\

\subsubsection{1612\,MHz maser associations}
We detect 72 sites with 1612\,MHz maser emission. Sixty-four per cent (46)
are associated with evolved stars, thirteen per cent (9) are associated with star
formation and twenty-four per cent (17) are unknown associations. Seventy-six per cent of 1612\,MHz
maser sites (55) only show emission in the 1612\,MHz line. These numbers
demonstrate the close association between 1612\,MHz maser sites and evolved
stars. Further to this, we find that every maser site associated with an
evolved star contains 1612\,MHz maser emission.\\

There are five 1612\,MHz maser sites that are associated with evolved stars,
but do not show any significant emission in the GLIMPSE images (G29.392-1.013,
G29.491-0.529, G29.530-0.300, G29.609+0.250 and G30.795+0.184). Indeed, none
of these sites show emission at 24\,$\mu$m in MIPSGAL \citep{carey09} images
either. If these
maser sites are indeed associated with evolved stars, then they must have
very thick envelopes that have extremely high extinction, rendering the
central star undetectable in the GLIMPSE images. Such thick envelopes would
imply very high mass-loss rates.\\

\subsubsection{1665/1667\,MHz maser associations}
Out of 11 maser sites that show only 1665\,MHz emission, we find that eight of them
are associated with star formation, with the rest remaining unknown. In
contrast to this, there are five maser sites that show only 1667\,MHz emission
but all of them have unknown associations. We also find 11 maser sites with
both 1665 and 1667\,MHz maser emission and all of these are associated with
star formation. It is interesting that there are differences in the association
rate. In particular that most of the 1665-only masers are associated with
star formation, whereas none of the 1667-only masers have known associations.
However, we caution that our sample sizes are too small to draw strong
conclusions on these differences and await for greater numbers from the full
THOR survey to determine if these differences are significant.\\

Of all the main-line maser sites (ie. any maser site that
includes 1665 and/or 1667\,MHz masers -- forty-six in total), fifty-four per cent (25) are
associated with star formation, fifteen per cent (7) are associated with evolved stars
and thirty per cent (17) have unknown associations. Thus, our results are in accordance
that the main-line maser sites are typically associated with star formation.
We also note that in this sample of forty-six main-line maser sites, ninety-six per cent (44)
appear to have an infrared counterpart in the GLIMPSE image (ie. bright or red
star, IRDC or co-spatial extended emission). Of the two that
have no GLIMPSE counterpart, one is a 1665\,MHz maser site: G31.247+0.322
and one is a 1667\,MHz maser site: G31.217+0.355. This high
detection rate of GLIMPSE counterparts is similar to (and possibly slightly
higher than) the detection rate of GLIMPSE counterparts to water masers,
reported by \citet{wal14}: ninety-two per cent. The slightly higher detection rate
towards OH masers may be because OH masers associated with star formation
are present at slightly later stages of evolution than water masers
\citep{breen10}. At later
stages of evolution, we expect that infrared sources will be more easily
detectable. However, we again caution that our sample size is too small to
draw a strong conclusion.\\

\subsection{Comparison with other OH maser surveys}

\subsubsection{Main line masers}
\begin{figure}
\includegraphics[width=0.45\textwidth]{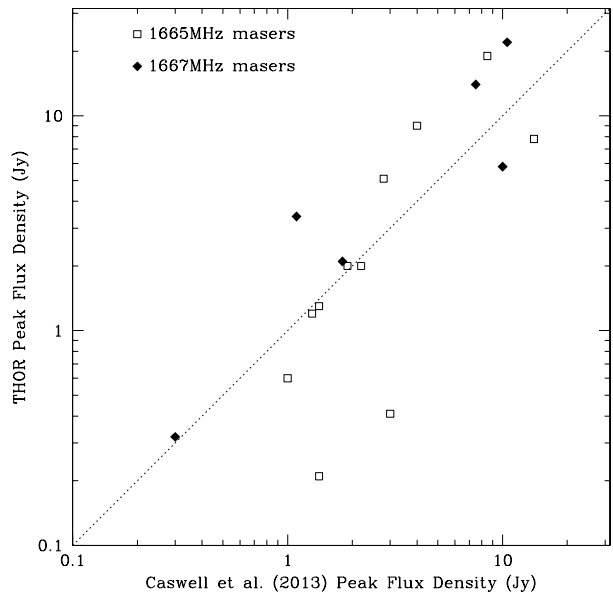}
\caption{Distribution of main-line OH maser peak flux densities, compared
between our THOR data and data presented in \citet{caswell13}. 1665\,MHz
masers are shown as open squares and 1667\,MHz masers are shown as
filled diamonds. The dotted line shows equality between the two
datasets.}
\label{caswell}
\end{figure}

Recently, \citet{caswell13} reported previous and new observations of main-line
OH masers, using the Parkes radio telescope. They found twelve OH maser sites
within the THOR pilot region, all of which
were also detected in our observations. In Figure \ref{caswell} we show a
comparison of the peak flux densities for these masers, by comparing data
taken in 2005 to our data -- a time difference of approximately 6.5 years.
The Figure shows some scatter, but good general agreement between the two
datasets. This suggests that whilst there may be some variability in individual
objects, these are not large variations of more than a factor of a few. The
variations are larger than expected from errors in the flux calibration
scales between the different telescopes, since these errors are typically less
than 50 per cent for any one measurement. The good general correspondence also
shows consistency in the calibration of our data, when compared to that of
\citet{caswell13}.\\

\subsubsection{1612\,MHz masers}
In the THOR pilot data we detect 72 1612\,MHz OH masers.
\citet{sev01} (hereafter S01) conducted an untargetted survey of 1612\,MHz
masers, covering
the THOR pilot survey region, but with a number of small holes in their
coverage. In the region covered by the THOR pilot survey, S01 detected 21 maser
sites, of which all but one are also detected in this work. The one undetected
maser in this work is located at G31.091-0.686. We find no evidence for a maser here above
a 3\,$\sigma$ noise level of 48\,mJy. We note that the coordinates given by
S01 are exactly one degree away from a very strong maser at
G30.091-0.686 which was detected by both S01 and us. Also, the
shape of the maser spectra, shown by S01 for both maser sites is
virtually identical, except for the intensity scale. Finally, inspection of
GLIMPSE images show nothing obvious at the position of G31.091-0.686.
We therefore suspect that G31.091-0.686 is a spurious detection, although we
cannot rule out the possibility that there is a real maser at this position
and intrinsic
variability has rendered it undetectable during our observations. Thus, we
consider 20 maser sites detected by S01 within the THOR pilot area,
all of which were also detected in our observations.\\

\begin{figure}
\includegraphics[width=0.45\textwidth]{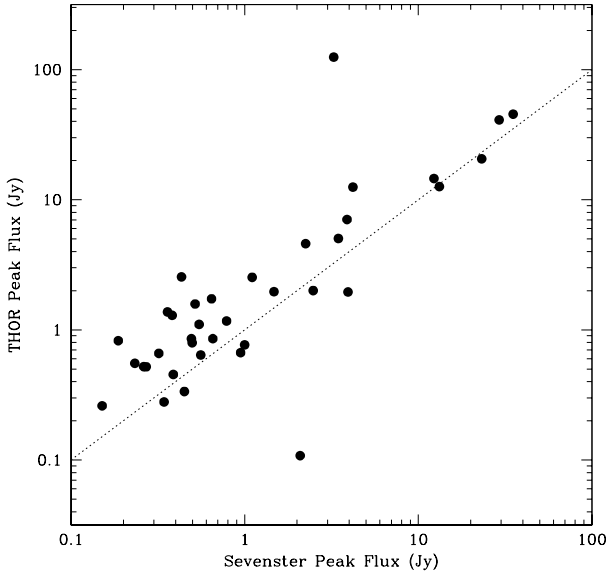}
\caption{Distribution of 1612\,MHz OH maser peak flux densities, compared
between our THOR data and data from S01. Only maser sites that
were detected in both surveys are shown. The dotted line shows equality between
the two datasets. Note that we have included one data point for each peak
in the spectrum. Since these 1612\,MHz OH maser sites all exhibit a
double-horned profile, there are 40 data points shown - twice the number of
maser sites. The dotted line shows equality between the two datasets.}
\label{sevplot1}
\end{figure}

In Figure \ref{sevplot1}, we show the distribution of peak flux densities
detected in both surveys. The Figure shows a scatter in the data points, which
may be attributed to instrinsic variability, which is rarely larger than a
factor of a few. However, nearly all the data points are found above the line
of equality, indicating that the THOR peak flux densities are measured
consistently higher than those by S01. One maser spot from
G30.944+0.035 has a significantly lower flux measured in THOR (0.1\,Jy),
compared to S01 (2.085\,Jy), which makes this data point stand
out well below the line in Figure \ref{sevplot1}. Since the other maser spot
in this site shows similar peaks in both datasets, we consider this
strong evidence for variability in this one maser spot peak. Discounting
this one spot, we can calculate the average ratio of THOR to S01 peak flux
densities, where we find the S01 peak is 28 per cent of the THOR peak flux.\\

A similar discrepancy was found between OH maser surveys reported by
\citet{dawson14}. They reported that consistently lower fluxes of OH masers
measured by \citet{sev97} were due to the large channel widths used by
\citet{sev97} which tend to reduce the peak flux densities of narrow maser
spots. We believe that the large channel width used by S01 (2.27\,km\,s$^{-1}$)
provides a similar explanation for the discrepancy in Figure \ref{sevplot1}.
Indeed, S01 estimate that their peak flux densities are about 35 per cent of
their true value, due to the wide bins used in their observations.\\

\begin{figure}
\includegraphics[width=0.45\textwidth]{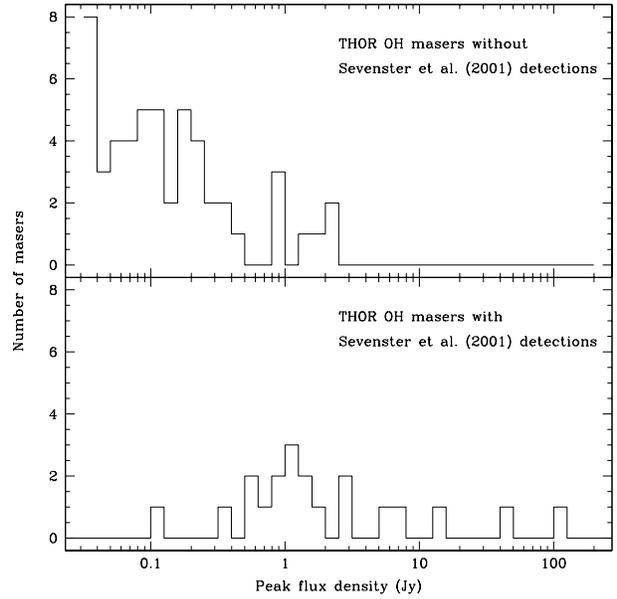}
\caption{Distributions of 1612\,MHz OH maser peak flux densities, detected
in THOR. The upper distribution shows those OH masers that were not detected
by S01 and the lower distribution shows those OH masers that were also detected
by S01.}
\label{sevplot2}
\end{figure}

As mentioned above, in the THOR pilot data we detect 72 1612\,MHz OH masers.
Therefore, we have detected seventy-one per cent (51) of maser sites that were
not found by S01. In Figure
\ref{sevplot2}, we show distributions of the THOR peak fluxes for masers that
were and were not found by S01. The S01 99 per cent completeness level is
500\,mJy, whereas the THOR completeness level is around 250\,mJy
a factor of about two times more sensitive than S01.
Thus, we expect that many of the peaks only
detected in THOR will be weak, rendering them undetectable by S01. This appears
to be the case in Figure \ref{sevplot2}, where most peaks are below 0.3\,Jy.
But there are also some strong peaks above 1\,Jy in the THOR data which were
not detected by S01. It is likely that these peaks have undergone intrinsic
variability between the observations for the two surveys. Such long term
variability has been detected in OH masers and appears to be common
(eg. \citealt{caswell13}).\\

\subsection{Comparison with dust emission}

We have compared the positions of the OH maser sites with dust emission traced
by ATLASGAL, which is an unbiased sub-millimetre survey of the inner Galactic
plane (\citealt{schuller09}). Matching
the masers with the ATLASGAL Compact Source Catalogue
(\citealt{contreras13, urquhart14}) we identify 24 OH maser sites that are
co-spatial with compact dust emission. Of these, 23 OH maser sites have already
been associated with star formation as described in \S\,5.2, including three
of the 1720\,MHz OH masers associated with star formation. These associations
are further strengthened by the good correlation between the velocities of the
thermal lines (eq., CO, NH$_3$) and the velocities of the peak maser emission
(typically better than a few \kms~for all associations).\\

The one OH maser site not previously associated with star formation is
G30.823$-$0.156. This source has been associated with an evolved star, due to
the presence of the distinctive double-horned profile seen in the 1612\,MHz
transition and a bright infrared star seen in the GLIMPSE
images that is co-spatial with the maser site. This maser
site is one of the most active with a total 12 other 1665 and 1667\,MHz OH
masers, strong CO and dust emission and extended far-infrared emission seen in
Hi-GAL images (\citealt{molinari10}). All of these tracers are normally
associated with star forming regions but may also be detected in evolved stars
with thick circumstellar envelopes. Further tests are needed to convincingly
resolve the nature of this site.\\

We note that 8 of the sources identified in \S\,5.2 as being associated with
star formation are not associated with an ATLASGAL source (G29.446-0.677,
G29.915-0.023, G30.249-0.232, G30.307-0.549, G30.311+0.659, G30.397+0.135, 
G30.941-0.157 and G31.213-0.180). This is somewhat
surprising since the dust emission is tracing the distribution of dense gas,
which is the raw material for star formation. These maser sites may simply be
associated with more distant star formation regions that fall below the
detection threshold of the ATLASGAL survey
(5$\sigma \sim 300$\,mJy\,beam$^{-1}$), which corresponds to
$\sim$100\,M$_\odot$ at a distance of 5\,kpc.\\

\subsection{Star formation maser sites and continuum emission}
The presence of radio continuum emission in sites of star formation is
usually a signature of a young \hii region (eg. an ultracompact \hii region).
These regions arise from young high-mass stars, once they have created
a sufficient ionising photon flux to dominate the star's immediate surroundings.
Thus, the presence of radio continuum emission indicates a reasonably well
developed star forming region. Masers
may be seen in earlier stages than the \hii~regions, although there is some overlap
(eg. \citealt{forster00}). Recent work \citep{breen10} has suggested that OH masers
may appear at a slightly later stage of evolution than other masers, such as
water or methanol masers. Evidence for this is the greater prevalence of OH
maser sites associated with radio continuum emission, compared to water or
methanol maser sites.\\

\citet{fc89} found that 23 out of 74 OH maser sites (31 per cent) were
associated with continuum emission. However, their sample was targeted towards
known regions of star formation at the time, which may bias the sample to more
evolved regions. Our observations remove this bias by covering an untargetted
section of the Galactic plane. We can compare continuum emission (Bihr et al.
{\em in preparation}) detected in THOR with the incidence of OH maser
sites associated with star formation. There are 31 OH maser sites associated
with star formation and of these, 18 are associated with continuum emission
(58 per cent). Thus, we find a higher proportion of OH masers associated with
continuum sources. We believe this is due to the more sensitive continuum
data from THOR, compared to those of \citet{fc89}. But this may suggest that
timelines of high-mass star formation may need to be revised to account for
a slightly longer OH maser site phase. We caution that this result is based
on only a small number of OH maser sites. More robust statistics will be
available from the full THOR survey, as well as from SPLASH.\\

\subsection{Diffuse OH emission and absorption}
%OH can be detected either as masers or as what is commonly referred to as
%thermal emission or absorption. However, as noted by \citet{dawson14}, the
%line profiles normally considered thermal may well be the result of weak masing,
%but with low gain. Such low gain spectral line profiles commonly follow expected
%ratios between the transitions. In particular that the satellite lines (1612 and
%1720\,MHz) appear inverted when compared to each other. Therefore, rather than
%referring to such line profiles as thermal emission or absorption, we refer
%to them as diffuse OH. Such diffuse OH can be distinguished from the strong
%masers referred to above by comparing the OH satellite line profiles. Diffuse
%OH may also be identified by showing only lines in absorption against strong
%continuum emission.

OH can be detected either as masers or as what is sometimes referred to as
``thermal'' emission or absorption. However, as noted by \citet{dawson14}, the
line profiles often considered thermal may show departures from the permitted
LTE line ratios, even in the 1667 and 1665 MHz main lines
\citep[see also][]{crutcher79}. Similarly, it is very common for the 1612 and
1720\,MHz satellite lines to be anomalously excited, with excitation
temperatures that are subthermal in one line, and either very high or
negative in the other -- the latter case being indicative of weak, low-gain
maser action. Therefore, rather than referring to such line profiles as thermal
emission or absorption, we refer to them as diffuse OH. The typical signature
is broad, weak profiles, with all transitions showing similar line widths. In
regions where the continuum background is significantly elevated above the CMB,
such diffuse OH profiles typically show an almost symmetrical pattern of
emission/absorption in the satellite lines, and main lines that are either both
in emission or both in absorption \citep[see e.g.][]{dawson14}. \\

\begin{table*}
\caption{Diffuse OH sources. Details, following the format of Table
\ref{maintable} are reported for those regions that show spectral profiles
indicative of diffuse OH, rather than strong maser emission. Such sources
show either inverted profiles in the two satellite lines or lines only in
absorption. Details are reported for the strongest OH emission or absorption
features. The peak flux density value is based on either the highest (most
positive) or lowest (most negative) channel in each spot.}
\label{diffuseOHtab}
\begin{tabular}{lccccccccl}
\hline
Name                &     RA       & Dec          & Peak Flux    & \multicolumn{3}{c}{Velocity (\kms)} & \multicolumn{2}{c}{Position Uncertainty} & Comments\\
                    & (J2000)      &     (J2000)  & Density      & Peak & Min. & Max. & Lon. & Lat.&\\
                    & ($^{\rm h\,m\,s}$)& ($^{\circ~\prime~\prime\prime}$)& (Jy)         &      &      &      & ($^{\prime\prime}$) & ($^{\prime\prime}$)&\\
\hline
G29.937-0.052-1612A & 18 46 09.859 & -02 41 16.99 & 0.14$\pm$0.03 & 101.2 & 99.3 & 102.4 & 2.3 & 2.1 & SF - Urq09\\
G29.937-0.052-1665A & 18 46 09.500 & -02 41 18.72 & -0.19$\pm$0.02 & 102.0 & 96.8 & 102.4 & 2.1 & 2.0 &\\
G29.937-0.052-1667A & 18 46 09.500 & -02 41 18.72 & -0.23$\pm$0.02 & 102.0 & 96.8 & 102.4 & 2.2 & 2.2 &\\
G29.937-0.052-1720A & 18 46 08.897 & -02 41 27.58 & -0.10$\pm$0.02 & 102.1 & 99.3 & 102.4 & 5.3 & 6.1 &\\
\\
G29.956-0.016-1612A & 18 46 03.982 & -02 39 20.69 & 0.11$\pm$0.03 & 99.8 & 100.5 & 101.9 & 2 & 1.9 & SF - Tho06\\
G29.956-0.016-1665A & 18 46 03.730 & -02 39 22.12 & 0.21$\pm$0.03 & 98.3 & 97.8 & 98.5 & 1 & 1&\\
G29.956-0.016-1665B & 18 46 03.750 & -02 39 23.04 & 0.10$\pm$0.03 & 102.0 & 101.3 & 103.4 & 2.4 & 2.3&\\
G29.956-0.016-1720A & 18 46 04.086 & -02 39 25.18 & -0.09$\pm$0.04 & 99.8 & 100.5 & 101.9 & 4.1 & 4.0&\\
\\
G30.533+0.021-1665A & 18 46 59.359 & -02 07 23.41 & -0.04$\pm$0.01 & 95.0 & 90.0 & 95.0 & 6.3 & 5.2 & SF - Urq09\\
G30.533+0.021-1667A & 18 46 59.116 & -02 07 30.53 & -0.04$\pm$0.01 & 93.1 & 90.0 & 95.0 & 2.3 & 2.1 &\\
\\
G30.688-0.071-1612A & 18 47 36.222 & -02 01 50.90 & -0.24$\pm$0.03 & 92.5 & 89.2 & 92.5 & 3.1 & 3.1 & SF - Pes05\\
G30.688-0.071-1665A & 18 47 35.984 & -02 01 49.07 & -0.14$\pm$0.02 & 95.7 & 89.2 & 95.7 & 3.1 & 3.1 &\\
G30.688-0.071-1667A & 18 47 36.027 & -02 01 49.40 & -0.23$\pm$0.03 & 95.7 & 89.2 & 95.7 & 3.1 & 3.1 &\\
G30.688-0.071-1720A & 18 47 36.032 & -02 01 49.66 & 0.22$\pm$0.06 & 91.3 & 89.2 & 92.5 & 3.1 & 3.1 &\\
\\
G30.689-0.043-1612A & 18 47 29.965 & -02 00 56.24 & -0.15$\pm$0.02 & 91.7 & 85.7 & 91.9 & 3.1 & 3.2 & SF - Mot03\\
G30.689-0.043-1665A & 18 47 30.644 & -02 00 54.54 & -0.25$\pm$0.05 & 92.9 & 87.1 & 93.3 & 4.0 & 3.8 &\\
G30.689-0.043-1667A & 18 47 30.055 & -02 00 54.51 & -0.22$\pm$0.02 & 94.3 & 85.7 & 94.7 & 3.1 & 3.1 &\\
G30.689-0.043-1720A & 18 47 29.916 & -02 00 57.68 & 0.23$\pm$0.06 & 90.1 & 87.1 & 91.9 & 3.0  & 3.0 &\\
\\
G30.720-0.083-1612A & 18 47 41.742 & -02 00 22.39 & 0.16$\pm$0.03 & 93.3 & 92.5 & 93.9 & 1.3 & 1.3 & SF - Mot03\\
G30.720-0.083-1612B & 18 47 41.733 & -02 00 23.62 & 0.43$\pm$0.07 & 97.1 & 96.1 & 98.3 & 1.1 & 1.1&\\
G30.720-0.083-1665A & 18 47 41.652 & -02 00 22.27 & -0.22$\pm$0.01 & 95.7 & 89.7 & 96.2 & 1.3 & 1.3&\\
G30.720-0.083-1667A & 18 47 41.652 & -02 00 22.27 & -0.33$\pm$0.01 & 97.1 & 90.4 & 97.6 & 1.1 & 1.1&\\
G30.720-0.083-1720A & 18 47 41.652 & -02 00 22.27 & -0.25$\pm$0.01 & 93.9 & 90.8 & 93.9 & 1.1 & 1.1&\\
G30.720-0.083-1720B & 18 47 41.652 & -02 00 22.27 & -0.25$\pm$0.07 & 97.1 & 96.1 & 98.3 & 1.1 & 1.1&\\
\\
G30.735-0.056-1612A & 18 47 37.965 & -01 58 55.07 & -0.13$\pm$0.02 & 93.2 & 87.8 & 93.2 & 3.1 & 3.1 & SF - Ros10\\
G30.735-0.056-1665A & 18 47 38.242 & -01 59 06.62 & -0.21$\pm$0.02 & 94.3 & 89.2 & 94.8 & 3.8 & 3.8 &\\
G30.735-0.056-1667A & 18 47 38.613 & -01 58 46.99 & -0.15$\pm$0.01 & 94.2 & 89.9 & 94.8 & 2.6 & 2.6 &\\
G30.735-0.056-1720A & 18 47 38.607 & -01 58 55.94 & 0.07$\pm$0.01 & 91.5 & 87.8 & 93.2 & 3.6 & 3.6 &\\
\\
G30.742+0.006-1612A & 18 47 25.092 & -01 56 45.09 & -0.20$\pm$0.01 & 91.8 & 89.1 & 91.9 & 1.6 & 1.6 & SF - Mot03\\
G30.742+0.006-1665A & 18 47 25.585 & -01 56 21.90 & -0.10$\pm$0.02 & 82.3 & 81.0 & 82.8 & 4.6 & 4.6 &\\
G30.742+0.006-1665B & 18 47 25.457 & -01 56 34.41 & -0.24$\pm$0.02 & 91.8 & 89.8 & 93.3 & 2.6 & 2.6 &\\
G30.742+0.006-1667A & 18 47 25.451 & -01 56 43.35 & -0.19$\pm$0.01 & 82.3 & 81.0 & 82.8 & 3.3 & 3.3 &\\
G30.742+0.006-1667B & 18 47 25.329 & -01 56 46.91 & -0.30$\pm$0.01 & 91.6 & 89.8 & 93.8 & 1.9 & 1.9 &\\
G30.742+0.006-1720A & 18 47 25.261 & -01 56 47.37 & 0.22$\pm$0.01 & 90.6 & 89.8 & 91.2 & 1.6 & 1.6 &\\
\\
G30.752-0.057-1612A & 18 47 39.816 & -01 57 57.41 & 0.75$\pm$0.02 & 96.9 & 96.1 & 97.6 & 1.2 & 1.2 & SF - Cas95\\
G30.752-0.057-1665A & 18 47 38.534 & -01 57 47.95 & -0.23$\pm$0.03 & 97.8 & 88.1 & 98.3 & 3.2 & 3.1 &\\
G30.752-0.057-1667A & 18 47 38.656 & -01 57 44.39 & -0.45$\pm$0.02 & 96.4 & 89.8 & 96.9 & 2.3 & 2.3 &\\
G30.752-0.057-1720A & 18 47 39.843 & -01 57 53.51 & -0.21$\pm$0.02 & 97.3 & 93.9 & 97.6 & 3.2 & 3.2 &\\
\\
G30.815-0.053-1612A & 18 47 44.511 & -01 54 35.64 & -0.34$\pm$0.04 & 97.6 & 92.4 & 97.5 & 1.1 & 1.1 & SF - Cas95\\
G30.815-0.053-1665A & 18 47 44.986 & -01 54 39.29 & -0.50$\pm$0.03 & 98.5 & 92.0 & 98.5 & 1.2 & 1.2 &\\
G30.815-0.053-1667A & 18 47 44.748 & -01 54 37.47 & -0.64$\pm$0.03 & 99.9 & 92.0 & 100.0 & 1.1 & 1.1 &\\
G30.815-0.053-1720A & 18 47 44.870 & -01 54 33.91 & 0.47$\pm$0.03 & 96.3 & 92.0 & 100.0 & 1.1 & 1.1 &\\
\\
G30.854+0.151-1612A & 18 47 06.170 & -01 46 41.10 & -0.05$\pm$0.01 & 96.7 & 92.6 & 96.9 & 6.3 & 6.2 & SF - Hil05\\
G30.854+0.151-1665A & 18 47 06.280 & -01 46 55.43 & -0.11$\pm$0.01 & 96.4 & 92.6 & 96.9 & 2.3 & 2.3 &\\
G30.854+0.151-1667A & 18 47 06.523 & -01 46 48.32 & -0.11$\pm$0.01 & 96.4 & 92.6 & 96.9 & 2.2 & 2.2 &\\
\\
G31.388-0.382-1665A & 18 49 59.141 & -01 32 48.11 & -0.02$\pm$0.01 & -3.4 & -5.0 & -3.0 & 5.3 & 5.1 & U\\
G31.388-0.382-1665B & 18 49 58.781 & -01 32 49.84 & -0.02$\pm$0.01 & 18.1 & 17.2 & 19.7 & 2.3 & 2.1 &\\
G31.388-0.382-1667A & 18 49 58.776 & -01 32 58.79 & -0.05$\pm$0.01 & -3.3 & -6.0 & -3.0 & 4.2 & 4.2 &\\
G31.388-0.382-1667B & 18 49 58.897 & -01 32 55.22 & -0.10$\pm$0.01 & 18.2 & 17.2 & 19.7 & 2.3 & 2.2 &\\
\hline
\end{tabular}
\end{table*}

\begin{figure*}
\begin{tabular}{ccc}
\hspace{-1.5cm}
\begin{tabular}{c}
~~~~G29.937-0.052~~~~~~~~~~~~~~~~SF\\
\includegraphics[width=0.25\textwidth]{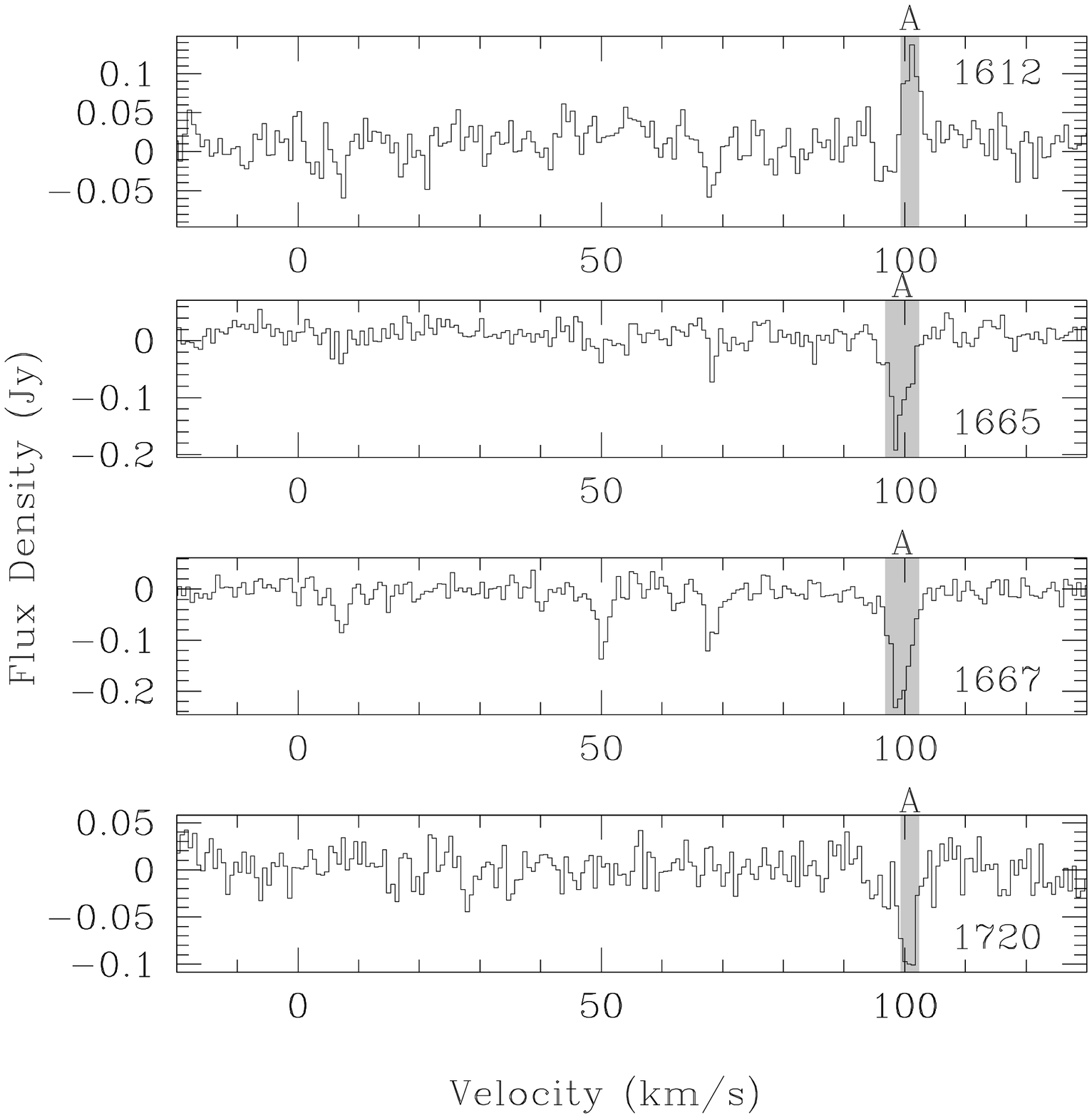}\\
\includegraphics[width=0.31\textwidth]{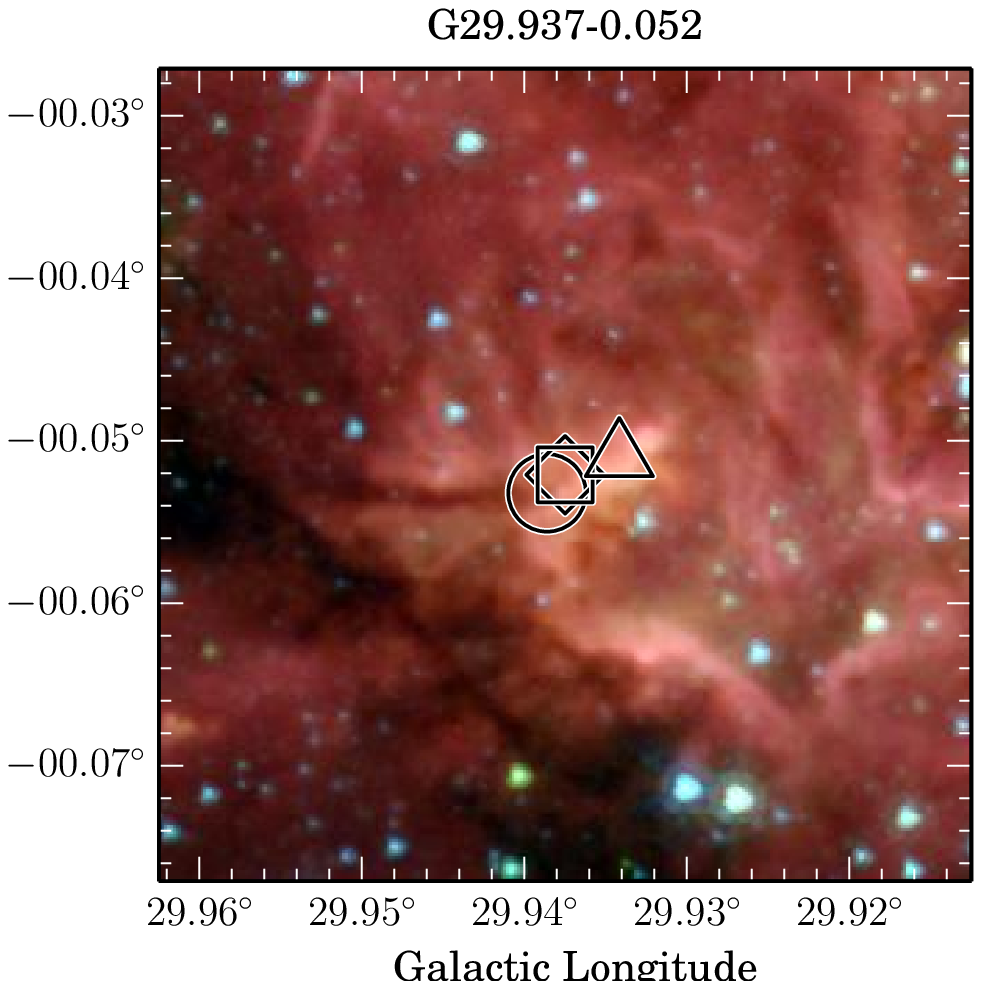}\\
\end{tabular}&
\hspace{-1.5cm}
\begin{tabular}{c}
~~~~G29.956-0.016~~~~~~~~~~~~~~~~SF\\
\includegraphics[width=0.25\textwidth]{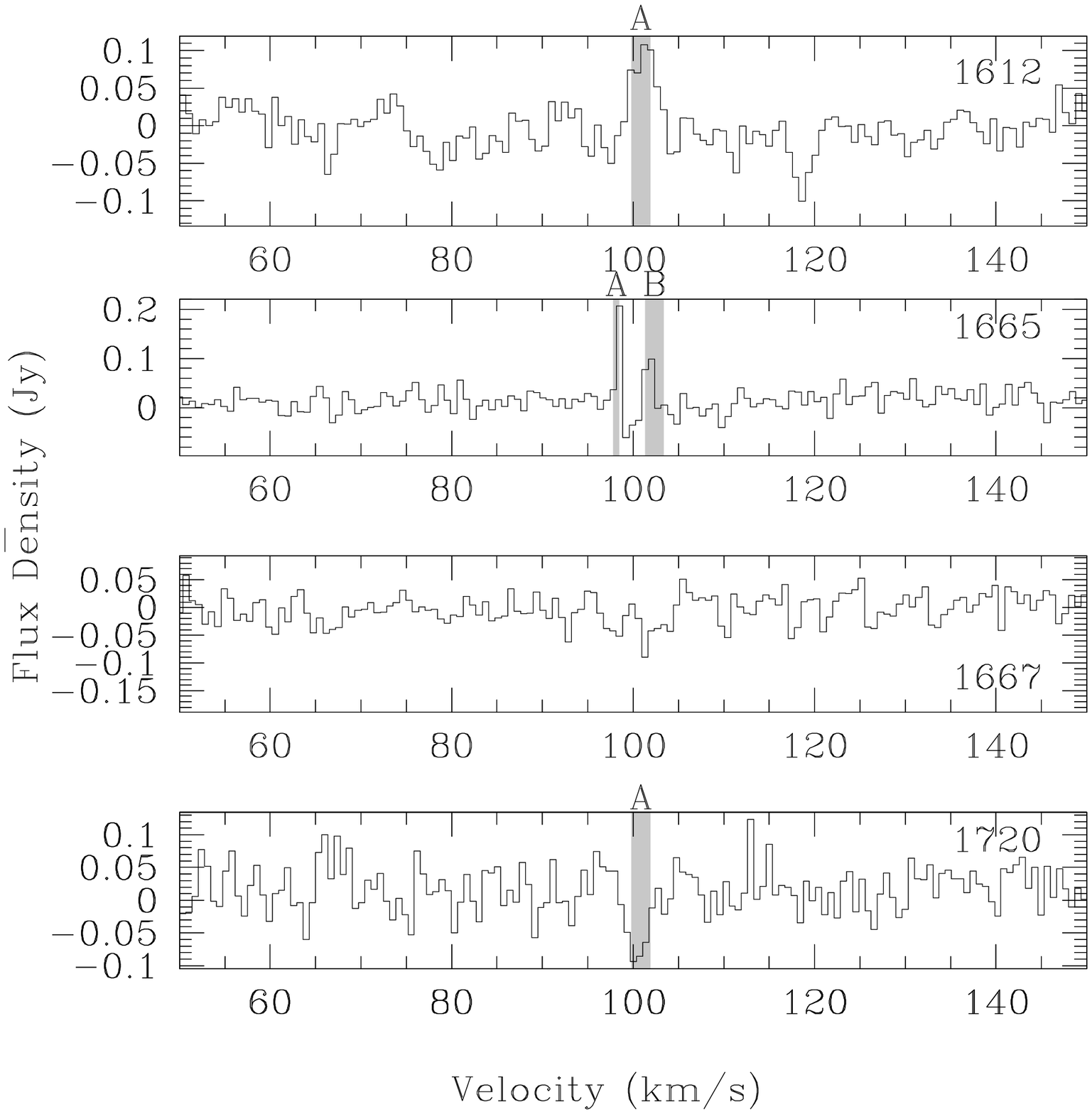}\\
\includegraphics[width=0.31\textwidth]{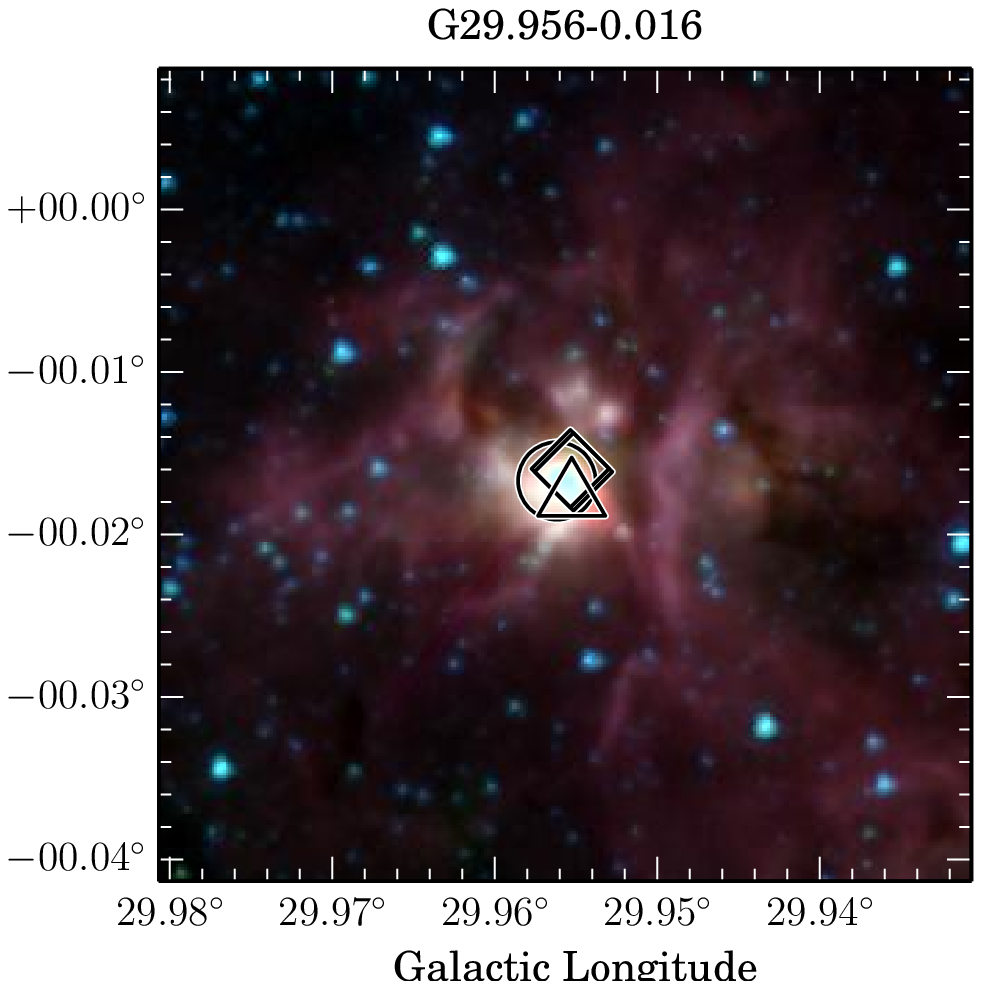}\\
\end{tabular}&
\hspace{-1.5cm}
\begin{tabular}{c}
~~~~G30.533+0.021~~~~~~~~~~~~~~~~SF\\
\includegraphics[width=0.25\textwidth]{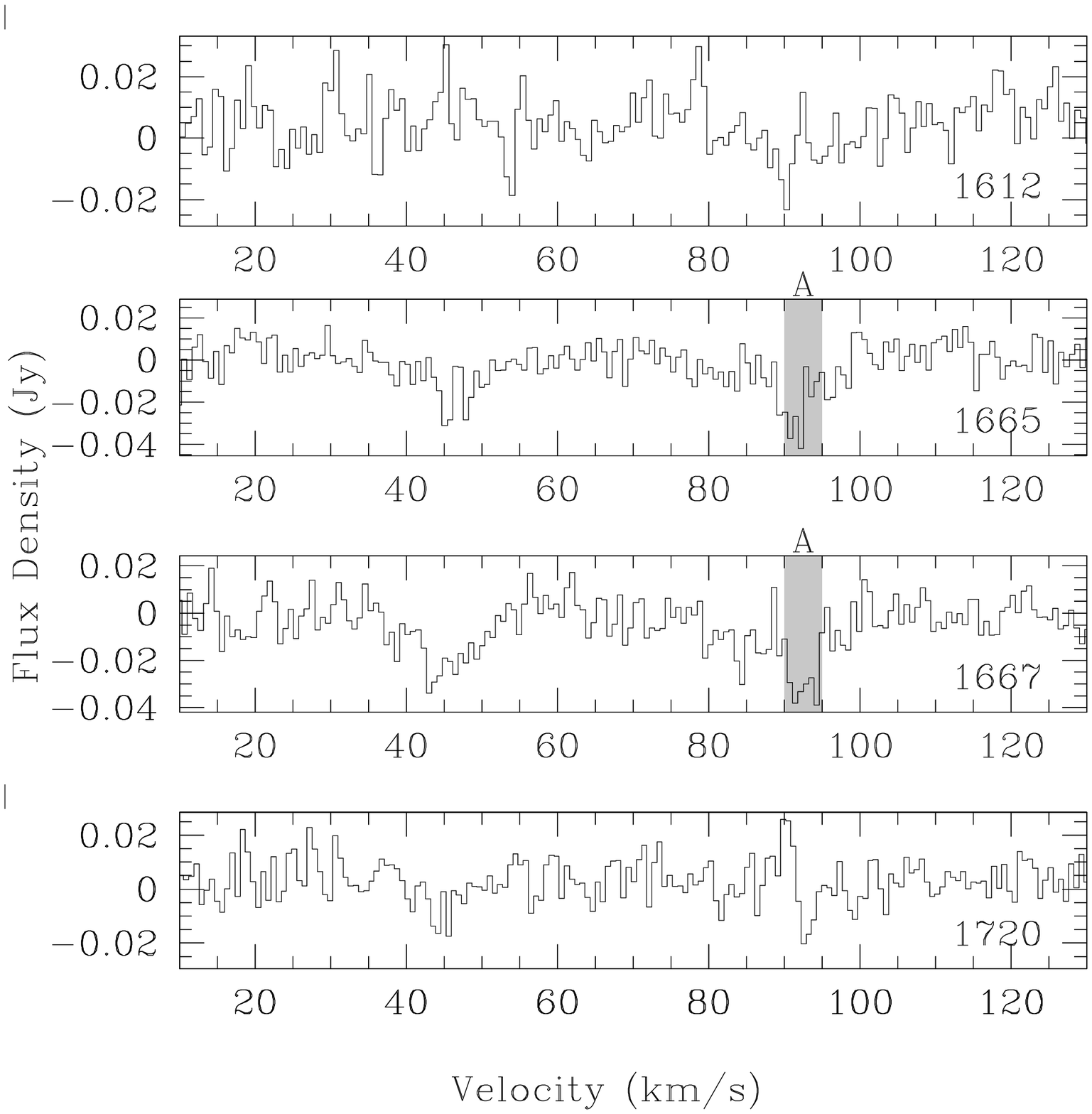}\\
\includegraphics[width=0.31\textwidth]{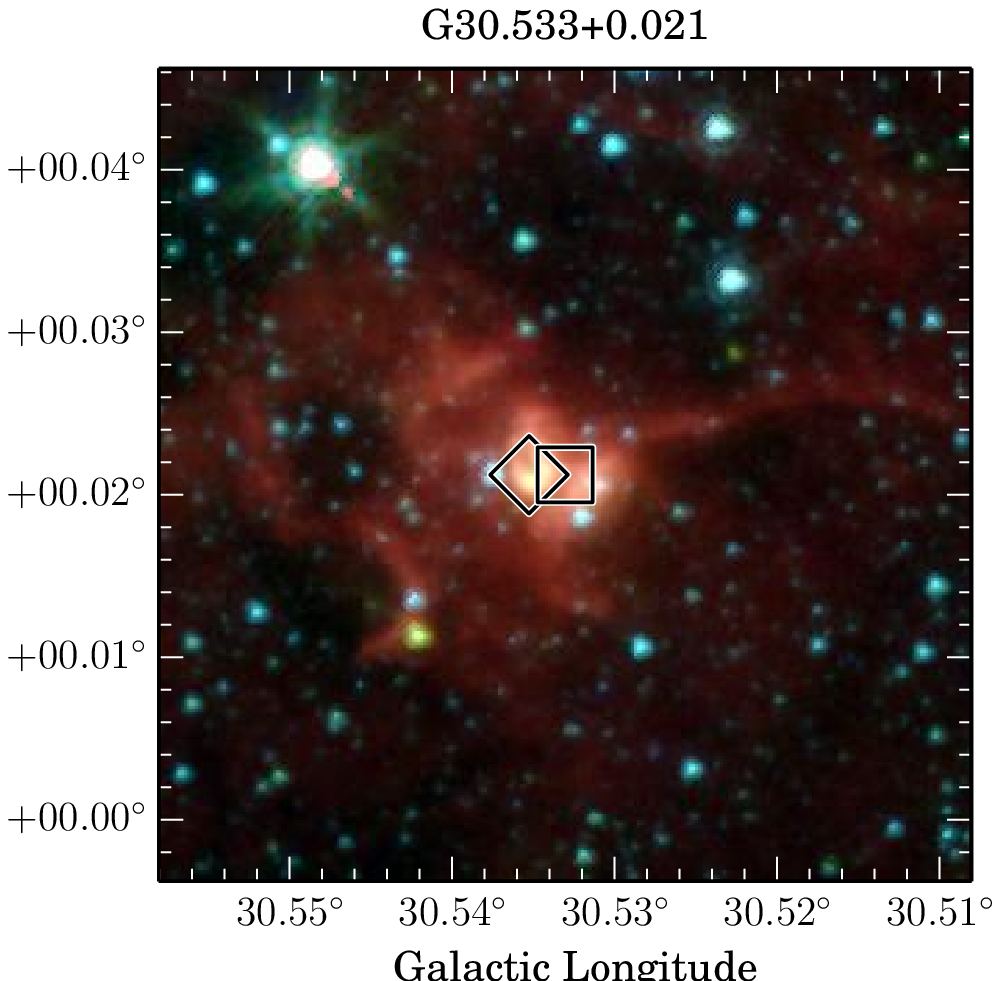}\\
\end{tabular}\\
\hspace{-1.5cm}
\begin{tabular}{c}
~~~~G30.688-0.071~~~~~~~~~~~~~~~~SF\\
\includegraphics[width=0.25\textwidth]{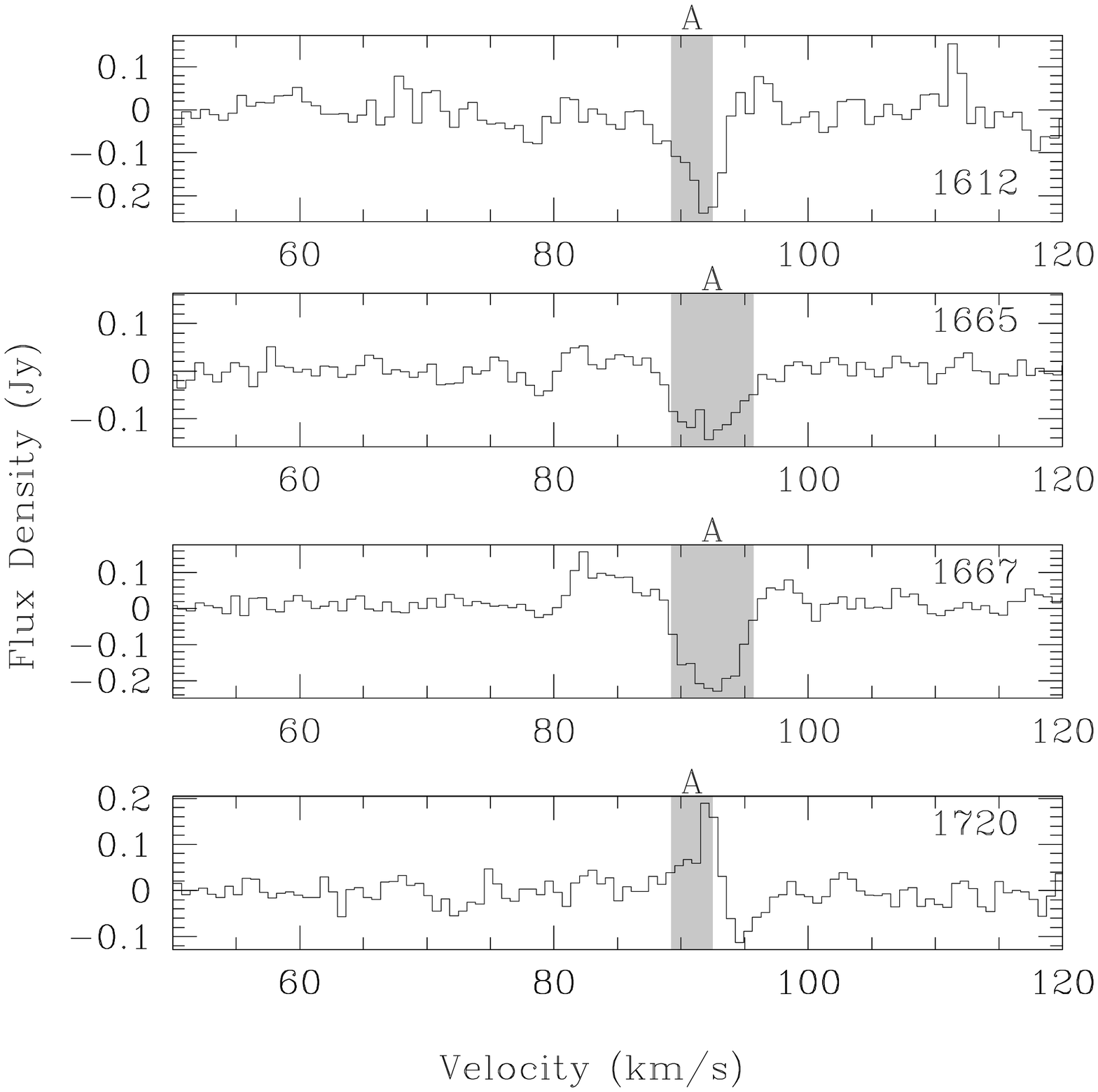}\\
\includegraphics[width=0.31\textwidth]{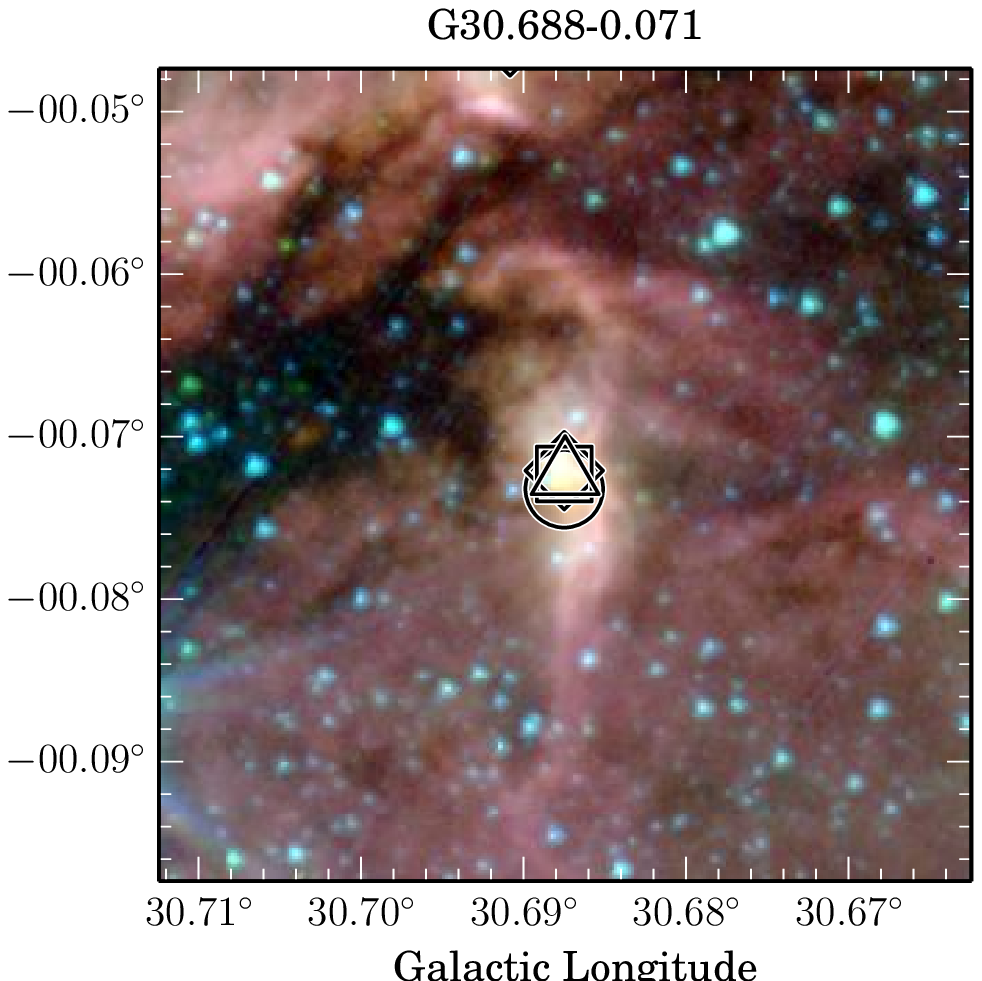}\\
\end{tabular}&
\hspace{-1.5cm}
\begin{tabular}{c}
~~~~G30.689-0.043~~~~~~~~~~~~~~~~SF\\
\includegraphics[width=0.25\textwidth]{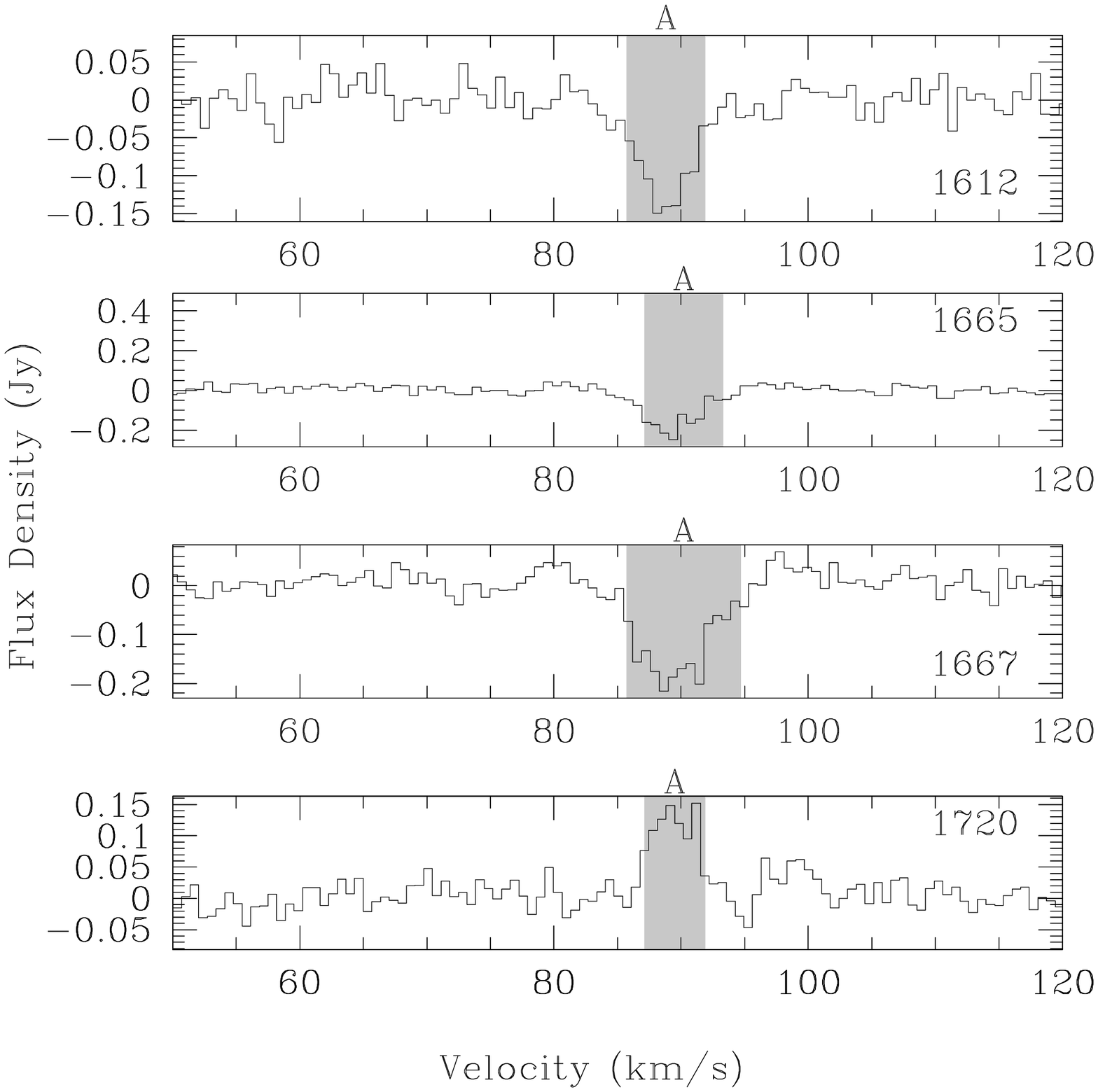}\\
\includegraphics[width=0.31\textwidth]{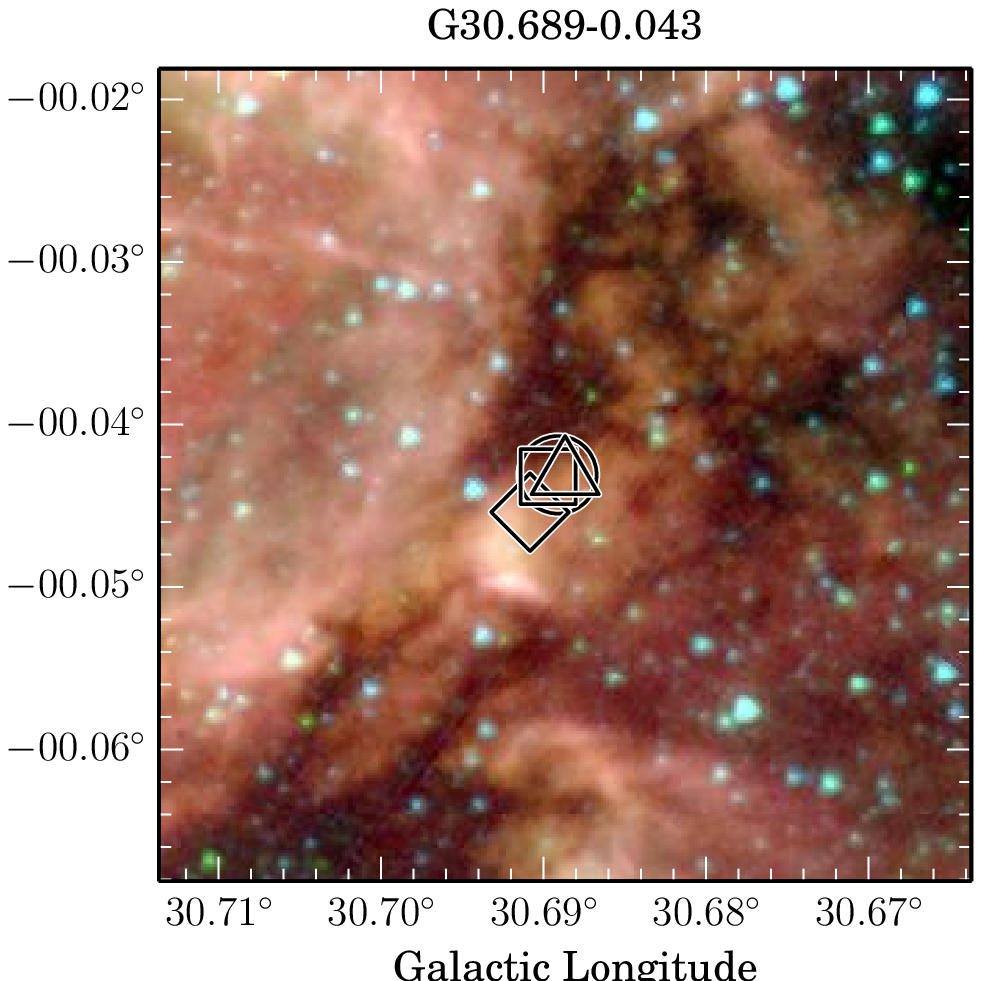}\\
\end{tabular}&
\hspace{-1.5cm}
\begin{tabular}{c}
~~~~G30.720-0.083~~~~~~~~~~~~~~~~SF\\
\includegraphics[width=0.25\textwidth]{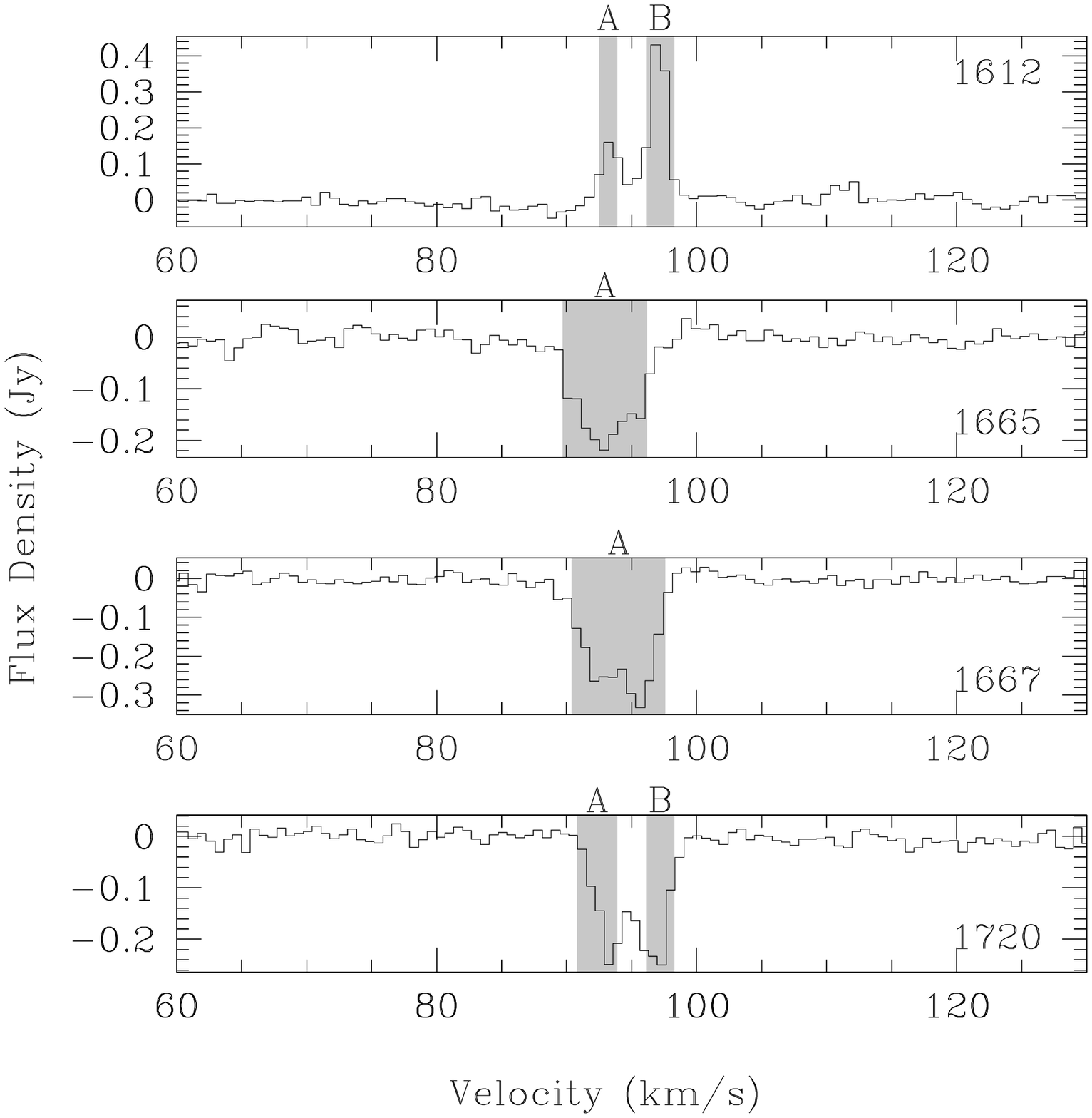}\\
\includegraphics[width=0.31\textwidth]{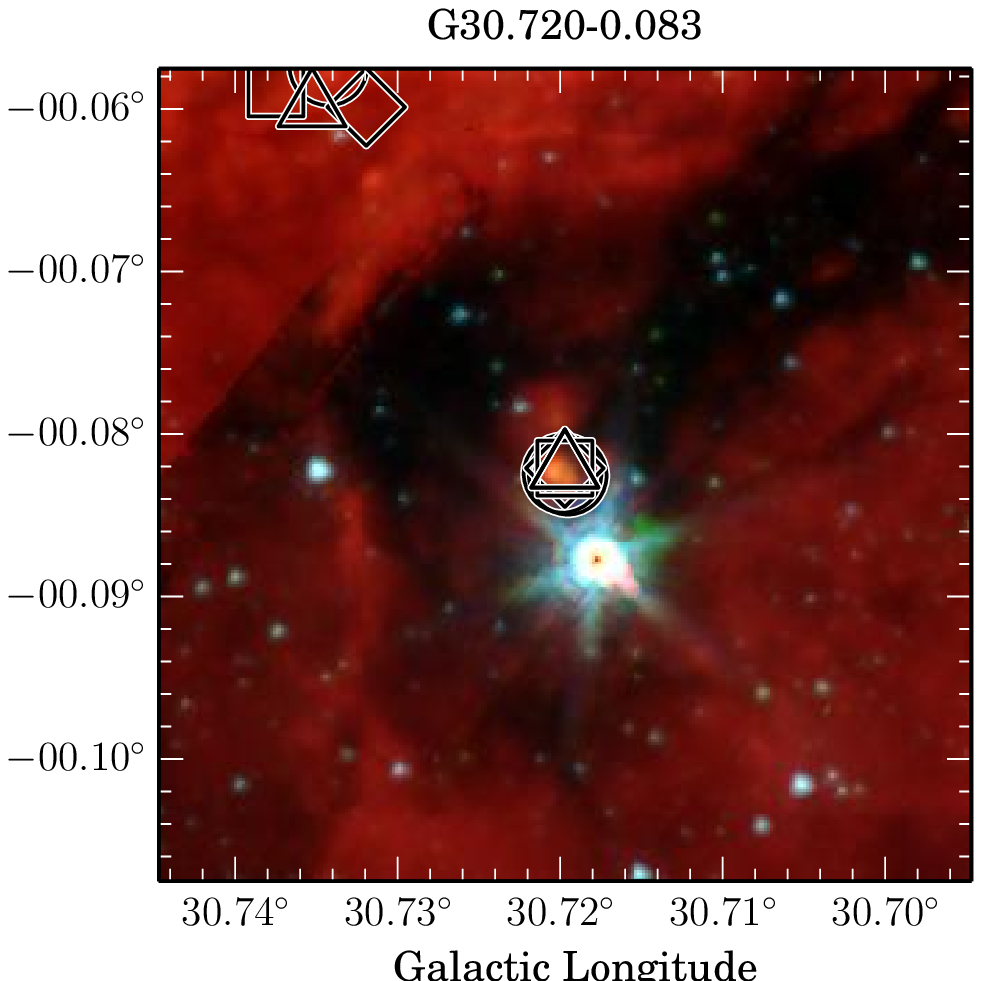}\\
\end{tabular}\\
\end{tabular}
\caption{Spectra and infrared images for sites of diffuse OH emission and
absorption. These sites are identified by the mirror-image 1612 and 1720\,MHz
OH spectra. GLIMPSE 3-colour images are
shown: blue = 3.5 $\mu$m, green = 4.5 $\mu$m and red = 8.0 $\mu$m.
Symbols represent the positions of strongest emission or absorption. 1612\,MHz
is shown as circles, 1665\,MHz as squares, 1667\,MHz as
diamonds and 1720\,MHz as triangles. The axes are in Galactic
coordinates. Designations for each region are shown in the upper-right
corner, where SF = star formation and U = unknown origin. The full Figure is available online.}
\label{diffuseOH}
\end{figure*}

\begin{figure*}
\begin{tabular}{ccc}
\hspace{-1.5cm}
\begin{tabular}{c}
~~~~G30.735-0.056~~~~~~~~~~~~~~~~SF\\
\includegraphics[width=0.25\textwidth]{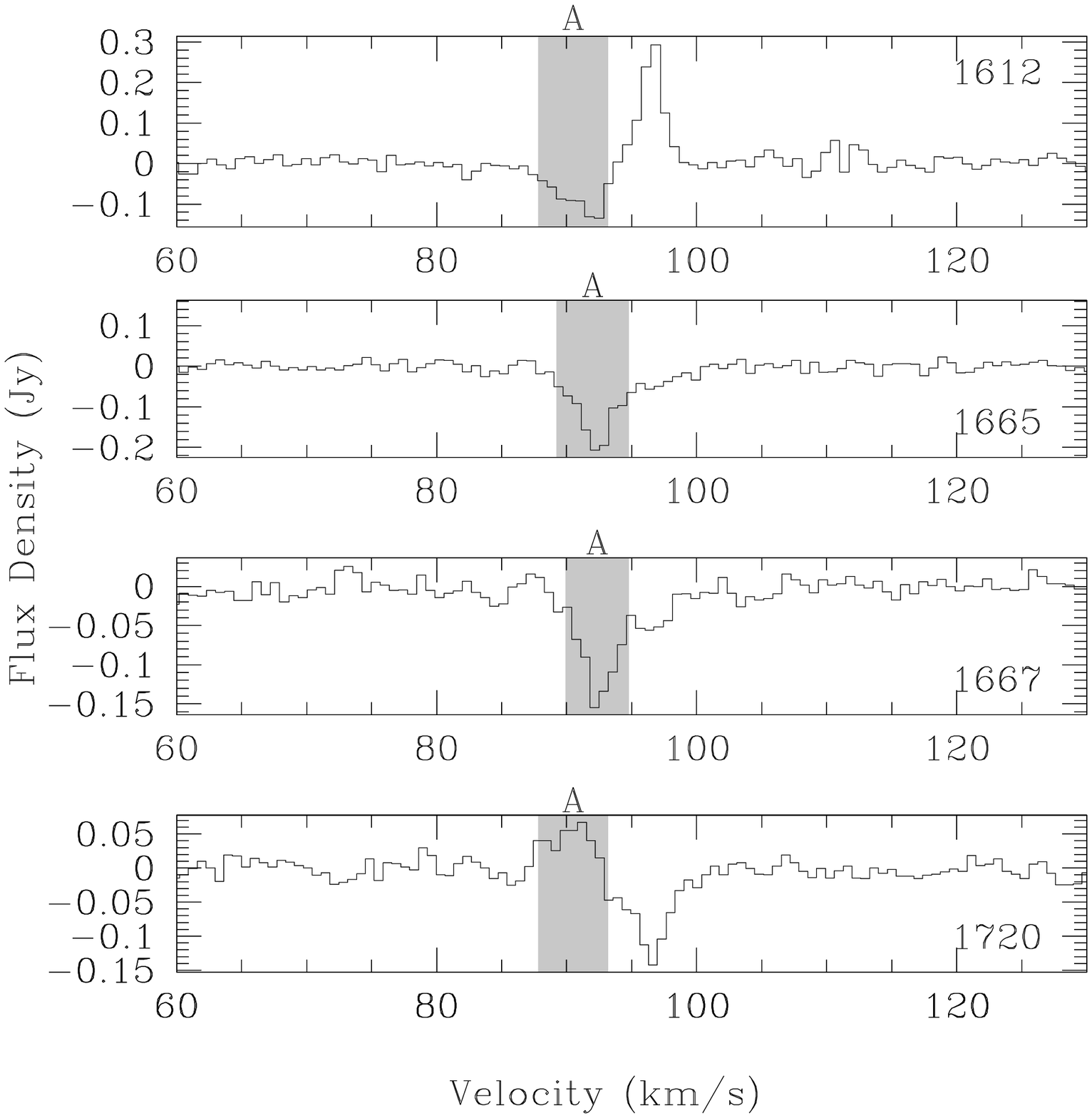}\\
\includegraphics[width=0.31\textwidth]{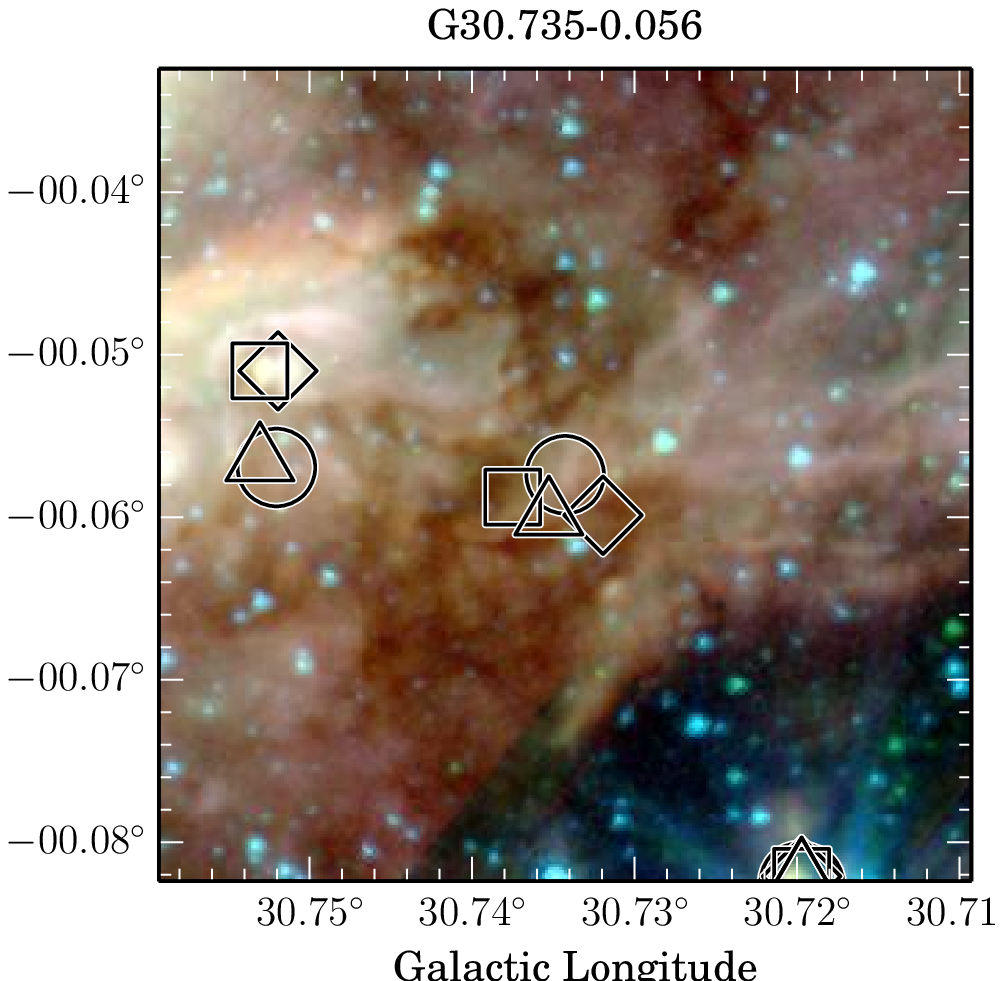}\\
\end{tabular}&
\hspace{-1.5cm}
\begin{tabular}{c}
~~~~G30.742+0.006~~~~~~~~~~~~~~~~SF\\
\includegraphics[width=0.25\textwidth]{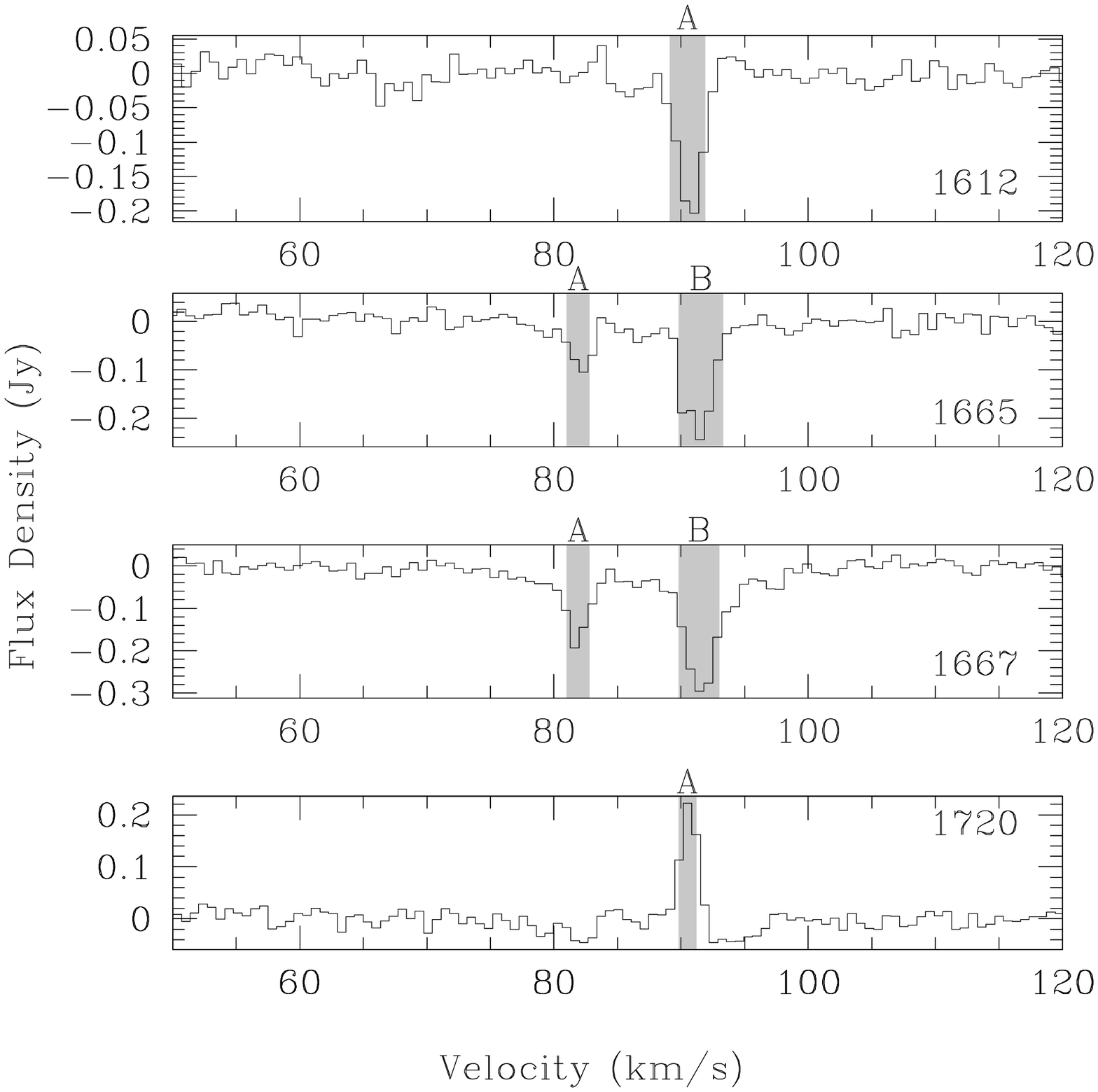}\\
\includegraphics[width=0.31\textwidth]{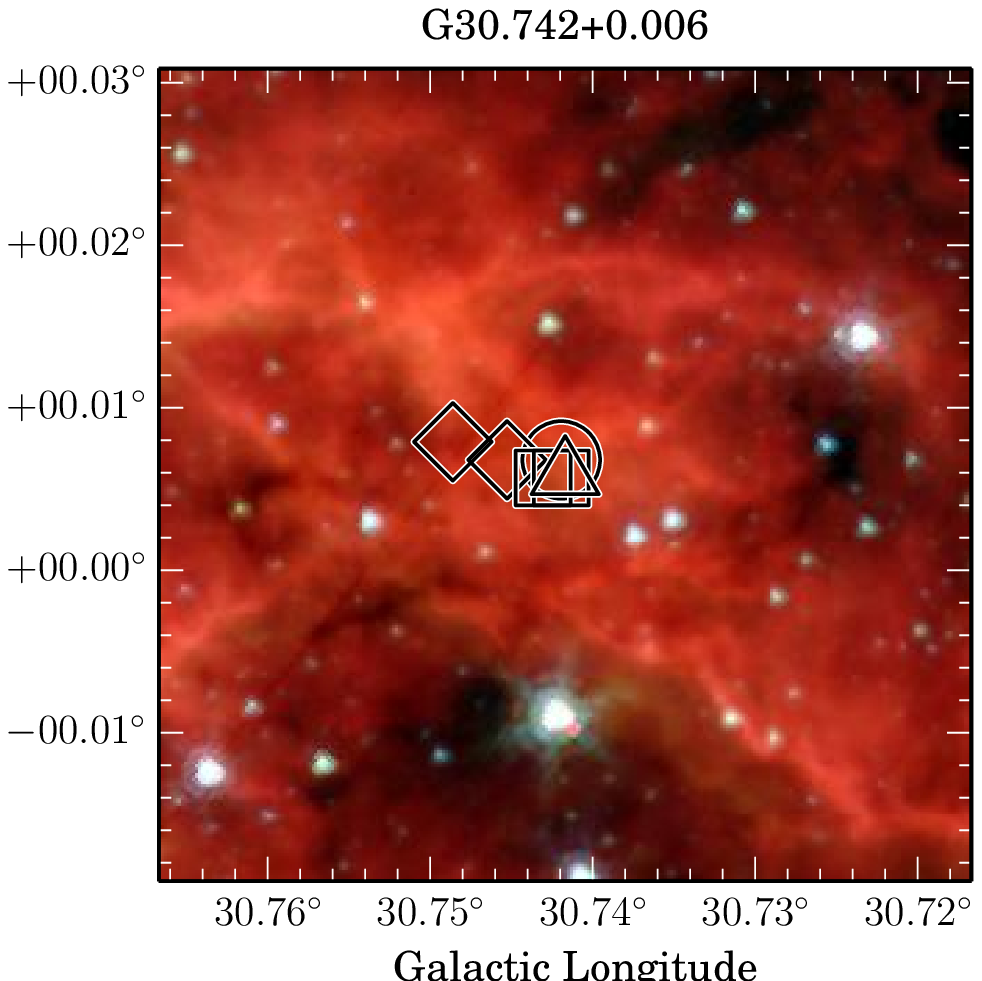}\\
\end{tabular}&
\hspace{-1.5cm}
\begin{tabular}{c}
~~~~G30.752-0.057~~~~~~~~~~~~~~~~SF\\
\includegraphics[width=0.25\textwidth]{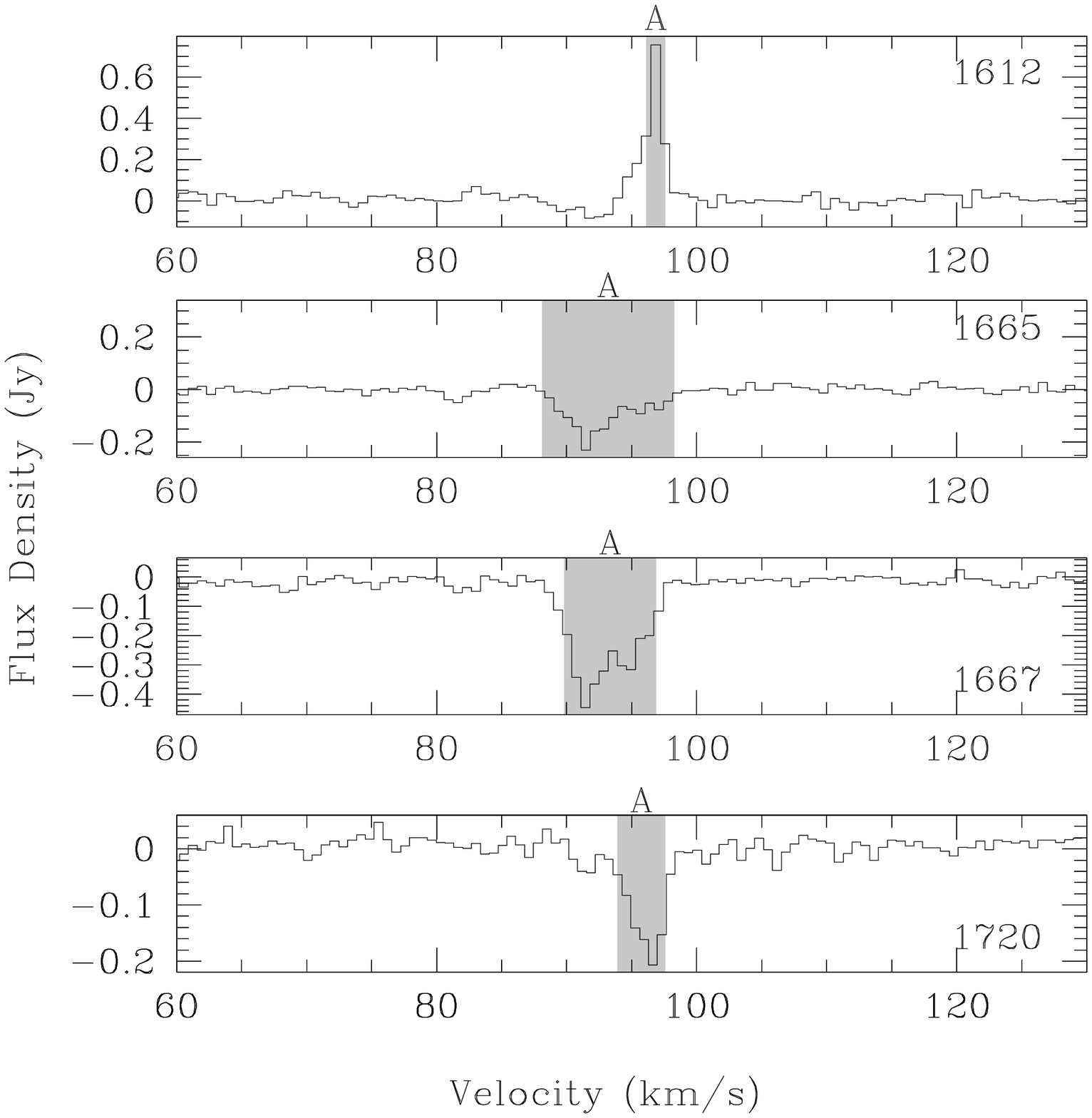}\\
\includegraphics[width=0.31\textwidth]{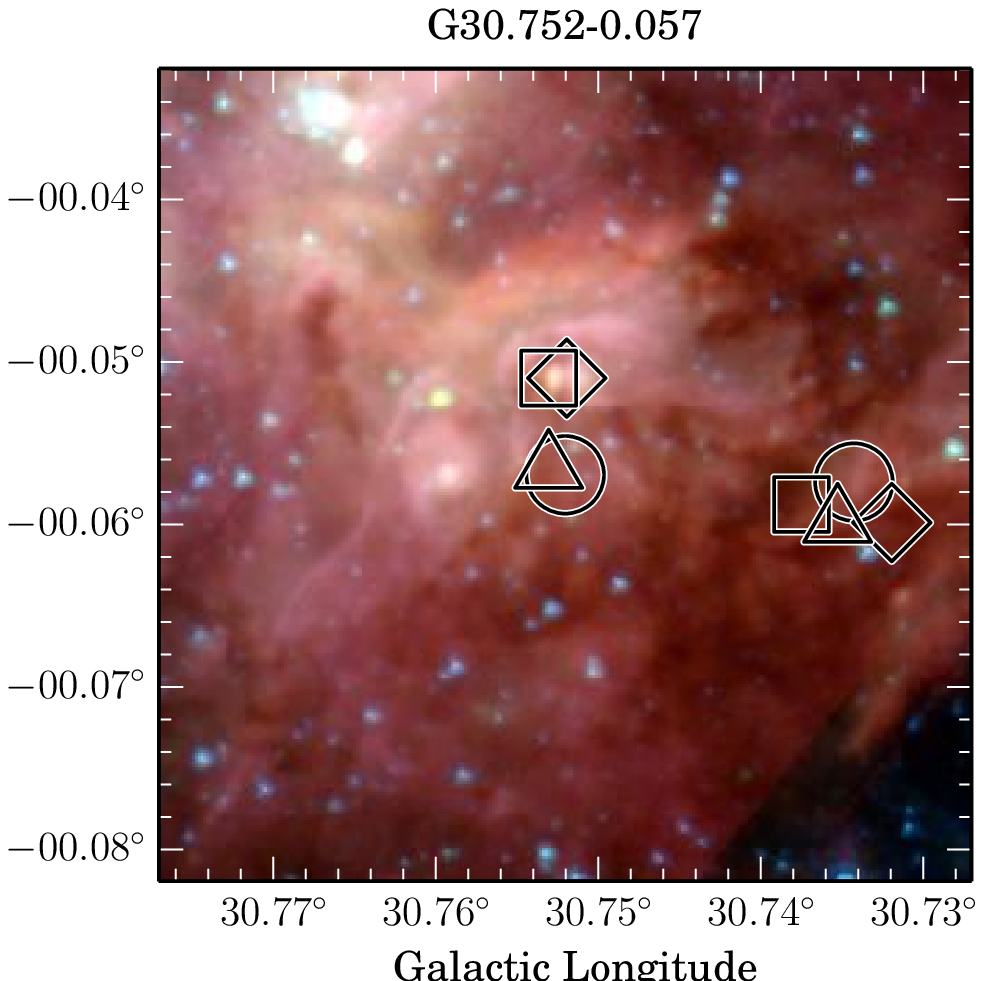}\\
\end{tabular}\\
\hspace{-1.5cm}
\begin{tabular}{c}
~~~~G30.815-0.053~~~~~~~~~~~~~~~~SF\\
\includegraphics[width=0.25\textwidth]{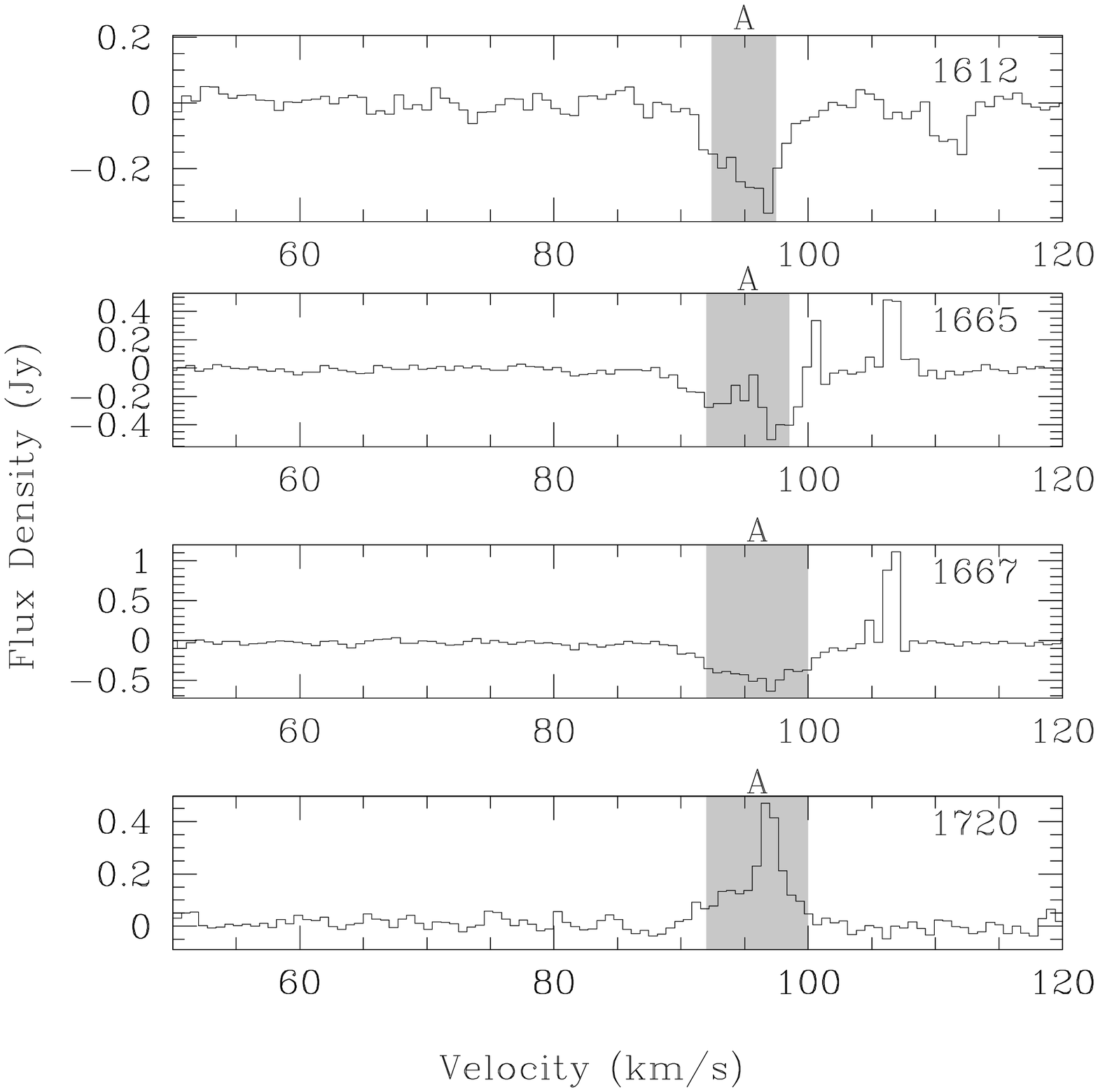}\\
\includegraphics[width=0.31\textwidth]{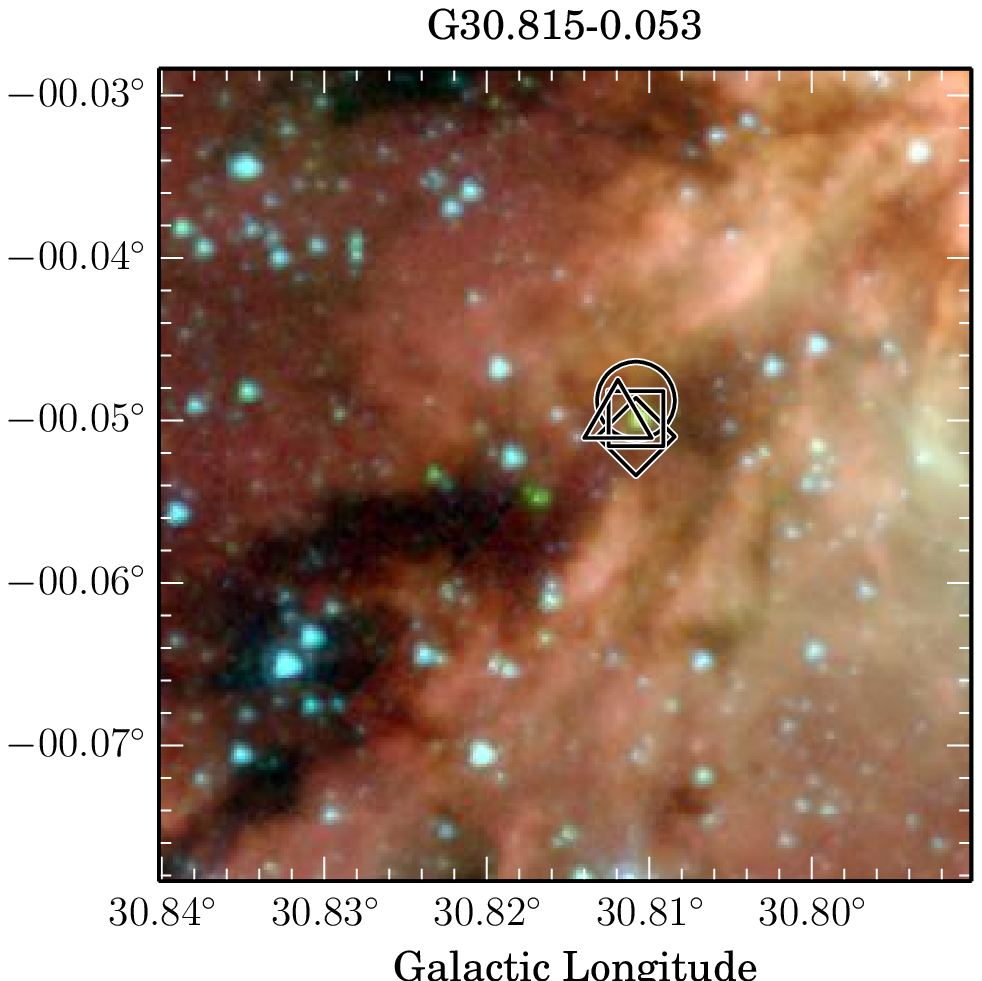}\\
\end{tabular}&
\hspace{-1.5cm}
\begin{tabular}{c}
~~~~G30.854+0.151~~~~~~~~~~~~~~~~SF\\
\includegraphics[width=0.25\textwidth]{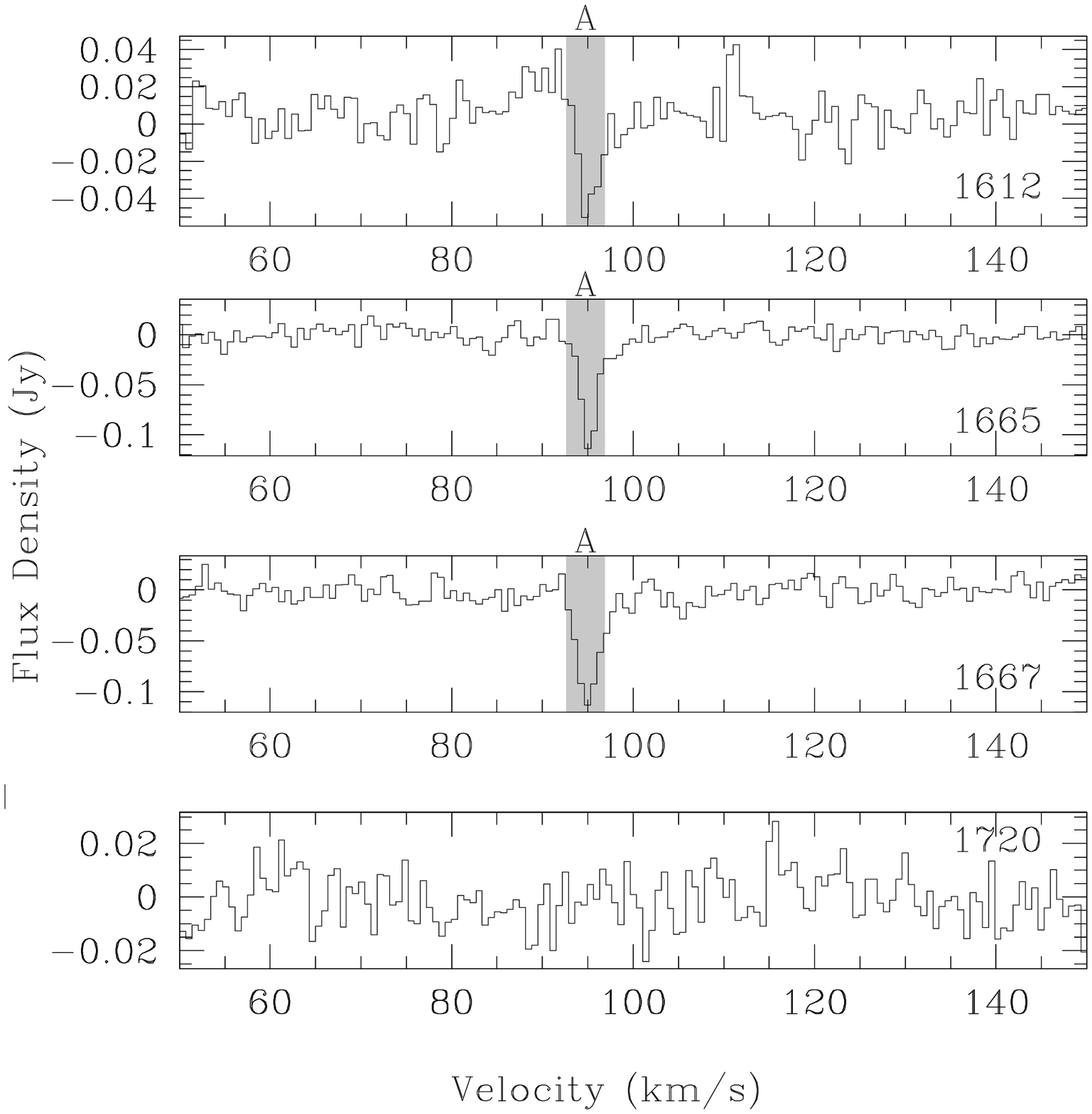}\\
\includegraphics[width=0.31\textwidth]{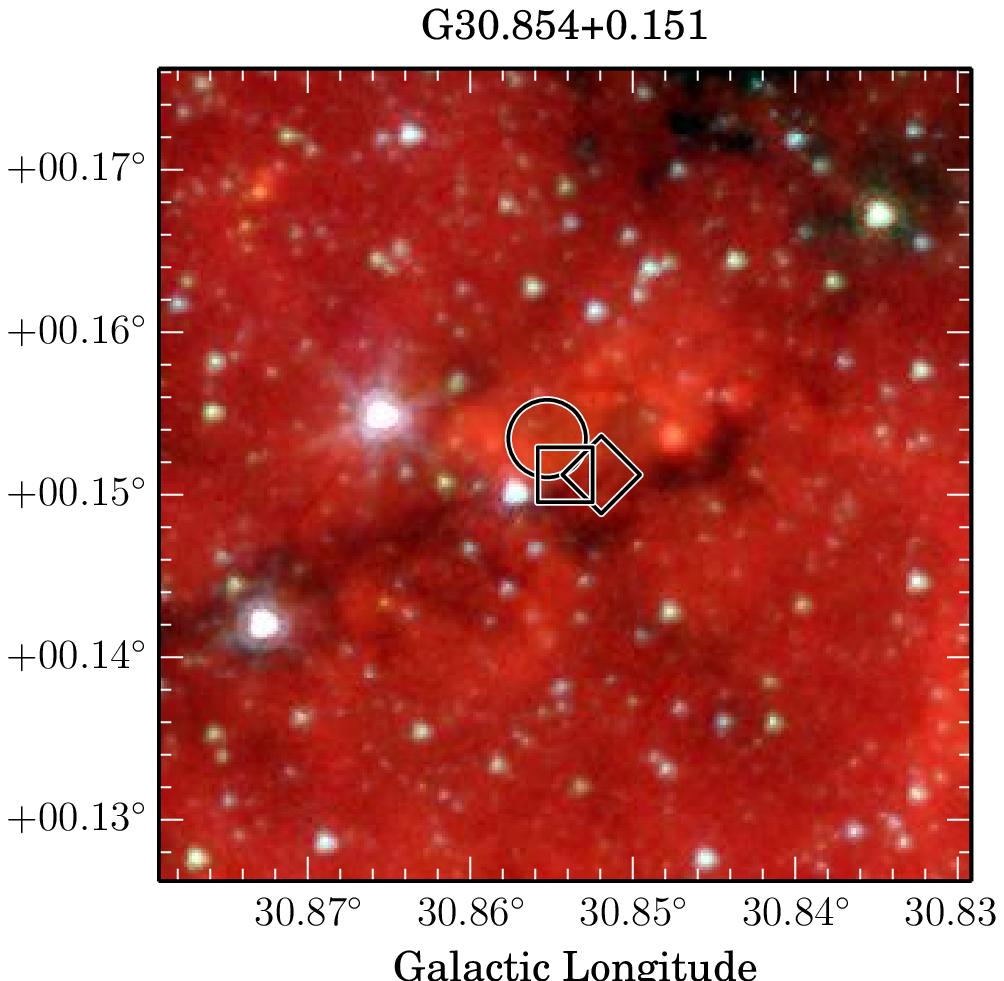}\\
\end{tabular}&
\hspace{-1.5cm}
\begin{tabular}{c}
~~~~G31.388-0.382~~~~~~~~~~~~~~~~~U\\
\includegraphics[width=0.25\textwidth]{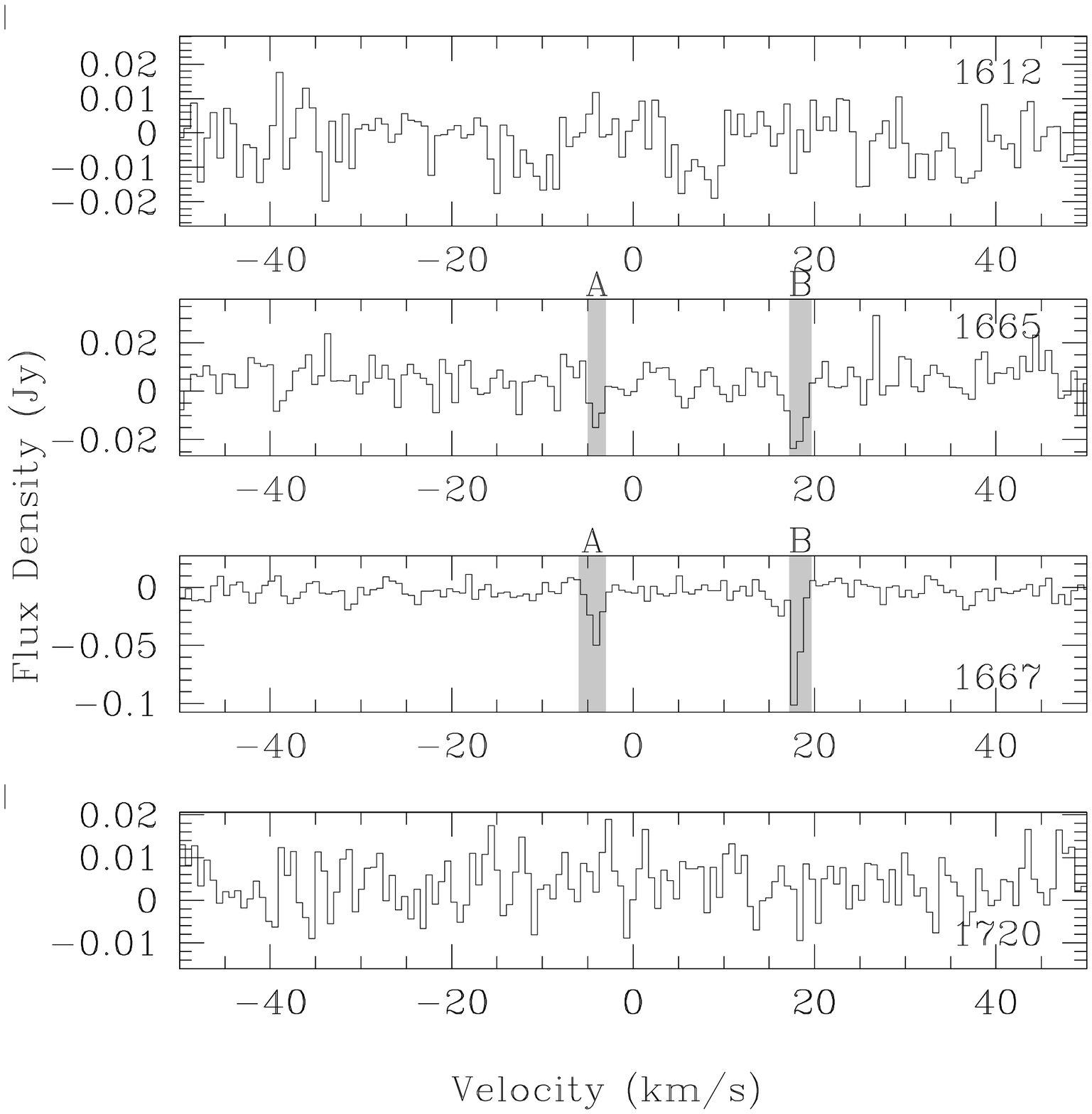}\\
\includegraphics[width=0.31\textwidth]{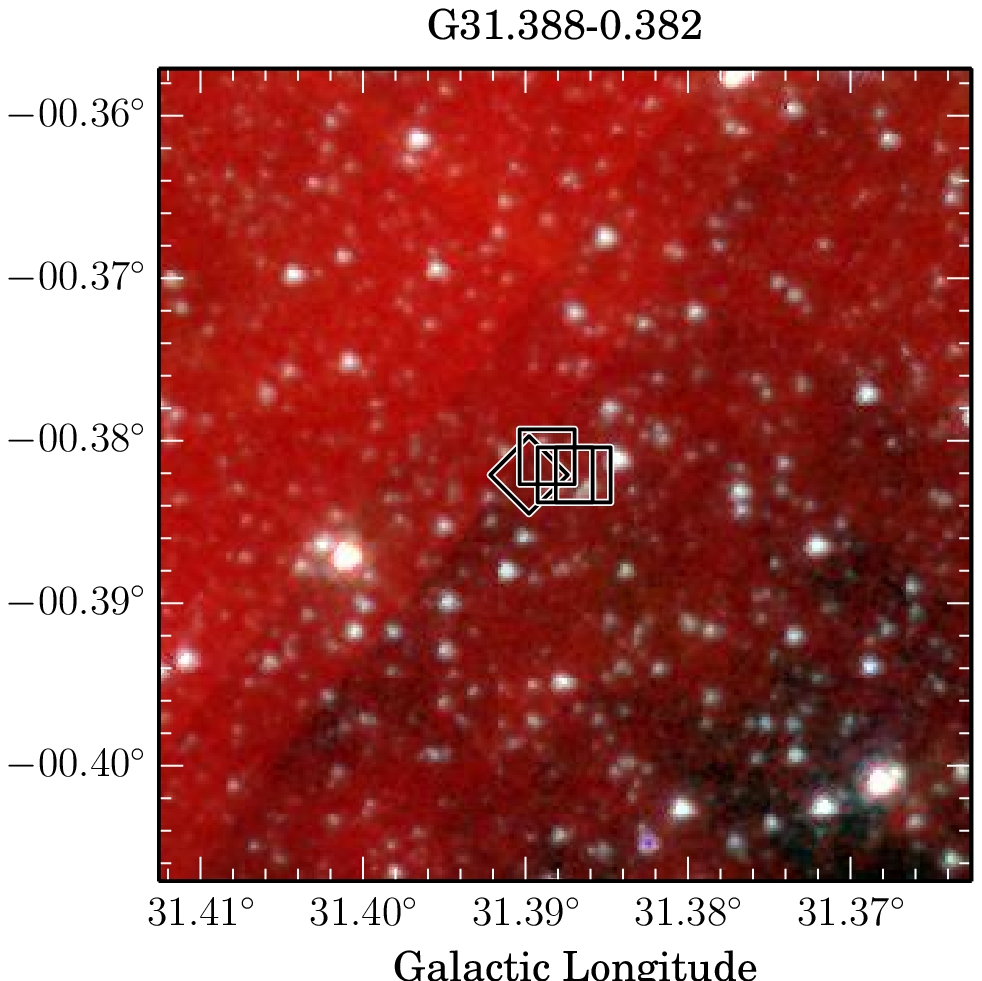}\\
\end{tabular}\\
\end{tabular}
\contcaption{}
\end{figure*}

During our search of OH masers, we identified 12 sites showing diffuse OH.
The details of the strongest emission and absorption features are given in
Table \ref{diffuseOHtab} and their line profiles and positions
are shown in Figure \ref{diffuseOH}. Of the 12 diffuse OH sites, all but three
(G30.533+0.021, G30.854+0.151 and G31.388-0.382) show main-line absorption and
well-detected satellite line features that are approximate mirror images of
each other. The remaining sites also show main-line absorption, but with weak
or absent detections in one or both of the satellites.\\

We note that the intensity ratios of the main-lines vary somewhat, with many
showing departures from the 5:9 ratio (1665:1667) expected from an optically thin
gas in LTE. This behaviour is expected to arise either from optical depth
effects or from a pattern of excitation temperatures that strongly
departs from thermal, with the actual line strengths also set by the difference
between these excitation temperatures and the continuum background
\cite[see also][]{dawson14}.\\

We also note that some of the spectral features in Figure \ref{diffuseOH} are
likely due to the blending of strong maser emission with diffuse OH features. 
For example, there are narrow emission features in both
the 1665 and 1667\,MHz spectra of G30.815-0.053 that do not have any
corresponding features in the satellite lines, suggesting that they are the
result of high-gain maser action from a localised parcel (or parcels) of gas
with strong velocity coherence, and not from the wider diffuse medium.\\

As shown in Table \ref{diffuseOHtab}, all but one of the diffuse OH detections
are co-spatial with sites of star formation, 
which may host the molecular gas that gives rise to the diffuse OH,
and may also provide the
bright continuum background against which the features may be clearly seen.
Indeed, most of the diffuse OH is associated with the well known star forming
region W43 Main\footnote{W43 Main is a small region of active star formation at
the centre of the W43 complex}
and only the features of G31.388-0.382 are not associated with an
established star formation site.\\

%\subsection{OH absorption}

%In the previous subsection, we identified a number of regions where we find
%both diffuse OH emission and absorption. Most of these regions are within W43.
%Towards W43, in all spectral lines, we find extended OH absorption against the
%continuum emission from W43. Figure \ref{absfig} shows the morphology of the 
%1667\,MHz OH absorption in this region. The Figure was produced by inverting
%the 1667 cube and then producing a peak temperature map (moment -2 map in {\sc
%miriad}). This way, we produce a sensitive map to wherever absorption is
%located and the intensity of the map corresponds to the depth of the absorption.

%The extended absorption towards W43 appears to be the only place in the THOR pilot
%region where we detect significantly extended OH absorption. The OH absorption
%sites reported outside of W43 in Table \ref{diffuseOHtab} and Figure
%\ref{diffuseOH} are all either spatially unresolved or partially resolved
%regions.

\begin{figure}
\includegraphics[width=0.45\textwidth]{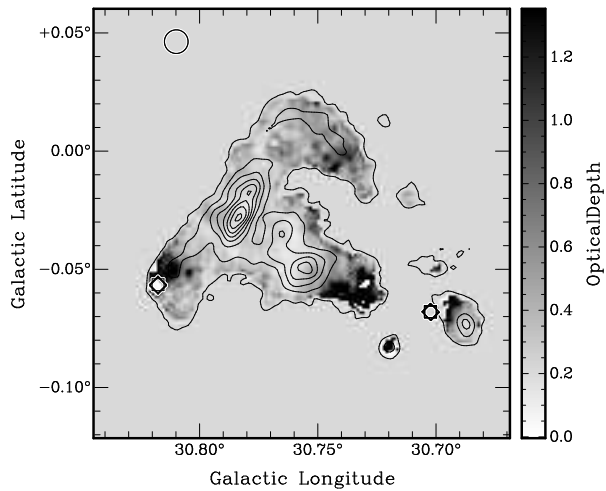}
\caption{Extended OH absorption towards W43 Main. The OH absorption image (shown in
greyscale) was created using the 1667\,MHz OH data.
This map shows the distribution of maximum optical depth at
each position. The map is masked to only include areas where the continuum
level is above 80\,mJy. The contours represent the continuum level, with the
lowest contour being 50\,mJy and contours increasing by 200\,mJy up to a maximum
of 1.45\,Jy. Three OH masers are found within this region. There is one 1612\,MHz
maser that appears outside the continuum contours (circle -- G30.810+0.047).
There are also two maser sites showing both 1665 and 1667\,MHz masers
(squares and diamonds -- G30.702-0.068 and G30.818-0.057) that occur close
to the extended absorption.}
\label{absfig}
\end{figure}

Figure \ref{absfig} shows the morphology of the 1667\,MHz OH absorption in the
W43 Main region, in which extended OH absorption is seen against the
continuum emission from W43 Main. The Figure shows the maximum optical depth
measured at each position. This is calculated from a data cube that contains
both the 1667\,MHz transition, as well as continuum emission (ie. not continuum
subtracted). Maximum optical depth is calculated from the velocity channel that
shows the maximum difference between the continuum level (measured at a
velocity where there is no OH emission or absorption) and the level at the
deepest absorption feature in the spectrum.\\

We note that the morphology of the optical depth does not match that of the
continuum emission. This is expected behaviour because the continuum
emission represents the morphology of W43 Main, but the OH that absorbs
the continuum originates in a cloud that does not
necessarily have the same morphology as the continuum emission. It is not clear if
this cloud is an unrelated foreground object, or related to the front envelope of
W43 Main. Given that the velocity of the absorption is close to that
expected of W43 Main, we consider it more likely that this is part of a foreground
envelope.\\

Most of the region with significant OH absorption shows optical
depths around 0.2, indicating that the OH is generally optically thin gas.
There are, however, small areas where the optical depth is significantly higher.
From this we assume that the OH gas is likely to be optically thin in all but
the densest regions, closely linked to star formation centres, like W43 Main.\\

The extended absorption towards W43 Main appears to be the only place in
the THOR pilot region where we detect significantly extended OH absorption; the
other detections in Table \ref{diffuseOHtab} and Figure \ref{diffuseOH} are all
either spatially unresolved or partially resolved regions. It is expected
to see only extended OH absorption towards W43 Main because this is the only region
within the field of view which shows strong, extended continuum emission and 
such continuum emission is required in order to see the OH absorption.
The optical depth in these unresolved
regions again appears to be about 0.2, indicating that optically thin OH is
typical.\\

\section{Conclusions}
In the THOR pilot region, we have conducted a search for OH emission and
absorption. We have adapted a previous simple search method to identify
sites of strong emission and successfully applied it to identify
maser sites. We found strong maser emission is common throughout the field,
with 1612\,MHz masers being the most common and 1720\,MHz masers being
the rarest. In all, we found 103 sites of strong maser emission within the
pilot survey region. \\

By comparing our results with previous OH maser observations, we find that
main-line OH masers occasionally show mild variability up to a factor of a
few over a period of years. OH masers in the 1612\,MHz satellite line also
show occasional variability on similar timescales, but we also identify
a small number of these maser that show strong variability with order of
magnitude changes in intensity over a period of a few years.\\

We have compared the positions of the maser sites to infrared images, as
well as the published literature, in order to understand the origins of the
masers. We found that three out of four 1720\,MHz maser sites are associated
with sites of star formation. There are 72 1612\,MHz maser sites, of which
65 per cent are associated with evolved stars and 11 per cent are associated
with star formation. There are 42 maser sites with main-line emission
(either 1665 or 1667\,MHz), of which 50 per cent are associated with star
formation and 17 per cent are associated with evolved stars.\\

The maser site G29.574+0.118 appears to be located close (within the
plane of the sky) to a candidate magnetar. However, we are not able to determine
if there is a physical link between the maser site and candidate magnetar.
We find that the maser emission from G31.128-1.015 is detected over the
entirety of a very large velocity range of 145.3\,\kms.\\

We also detect diffuse OH emission and absorption against locations of bright
background continuum. Extended OH absorption is seen towards the well known
star-forming region W43 Main, while unresolved or partially resolved detections are
seen elsewhere. The OH gas generally appears to be optically thin.
The diffuse OH line profiles show a characteristic pattern:
main lines are seen in absorption whereas the satellite-line profiles are
approximate mirror images in emission and absorption.  \\

We look forward to the full THOR survey, which will identify many more OH
masers and provide us with more robust statistics on their relative occurrence
in star forming regions and evolved stars. The
full survey will also identify more sites where we find diffuse OH absorption
and/or emission. Combining with the results of SPLASH, such a greatly expanded
dataset can be used to directly assess the OH maser population in the Galaxy,
as well as help identify a well-defined sequence for high-mass star formation.\\

\section{Acknowledgements}
We would like to thank the anonymous referee for helpful and insightful comments
that have greatly improved the quality of this paper.
This research has made use of the SIMBAD database, operated at CDS, Strasbourg,
France. The National Radio Astronomy Observatory is a facility of the National
Science Foundation operated under cooperative agreement by Associated
Universities, Inc. RSK acknowledges financial support by the Deutsche
Forschungsgemeinschaft (DFG) via the Heidelberg Sonderforschungsbereich
SFB 881 “The Milky Way System” (subprojects B1, B2, and B8) as well as the
Schwerpunktprogramm SPP 1573 "Physics of the Interstellar Medium". RSK
also thanks for funding from the European Research Council under
the European Community’s Seventh Framework Programme (FP7/2007-2013) via the
ERC Advanced Grant STARLIGHT (project number 339177). This research made use
of the Duchamp source finder, produced at the Australia Telescope National
Facility, CSIRO, by M. Whiting.

\label{lastpage}
\end{document}